\documentclass[aps,twocolumn,pra,floatfix]{revtex4}
\usepackage{graphicx}
\usepackage{amsfonts}
\usepackage{amssymb}
\usepackage{amsmath}
\usepackage{wasysym}
\usepackage{color}

\def\wfig{4.25cm}
\def\wfigt{4.00cm}
\def\wfigtwo{6.00cm}
\def\wfigthree{7.5cm}

\def\rootfigsmall{./}
\def\rootfig{./}

\begin{document}

\title{
Radially Symmetric Nonlinear States of Harmonically Trapped \\ 
Bose-Einstein Condensates}

\author{G. Herring$^{1}$,
L. D. Carr$^2$,
R. Carretero-Gonz{\'a}lez$^3$,
P. G.\ Kevrekidis$^{1}$,
and D. J. Frantzeskakis$^{4}$.
}
\affiliation{
$^{1}$ Department of Mathematics and Statistics, University of Massachusetts, Amherst MA 01003-4515, USA \\
$^{2}$ Department of Physics, Colorado School of Mines, Golden, Colorado 80401,
USA \\
$^3$ Nonlinear Dynamical Systems Group\footnote{%
URL: \texttt{http://nlds.sdsu.edu/}}, Department of Mathematics and
Statistics, and Computational Science Research Center\footnote{%
URL: \texttt{http://www.csrc.sdsu.edu/}}, San Diego State University, San
Diego CA, 92182-7720, USA \\
$^{4}$ Department of Physics, University of Athens, Panepistimiopolis, Zografos, Athens 15784, Greece
}

\date{To appear in {\em Phys.~Rev.~A}}

\begin{abstract}
Starting from the spectrum of the radially symmetric quantum
harmonic oscillator in two dimensions, we create a large set of nonlinear
solutions. The relevant three principal branches, with $n_r=0,1$ and $2$
radial nodes respectively, are systematically
continued as a function
of the chemical potential and their linear
stability is analyzed in   detail, in the
absence as well as in the presence of topological charge $m$, i.e., vorticity.
It is found that for
repulsive interatomic interactions {\it only} the ground
state
is {\it linearly stable} throughout the parameter range examined.
Furthermore, this is true for topological charges $m=0$ or $m=1$; solutions
with higher topological charge can be unstable  even in that case.
All higher excited states are found to be unstable in a wide parametric
regime.  However, for the focusing/attractive case the ground
state with $n_r=0$ and $m=0$ can only be
stable for a sufficiently low number of atoms.
Once again, excited states are found to be generically unstable.
For unstable profiles,
the dynamical evolution of the corresponding branches is also followed
to monitor the temporal development of the instability.
\end{abstract}

\maketitle

\section{Introduction
}

The study of trapped Bose-Einstein condensates (BECs) has had a
high impact in recent years in a number of fields, including
atomic, molecular,
and optical physics, nuclear physics, condensed matter physics,
chemical physics, applied mathematics, and nonlinear 
dynamics~\cite{stringari,pethick,BECBOOK}.
From the point of view
of the latter, the topic of particular interest here
is that at the mean field level, the inter-particle interaction
is represented as a classical, but nonlinear self-action \cite{dalfovo},
leading to the now famous Gross-Pitaevskii (GP) equation as a celebrated
model for BECs in appropriate settings. This has resulted in a large
volume of studies of nonlinear excitations, including 
the prediction of excited states \cite{yukalov_add},
the experimental observation of dark \cite{dark,dark1,dark2,dark3},
bright \cite{expb1,expb2}
and gap \cite{gap} solitons in quasi-one-dimensional systems,
as well as theoretical and experimental investigations of
vortices,
vortex lattices \cite{review1,review2}, and ring solitons \cite{djf,brand,carr}
in quasi-two-dimensional systems.

Apart from purely nonlinear dynamical techniques, such as the perturbation 
theory for solitons employed in Ref.~\cite{djf},
other methods based on the corresponding linear Schr\"{o}dinger 
problem may also be employed for the study of excited BECs.
In particular, the underlying linear system for a harmonic
external trapping potential is the quantum harmonic oscillator (QHO)
\cite{stringari,pethick,dalfovo}, whose eigenstates are well-known since
Schr\"{o}dinger's original treatment of the problem in 1926.
Beginning with the linear equation, solutions can be numerically
continued to encompass the presence of the nonlinear
representation of the inter-particle interaction. Then,
the QHO states can persist, bifurcate, or even disappear.
This path does not seem to have been exploited in great
detail in the literature. In one spatial dimension, it has been used in 
Ref.~\cite{tristram,jpb},
to illustrate the persistence \cite{tristram} and dynamical relevance
\cite{jpb} of the nonlinear generalization of the QHO eigenstates.
In higher dimensions, the work of Ref.~\cite{tristram2} 
illustrated the existence of solutions in a radially symmetric
setting. Further progress
has been hampered by the additional difficulties in
(a) examining the linear stability
and (b) converting the radial coordinate system to a Cartesian one
to study evolution dynamics, including nonlinear stability.
However, the mathematical tools for such an analysis exist, as we
discuss in more detail below, and have to a considerable extent
been used in Ref.~\cite{carr}, especially in connection
with ring-like structures with vorticity.

In this paper we present
a systematic analysis of the existence, linear stability, and
evolution dynamics of the states that exist in the spectrum of
harmonically confined condensates from the linear limit, and
persist in the nonlinear problem.
Our analysis shows that in the
case of repulsive interactions (defocusing nonlinearity), the only branch that is {\it linearly stable}
consists of the ground state solution, i.e., a single hump with
$n_r=0$ radial nodes. All higher excited states are linearly unstable
and break up in ways that we elucidate below, if evolved
dynamically in our system; typically this last stage involves
also the loss of radial symmetry to unstable azimuthal perturbations.
Here, we will treat explicitly the cases of $n_r=0,1$ and $2$
radial nodes; we have confirmed similar results for higher number of nodes.

On the other hand, in the
case of attractive interactions (focusing nonlinearity), 
the system is subject to collapse in the absence of the potential.
In fact, this constitutes the critical dimension for the underlying
nonlinear problem \cite{sulem} beyond which collapse is possible.
In this case, we observe that only the ground state branch with small
values of the chemical potential, i.e., a small number of atoms, can
be marginally stable. Once again, all excited states are
unstable and the typical scenario here involves the manifestation
of collapse-type catastrophic instabilities \cite{sulem}, as we elucidate
through numerical simulations.

Our presentation is structured as follows: in section II, we provide
the theoretical setup and numerical techniques. In section III, we
present and discuss the relevant results.  In section IV,
we summarize our findings and present our conclusions. 
Finally, in the appendix we provide relevant details regarding our numerical methods.

\section{Setup and Numerical Methods\label{sec:theo}}

We will use as our theoretical model the well-known
GP equation in a two-dimensional (2D) setting.
As is well-known, the ``effective'' 2D GP equation applies
to situations where the condensate has a nearly planar, i.e., ``pancake''
shape (see, e.g., Ref.~\cite{Beersheva} and references therein.)
We express the equation in harmonic oscillator units \cite{rupr}
in the form:
\begin{eqnarray}
iu_t  =  - \frac{1}{2}\Delta u +
\frac{\Lambda^2}{2}\left( {x^2  + y^2 } \right) u + \sigma \left| u \right|^2 u,
\label{GenEqn}
\end{eqnarray}
where
$u=u(x,y,t)$ is the 2D wave function (the subscript denotes partial derivative). The wave function
is the condensate order parameter, and has a straightforward
physical interpretation: $\rho=|u|^2$ is the local condensate
density.
The external potential
$V(r)=\Lambda^2 r^2/2$ assumes the typical harmonic form,
with $\Lambda$
being the effective frequency of the parabolic trap;
the latter can be
expressed as $\Lambda \equiv \omega_{r}/\omega_{z}$, where
$\omega_{r,z}$ are the confinement frequencies in the radial and
axial directions, respectively.  It is assumed that $\Lambda  \ll 1$ for
the pancake geometry considered herein; in particular, we choose
$\Lambda=0.1$ for our computations.
Finally, $\sigma=\pm 1$ is the normalized coefficient of the nonlinear term, which
is fixed to $-1$
for attractive interactions and to $+1$ for repulsive interactions.  Accordingly,
the squared $L^2$ norm is
\begin{equation}
\int {2\pi} r\,dr\,|u|^2 = N |U|,
\end{equation}
where $N$ is the number of atoms and $U=\sigma |U|$ is the usual
nonlinear coefficient in the GP equation in harmonic oscillator units, renormalized
to two dimensions appropriately~\cite{carr2005c}.

In order to numerically identify stationary nonlinear solutions of
Eq.~(\ref{GenEqn}), we use the standing wave ansatz
$u(x,y,t) = v(r)\, \exp[i(\mu t{+ m\theta})]$,
where we assume the density of the solution is radially symmetric with chemical potential
$\mu$ and topological charge (vorticity) $m$. This results
in the steady-state equation:
\begin{eqnarray}
\mu v =  - \frac{1}{2}\left( {v_{rr}  +
\frac{1}{r}v_r } {-\frac{m^2}{r^2} v} \right) +
\frac{\Lambda ^2}{2}r^2 v + \sigma \left| v \right|^2 v.
\label{RadEqn}
\end{eqnarray}
Equation~(\ref{RadEqn}) exhibits infinite branches of nonlinear bound states, each
branch stemming from the corresponding mode of the underlying linear
problem.
These branches are constructed and
followed using  the method of pseudo-arclength continuation
\cite{arclength1,arclength2}, in order to obtain subsequent points along
a branch, once a point on it has been identified.
The first point on each branch is found using a bound state of the
underlying linear equation
\begin{eqnarray}
Ev =  - \frac{1}{2}\left( v_{rr}  +
\frac{1}{r}v_r {-\frac{m^2}{r^2} v} \right) +
\frac{\Lambda ^2}{2}r^2 v.
\label{LinEqn}
\end{eqnarray}
The linear state corresponds to a solution of Eq.~(\ref{RadEqn}) with
parameters $\mu = E$, frequency $\Lambda$, vorticity $m$,
number of radial nodes $n_r$ and normalization
$\int 2\pi r \,dr {\left| v \right|^2 }$ tending to zero; in that limit,
the nonlinearity becomes
negligible and the linear solutions are the
well-known Gauss-Laguerre modes with $E=(2 n_r + m + 1) \Lambda$.
Using such an initial guess, a non-trivial solution is found through a
fixed-point iterative scheme for a slightly perturbed value of the
chemical potential. This is done on a grid of Chebyshev points
suited to the radial problem, following the approach of Ref.~\cite{trefethen}
for the Laplacian part of the equation,
as is explained in detail in the appendix.
The pseudo-arclength method, used to trace subsequent solutions,
works via the introduction of a pseudo-arclength parameter $s$ and an
additional equation $F(v,\mu,s)=0$ such that $F({\bar v},{\bar \mu},0)=0$
where $({\bar v},{\bar \mu})$ is a solution of Eq.~(\ref{RadEqn}).
We used for $F$:
\begin{eqnarray}
F(v,\mu ,s) = \left| {v - \bar v} \right|^2  + \left| {\mu  - \bar \mu } \right|^2  - s^2.
\label{EqnF}
\end{eqnarray}
Lastly, the new expanded system of equations is solved using a
predictor-corrector method.

\begin{figure}[tbp]
\includegraphics[width=\wfigtwo,angle=0,clip]{\rootfigsmall  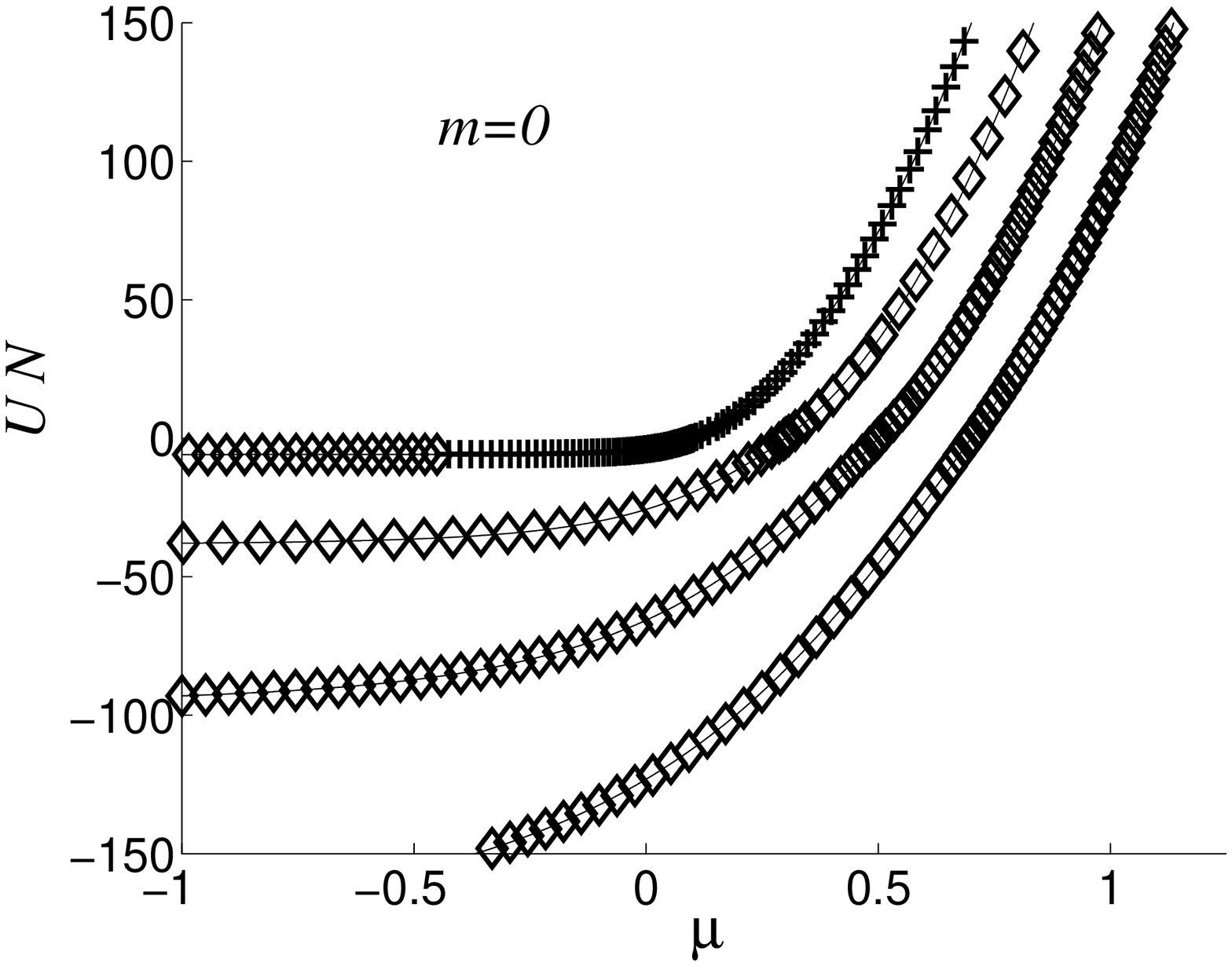}\\
\includegraphics[width=\wfigtwo,angle=0,clip]{\rootfigsmall  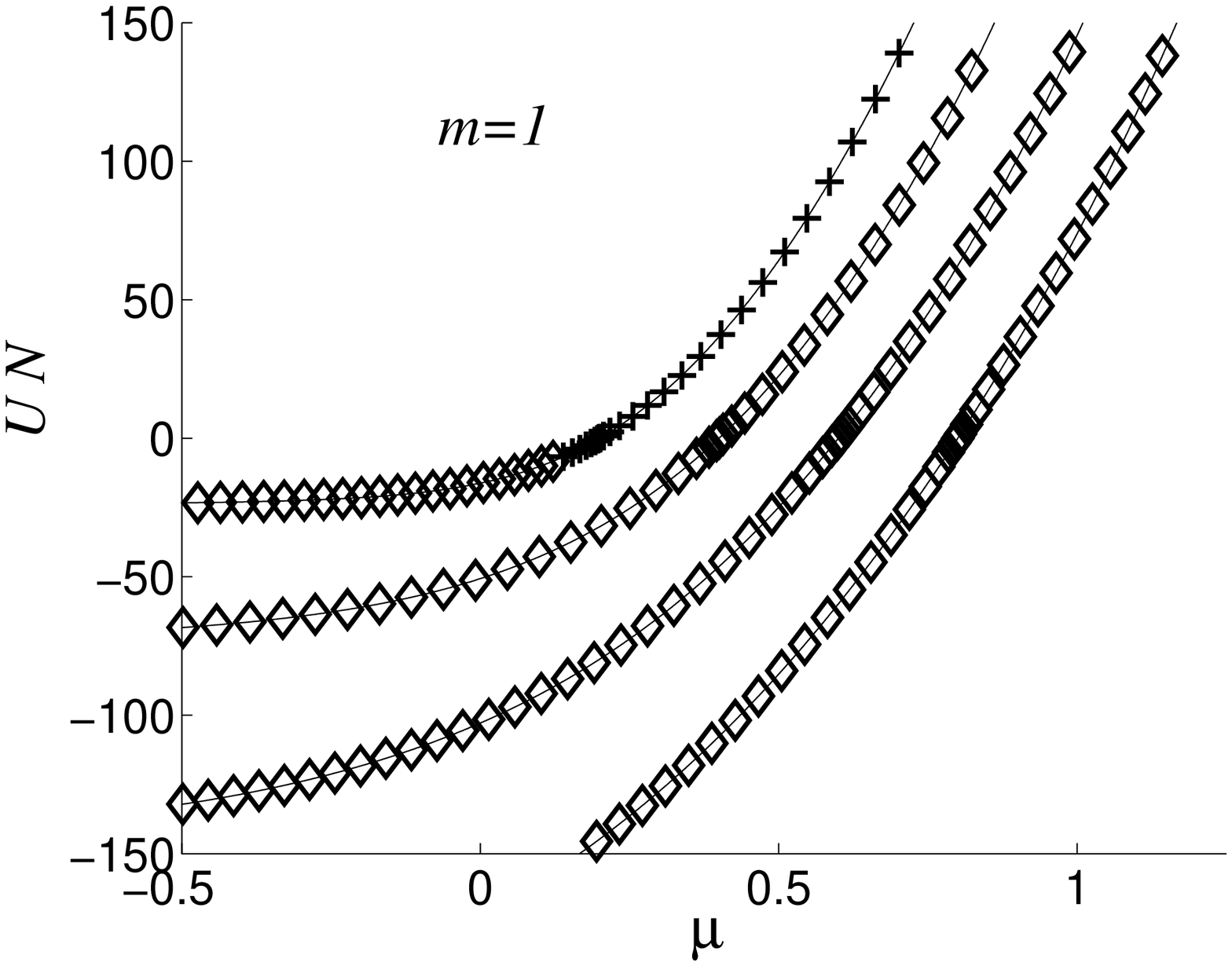}\\
\includegraphics[width=\wfigtwo,angle=0,clip]{\rootfigsmall  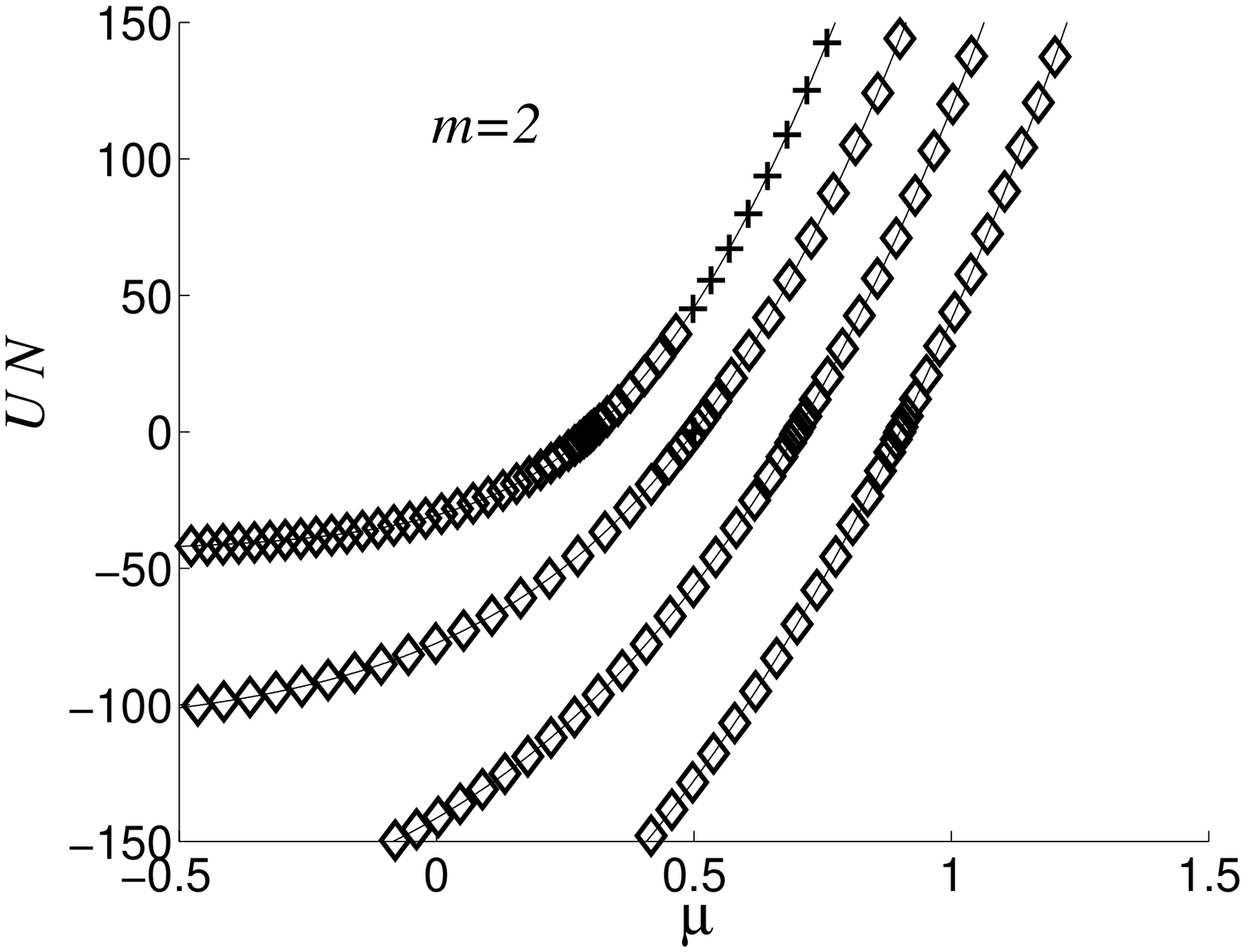}
\caption{
Linear stability analysis along the first four branches:
ground state (no radial nodes) and 1$^{\rm st}$--3$^{\rm rd}$
excited states ($n_r=1, 2$ and $3$ radial nodes)
depicted
in the curves from left to right in each panel.
The top, middle and bottom panels display $\sigma$ times the norm of the solution
as a function of the chemical potential for $m=0$ (no topological
charge), $m=1$ (singly
charged) and $m=2$ (doubly charged) solutions.
Each point on the branch is
depicted by crosses for stable solutions and diamonds for unstable solutions.
The areas of stability for $m=0,1$ appear only for $n_r=0$ and
only for lower powers for $\sigma = -1$ and all powers of the ground state
for $\sigma = 1$. In the case of
$m=2$, the ground state with $n_r=0$ is
only linearly stable for stronger nonlinearity/norm $NU$ in the case of $\sigma = 1$.
}
\label{Branches}
\end{figure}

The linear stability of the solutions is analyzed by using the following ansatz
for the perturbation:
\begin{equation}
u(r,t)=e^{i\mu t}\left[ v(r)+a(r)e^{i q \theta+\lambda t}+b^{\ast }(r)e^{-i q \theta+\lambda^{\star }t}\right],
\label{PertAnsatz}
\end{equation}
where the asterisk stands for the complex conjugation and $\theta$ is the
polar angle. One solves the resulting
linearized equations for the perturbation eigenmodes $\{a(r),b(r)\}=\{a_q(r),b_q(r)\}$ and
eigenvalues $\lambda=\lambda_q$ associated with them.  The key observation
that allows one to carry this task through is that the subspace of the
different azimuthal perturbation eigenmodes, i.e., subspaces of different
${q}$, decouple \cite{pego,carr}, leading each eigenmode $e^{i {q} \theta}$
to be coupled only with
its complex conjugate. This in turn allows the examination of the
stability of the 2D problem in the form of a denumerably infinite set
of one-dimensional radial eigenvalue problems that are solved
on the same grid of Chebyshev points as the original existence
problem of Eq.~(\ref{RadEqn}).

\begin{figure}[tbp]
\includegraphics[width=8.0cm,angle=0]{\rootfigsmall  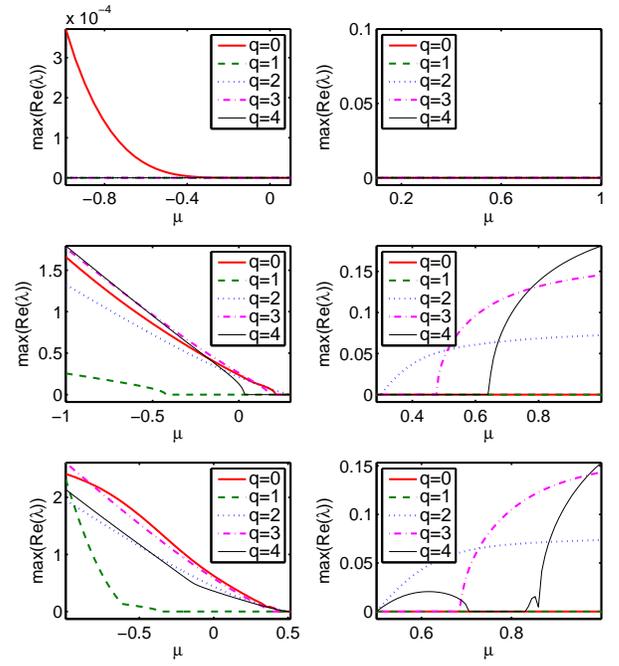}
\caption{(Color online)
Stability eigenvalues for chargeless ($m=0$) solutions.
The left column of plots
corresponds to the real part of the primary eigenvalue for $q=0,1,2,3,4$
for the ground state with $n_r=0$ (top panel) and first
and second excited states with $n_r=1$ and $2$
(middle and bottom panels, respectively)
for $\sigma = -1$; while the right column of plots corresponds to the case
of $\sigma = +1$.
The ground state for $\sigma = +1$ is omitted since the entire branch
is stable.
}
\label{Stability0}
\end{figure}

\begin{figure}[tbp]
\includegraphics[width=8.0cm,angle=0]{\rootfigsmall  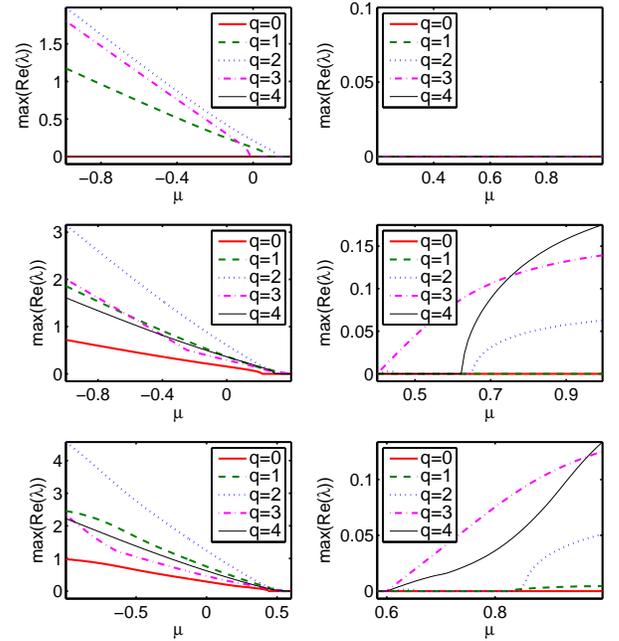}
\caption{(Color online)
Same as Fig.~\ref{Stability0} for singly charged ($m=1$) solutions.}
\label{Stability1}
\end{figure}

\begin{figure}[tbp]
\includegraphics[width=8.0cm,angle=0]{\rootfigsmall  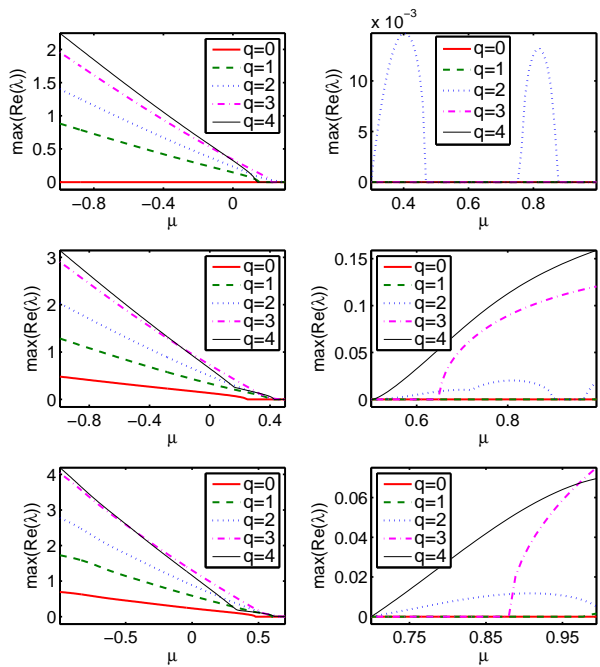}
\caption{(Color online)
Same as Fig.~\ref{Stability0} for doubly charged ($m=2$) solutions. Notice that in this case the $n_r=0$ branch can be unstable even for the
repulsive interactions, i.e., defocusing nonlinearity, see top right panel.}
\label{Stability2}
\end{figure}

In order to examine the dynamics of the cases found to be
unstable, to confirm the validity of the cases identified
as stable, and to test for nonlinear instabilities,
the 1D radial solutions of Eq.~(\ref{RadEqn})
were taken as an initial condition for a code which solved Eq.~(\ref{GenEqn})
with a Chebyshev spectral radial-polar method in space and
a fourth-order integrator in time.
%
The Chebyshev spectral radial-polar method
is the most natural choice for our setting, as it
a) avoids the conversion of the radial solution into an 
interpolated Cartesian grid, and, more importantly, 
b) avoids spurious effects associated with a mismatch between
the symmetry of the solution and that of the grid. 
For example, we have observed that
the use of a Cartesian grid artificially enhances the excitation
of modes that have a similar symmetry as the grid, and in particular
the $q=4$ mode.
%
The results of all of the above numerical techniques, i.e.,
existence, linear stability, and dynamical evolution, are reported
in what follows.

\section{Results}

Our steady state and stability results are summarized in Fig.~\ref{Branches}.
The panel shows the continuation from the linear limit of the
relevant states.  The nonlinearity leads to a decrease of the
chemical potential with increasing power for the focusing case and
to a corresponding increase for the defocusing case. Notice that some
of the branches presented here have also appeared in the earlier work of
Ref.~\cite{tristram2}. However, the important new ingredient of the present
work is the detailed examination, both through linear stability analysis
and through dynamical evolution, of the stability of these solutions.
The latter is encapsulated straightforwardly in the symbols associated
with each branch. The plus symbols denote branches which are stable, while
the diamond symbols correspond to unstable cases. We observe
that the only truly stable solution corresponds to the ground state
with $n_r=0$
of the defocusing problem, which for large number of atoms can be well
approximated by the Thomas-Fermi state \cite{stringari,pethick}.
Furthermore, this is true only for topological charges $m=0$ and
$m=1$; for higher topological charges $m \geq 2$, there are regions
of stability for $n_r=0$, as was originally shown in Refs.~\cite{pu}
and \cite{vortex_stability_add}.
All higher excited states with one, two or more nodes are directly
found to be unstable, even close to the corresponding linear limit
of these states.
In the focusing case, the situation is even more unstable due
to the catastrophic effects of self-focusing and wave collapse \cite{sulem,doddRJ1996}.
In particular, even the ground state with $n_r=0$ may be unstable
due to collapse (represented by the mode with $q=0$ in this case),
although the instability growth rate may be very small, as is the
case for $m=0$, see e.g., Fig.~\ref{Stability0}.
Excited states are always unstable for the focusing nonlinearity also;
in fact, they are more strongly so than in the defocusing case, again
due to the presence of the $q=0$ mode.

\begin{table}[h]
\begin{tabular}{|c|c|c|c|c|c|c|c|c|c|c|c|c|c|}
\hline
~$q$-value~ & \,0\, &\,1\, &\,2\, &\,3\, &\,4\, &\,5\, &
\,6\, &\,7\, &\,8\, &\,9\, & \,10\, & \,11\, & \,12--50\, \\[0.1ex]
\hline
~symbol~ & $\Circle$ & $\times$ & $+$ & $\varhexstar$ & $\Square$ &
$\Diamond$ & $\triangledown$ & $\vartriangle$ & $\vartriangleleft$ &
$\vartriangleright$ & $\pentagon$ & $\hexagon$ &
{\footnotesize\textbullet} \\[0.3ex]
\hline
\end{tabular}
\caption{Table of symbols used for the unstable eigenvalues in
Figs.~\ref{Ground}--\ref{SecondExm2}. For eigenvalues smaller than
$10^{-7}$ or $q>11$ we use a small black dot.
}
\label{mytable}
\end{table}

\begin{figure}[tbp]
\includegraphics[width=\wfig,angle=0,clip]{\rootfigsmall  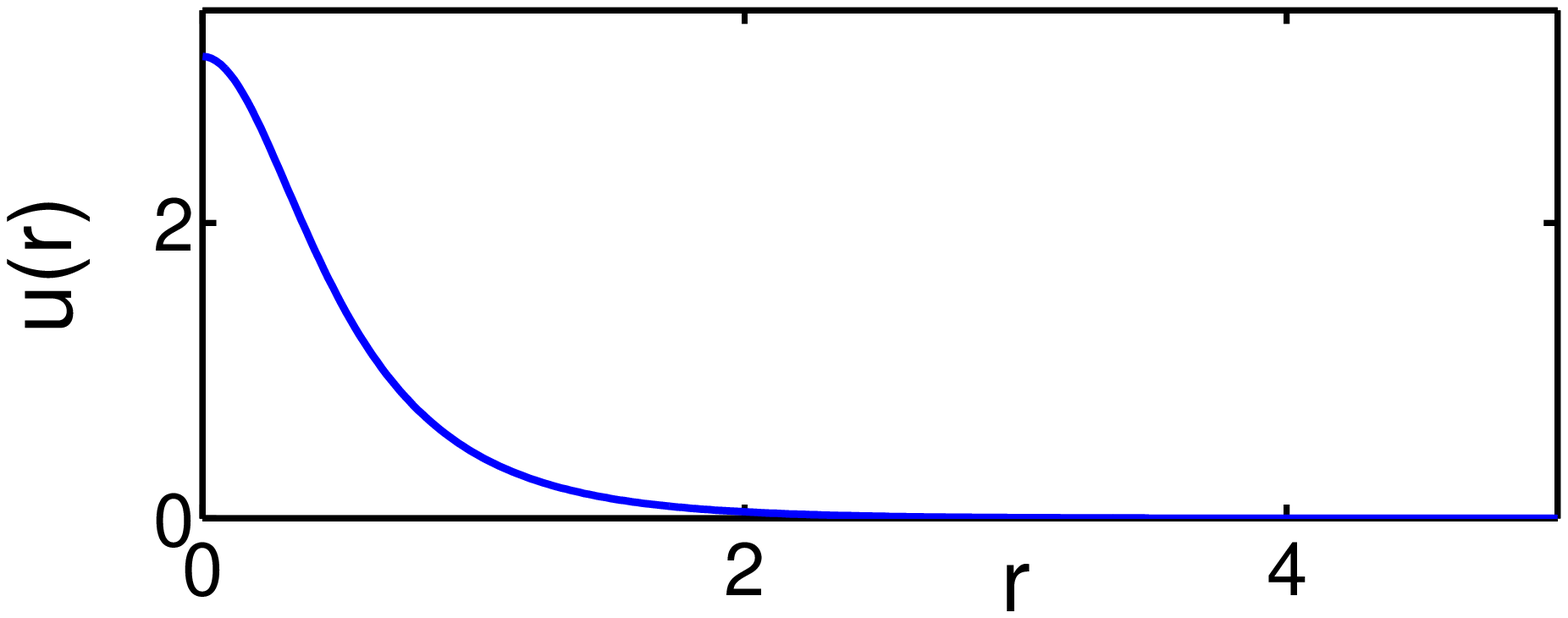}
\includegraphics[width=\wfig,angle=0,clip]{\rootfigsmall  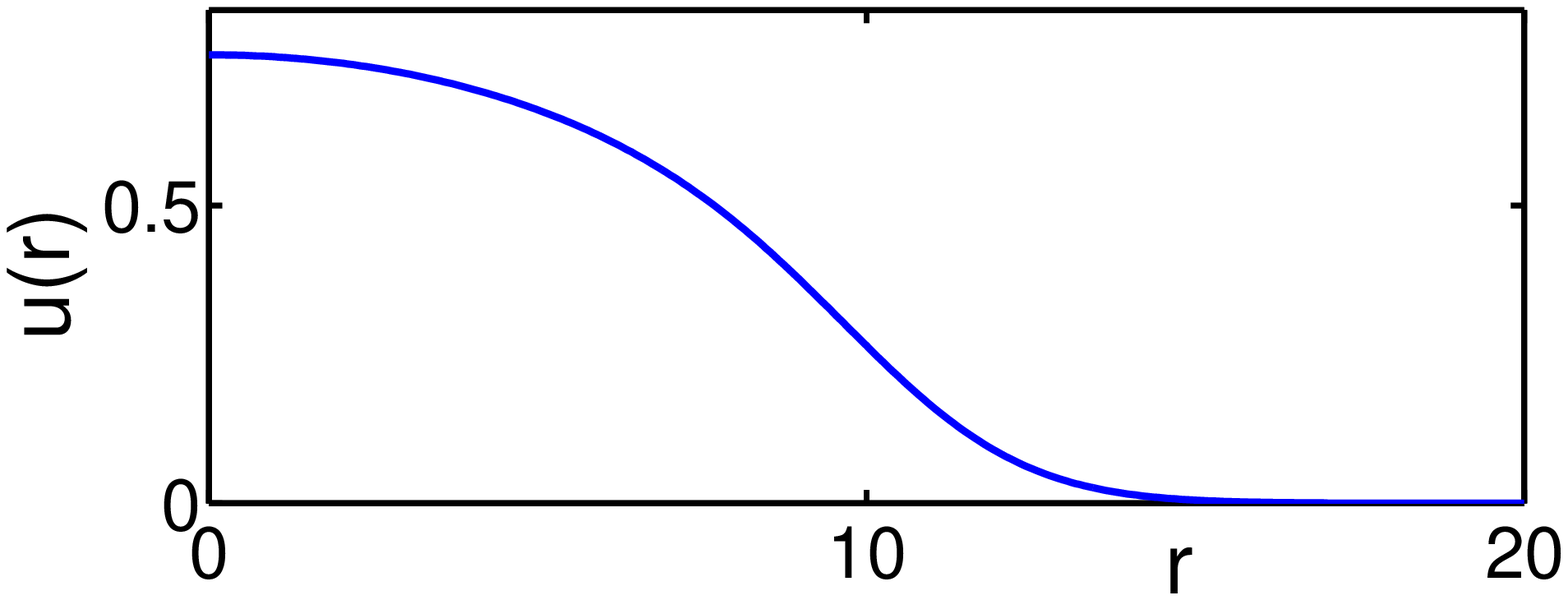}\\
\includegraphics[width=\wfig,angle=0,clip]{\rootfigsmall  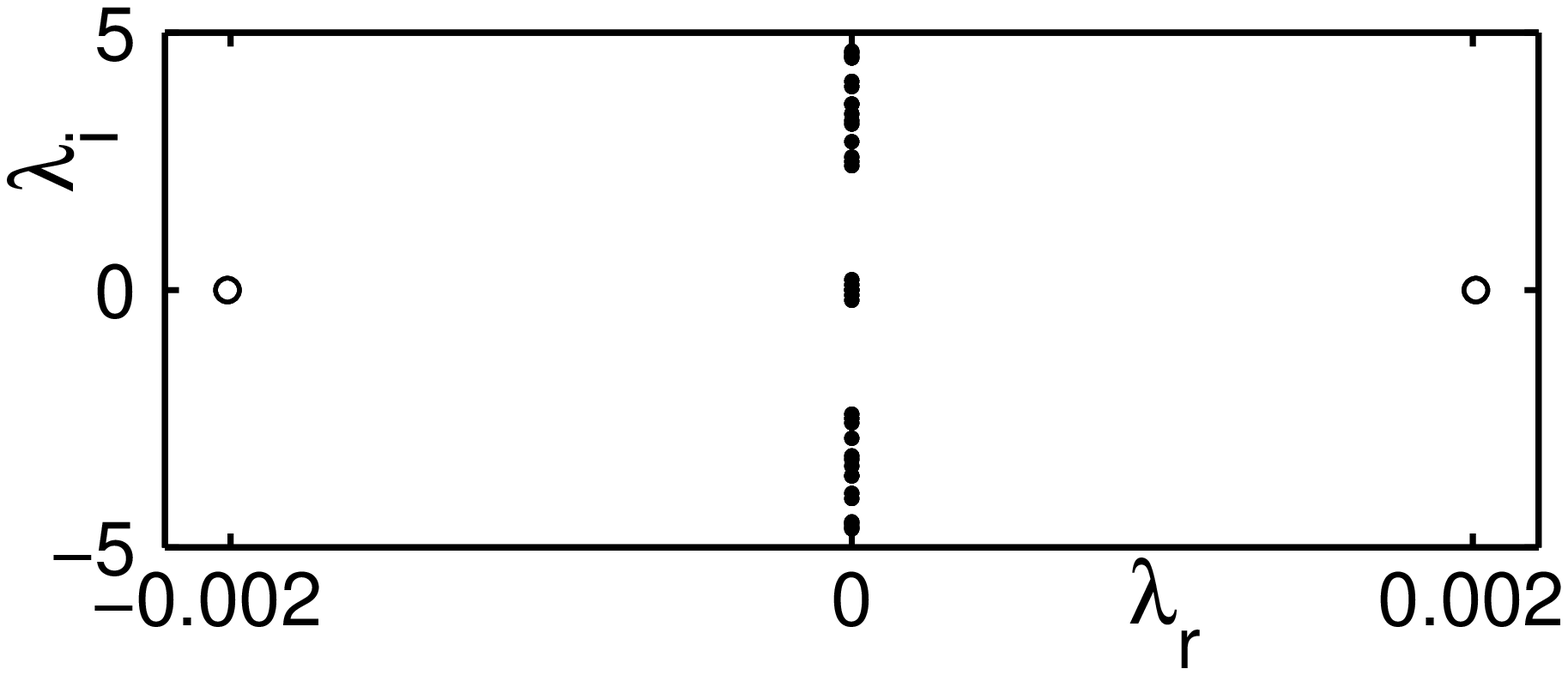}
\includegraphics[width=\wfig,angle=0,clip]{\rootfig 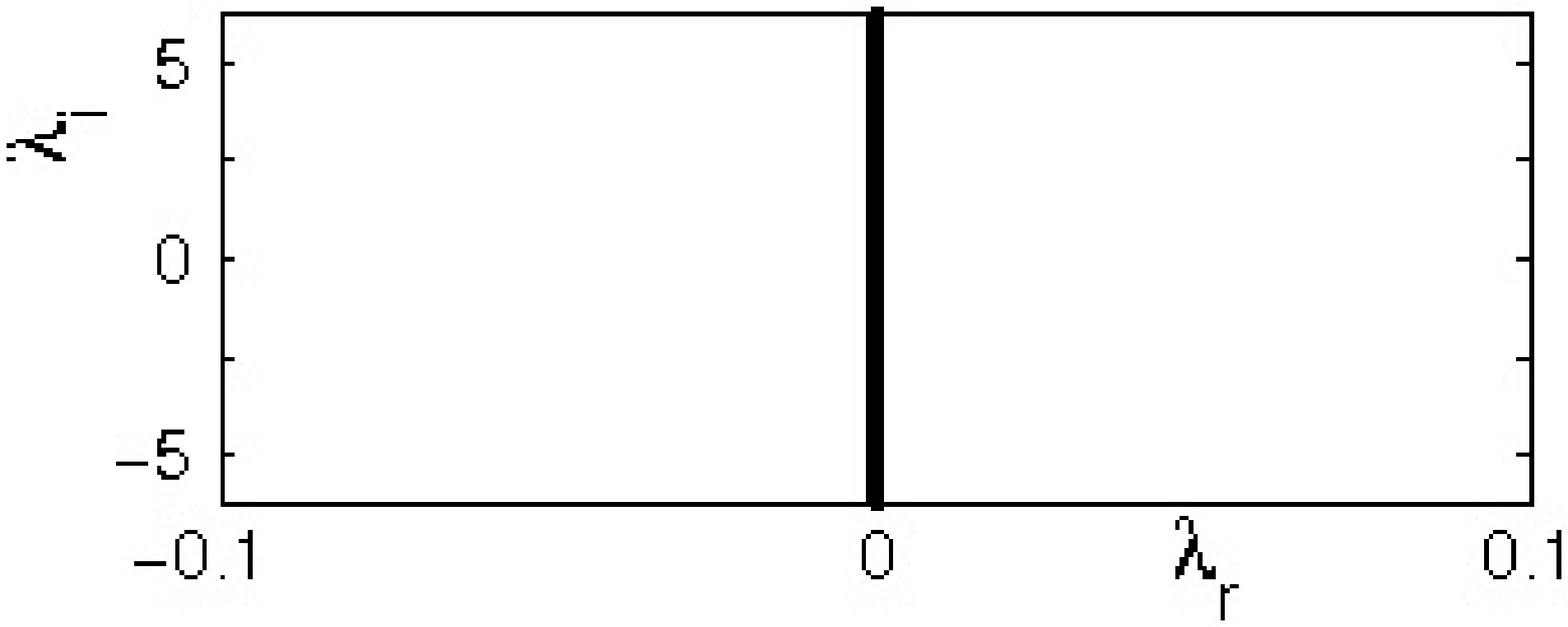}\\
~~\includegraphics[width=\wfigt,angle=0,clip]{\rootfig 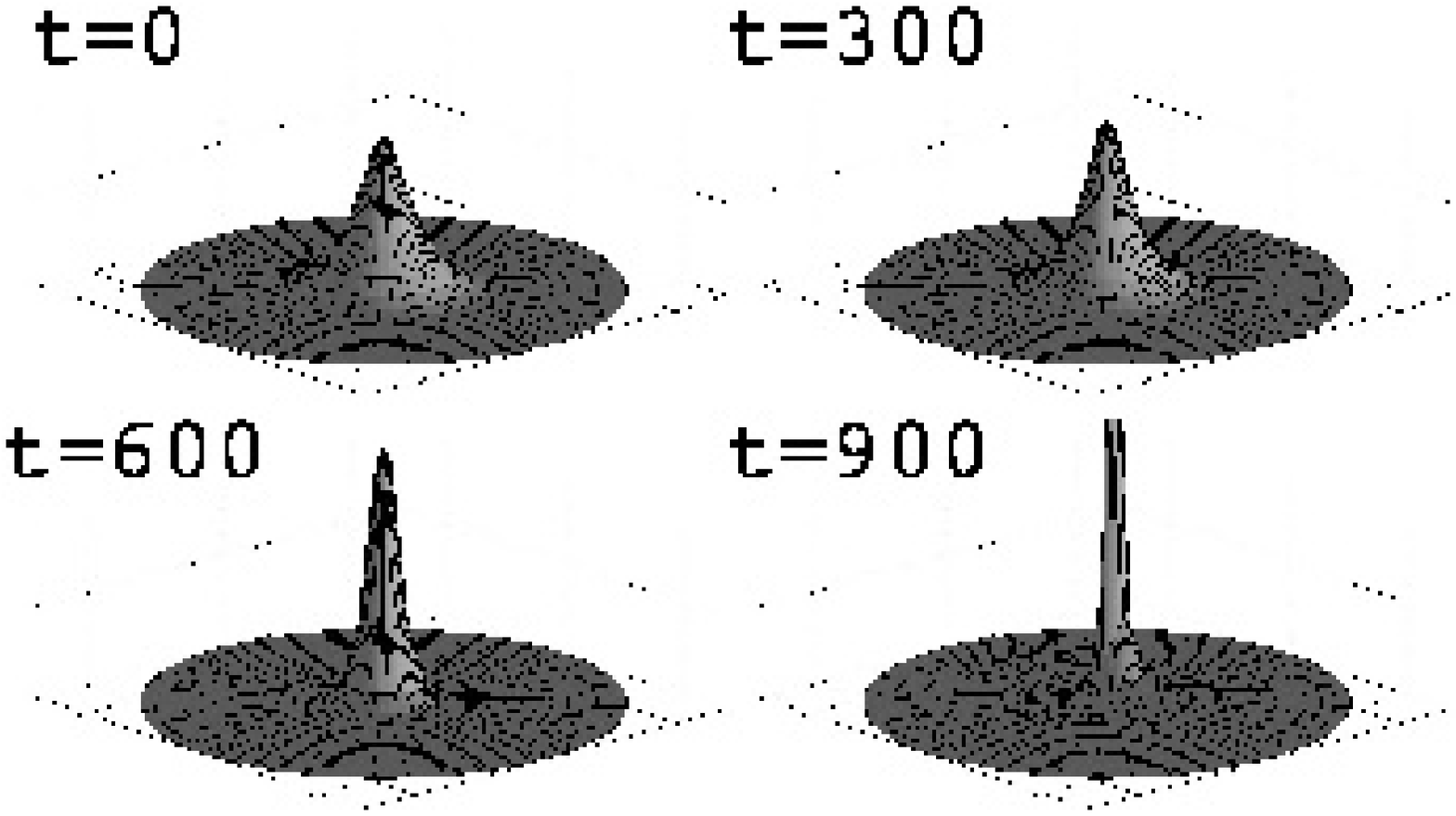}
~~~\includegraphics[width=\wfigt,angle=0,clip]{\rootfig 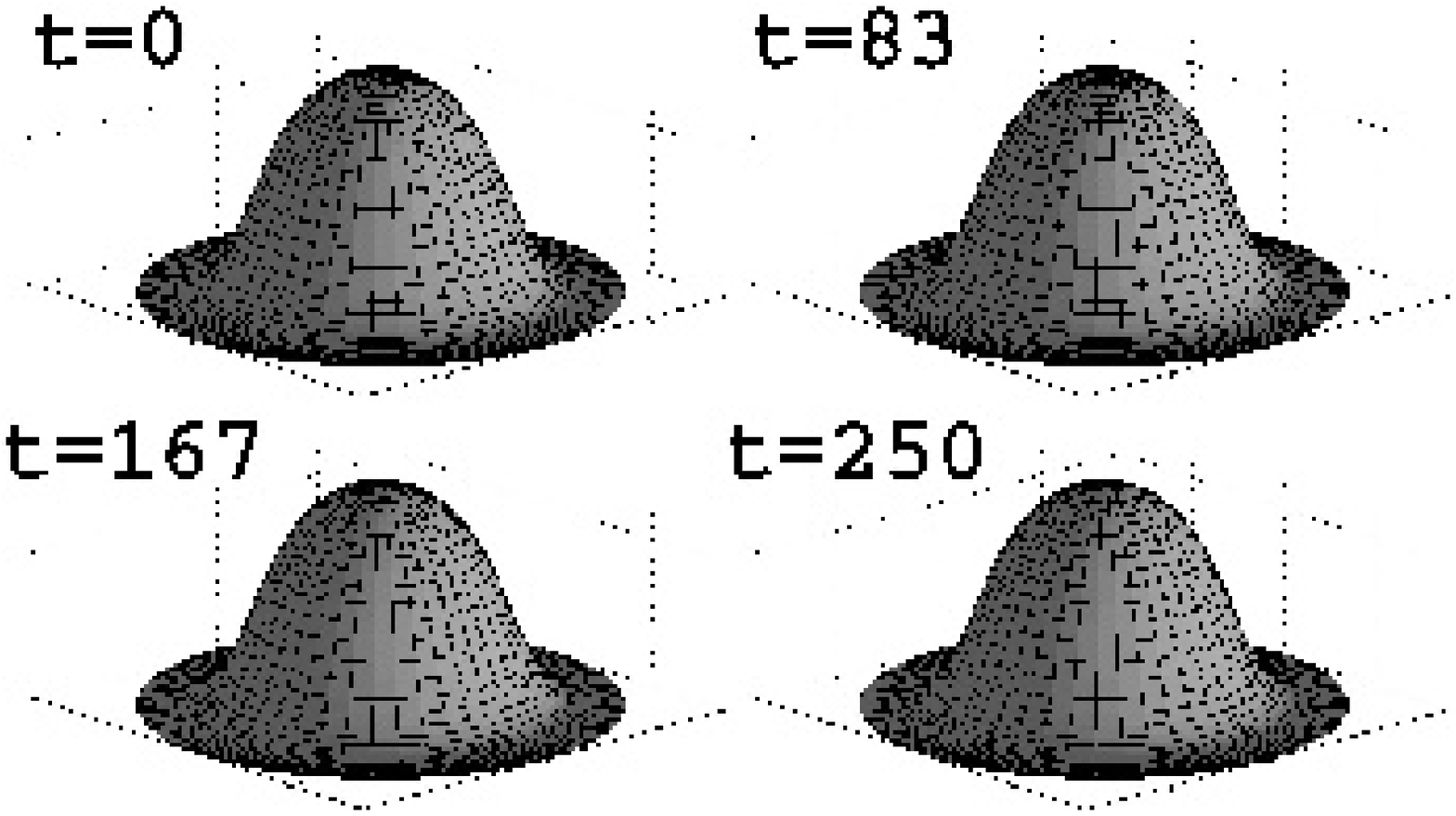}
\caption{
Data for the ground state ($n_r=0$) for $m=0$ 
for $\sigma = -1$ (left panels) and $\sigma = +1$ (right panels).
The top panels show the profile of the solution and the middle  
panels show the corresponding
eigenvalues for $q=0,...,50$ in Eq.~(\ref{PertAnsatz}) on the complex plane
($\lambda_r,\lambda_i$) of eigenvalues $\lambda=\lambda_r+i \lambda_i$;
see Table~\ref{mytable} for the correspondence between $q$-values and the
different symbols used to depict the eigenvalues.
The bottom panels depict the time evolution of the solution
{in harmonic oscillator units} after an initial random perturbation of $10^{-2}$.
Solutions in this figure correspond to $\mu=-2$ for $\sigma=-1$ and
norm {$N|U|=100$} for $\sigma=+1$.
For all other figures we use $\mu=-0.5$ when $\sigma=-1$.
}
\label{Ground}
\end{figure}

\begin{figure}[tbp]
\includegraphics[width=\wfig,angle=0,clip]{\rootfigsmall  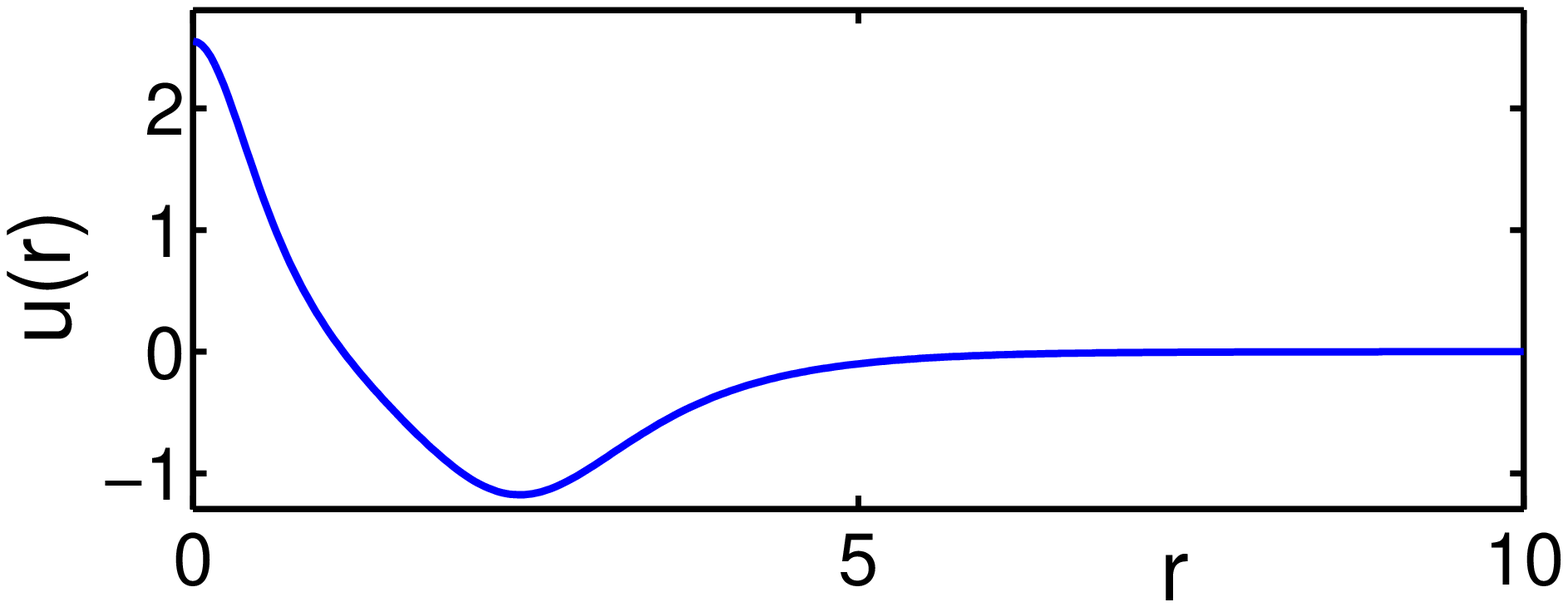}
\includegraphics[width=\wfig,angle=0,clip]{\rootfigsmall  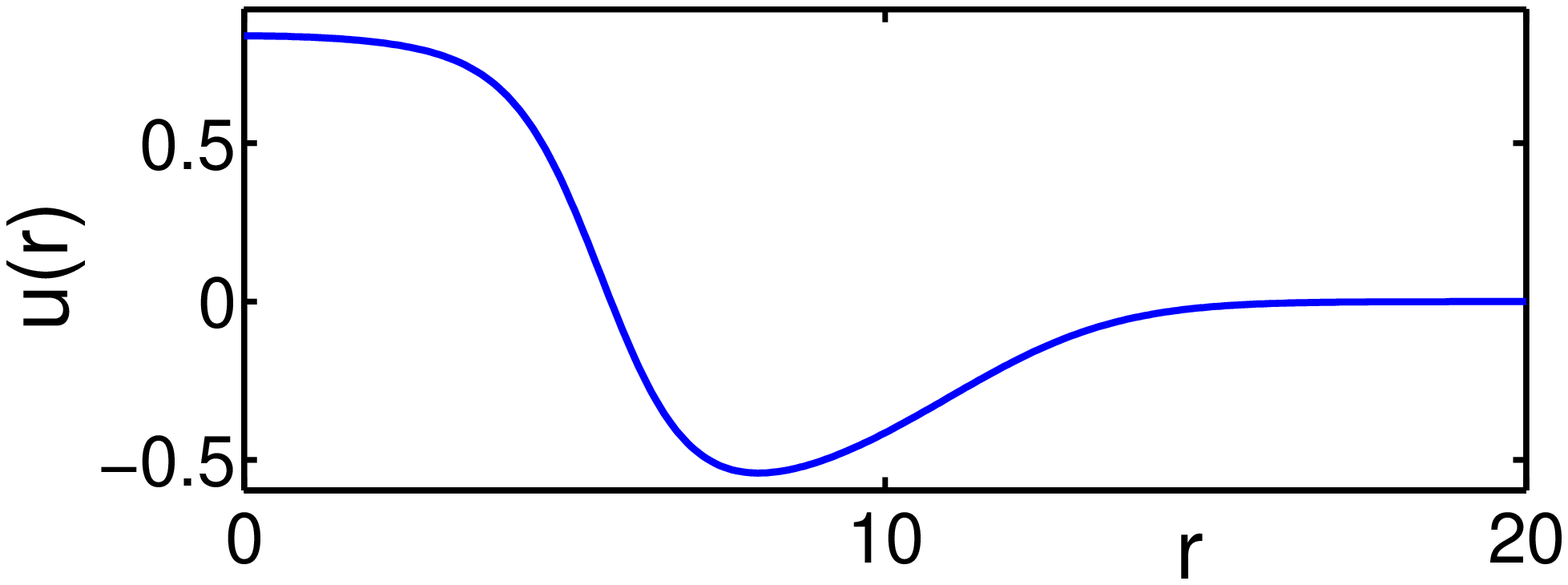}\\
\includegraphics[width=\wfig,angle=0,clip]{\rootfigsmall  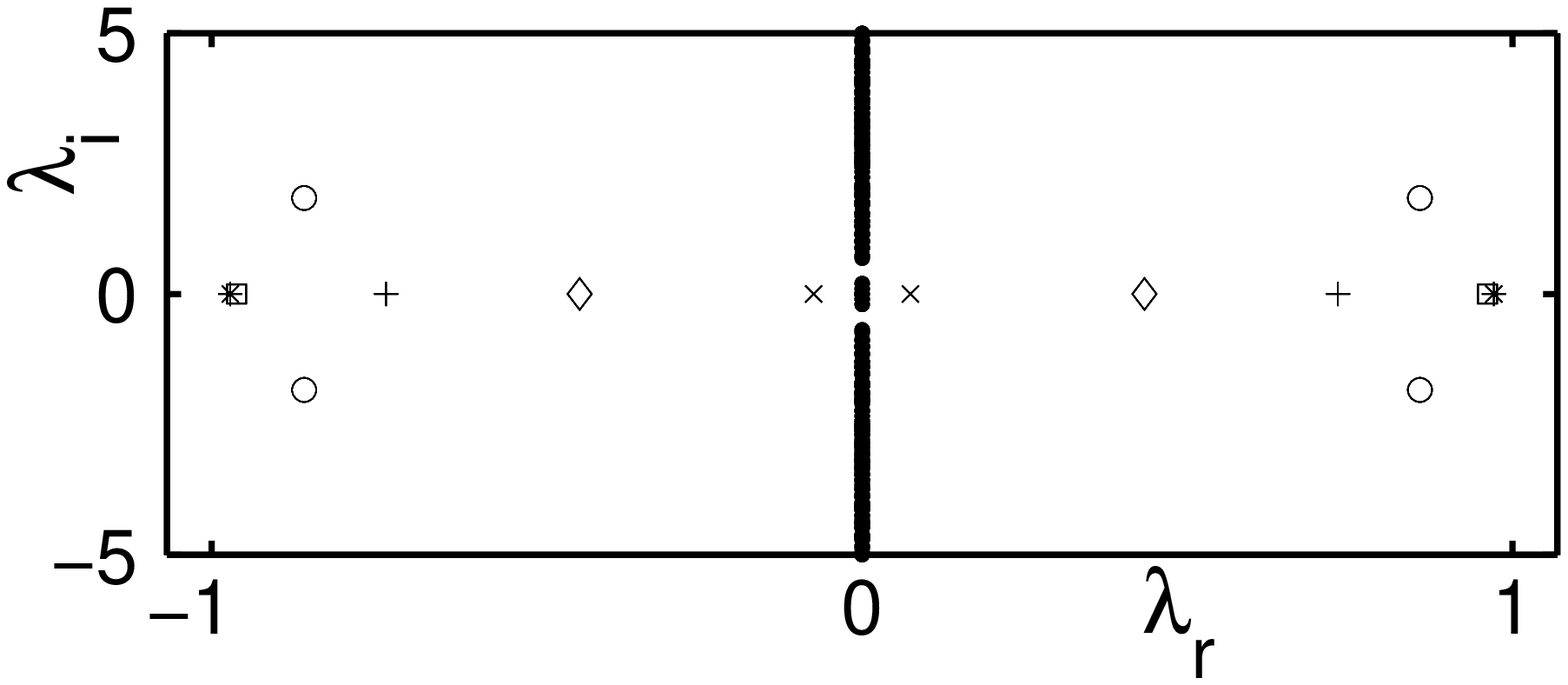}
\includegraphics[width=\wfig,angle=0,clip]{\rootfig 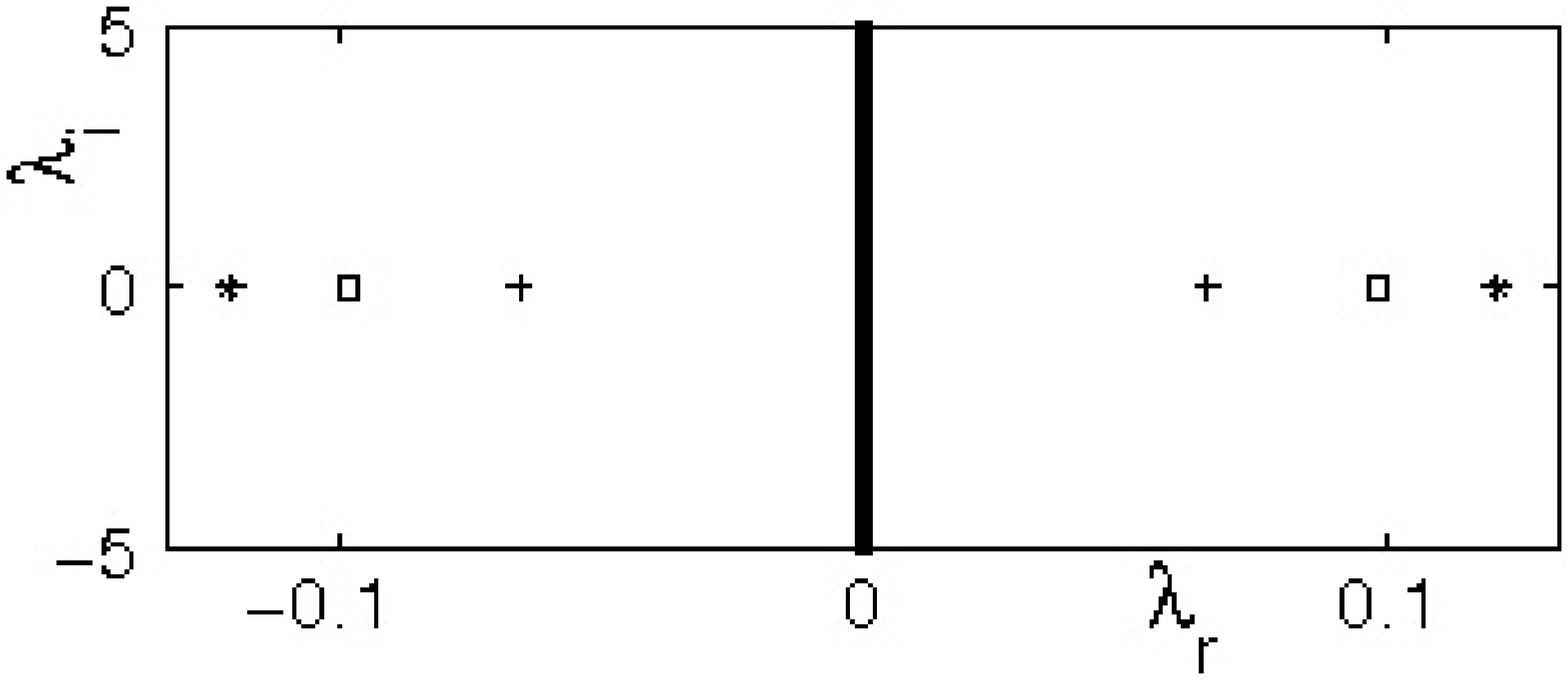}\\
~~\includegraphics[width=\wfigt,angle=0,clip]{\rootfig 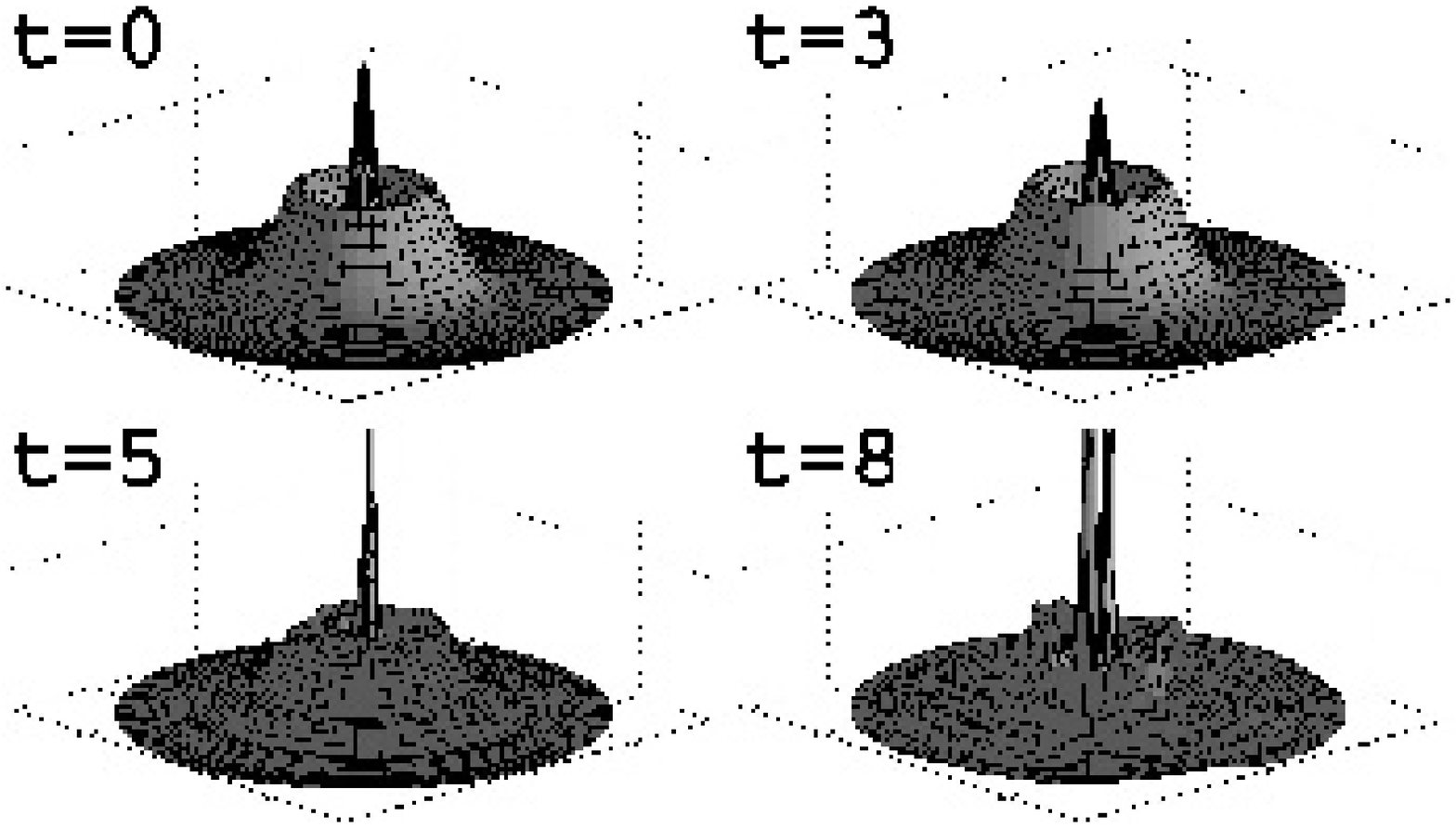}
~~~\includegraphics[width=\wfigt,angle=0,clip]{\rootfig 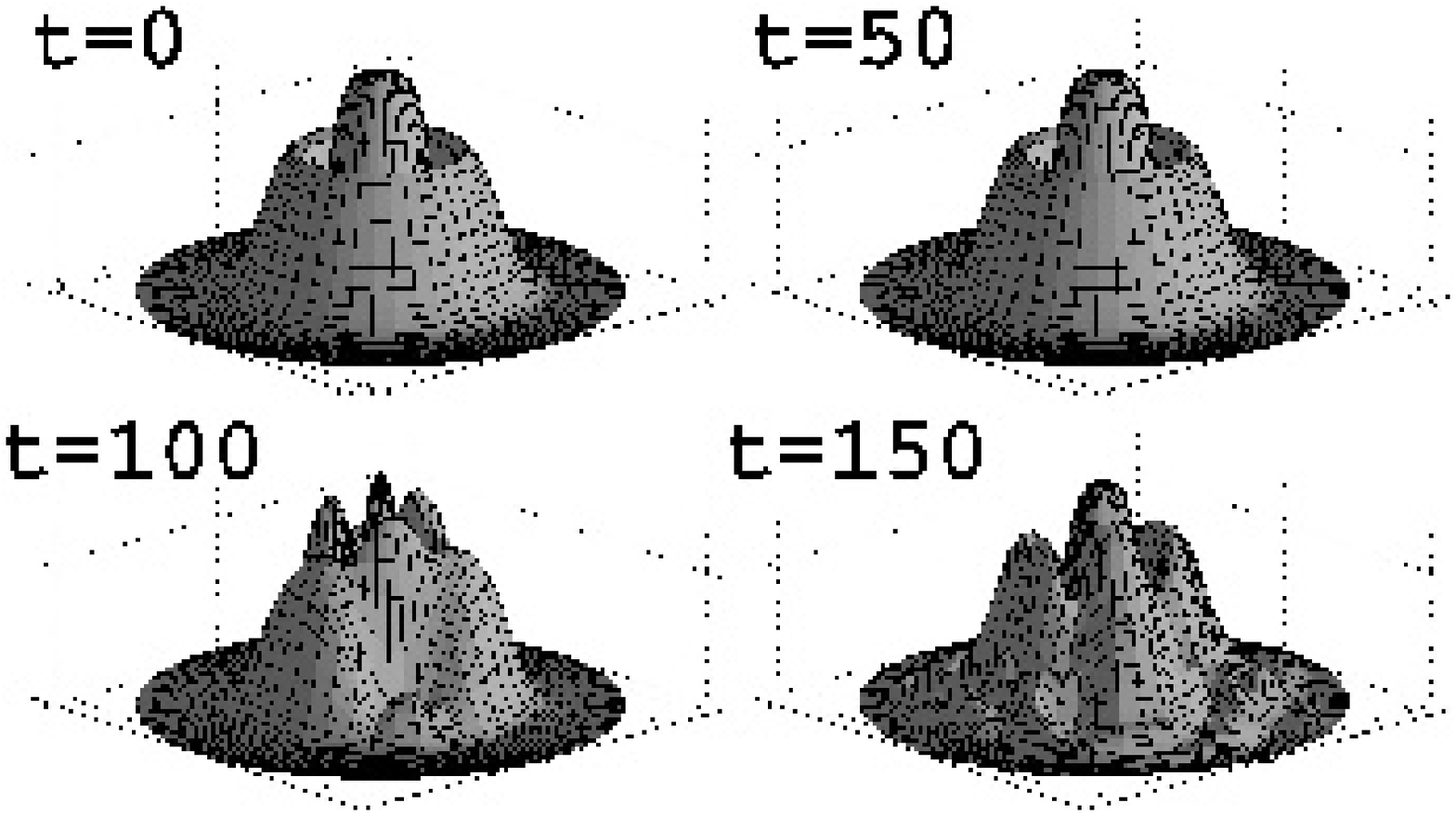}
\caption{Same as Fig.~\ref{Ground} ($m=0$) for the first excited state
($n_r=1$).}
\label{FirstEx}
\end{figure}

\begin{figure}[tnp]
\includegraphics[width=\wfig,angle=0,clip]{\rootfigsmall  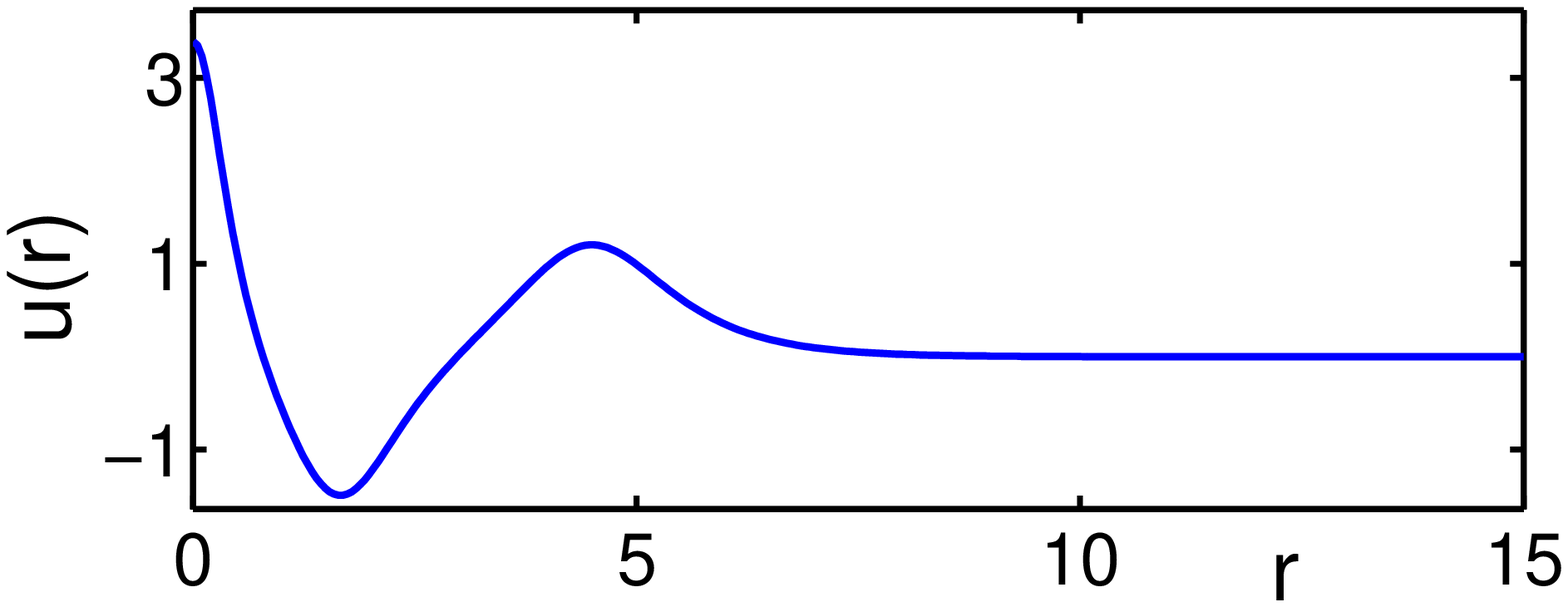}
\includegraphics[width=\wfig,angle=0,clip]{\rootfigsmall  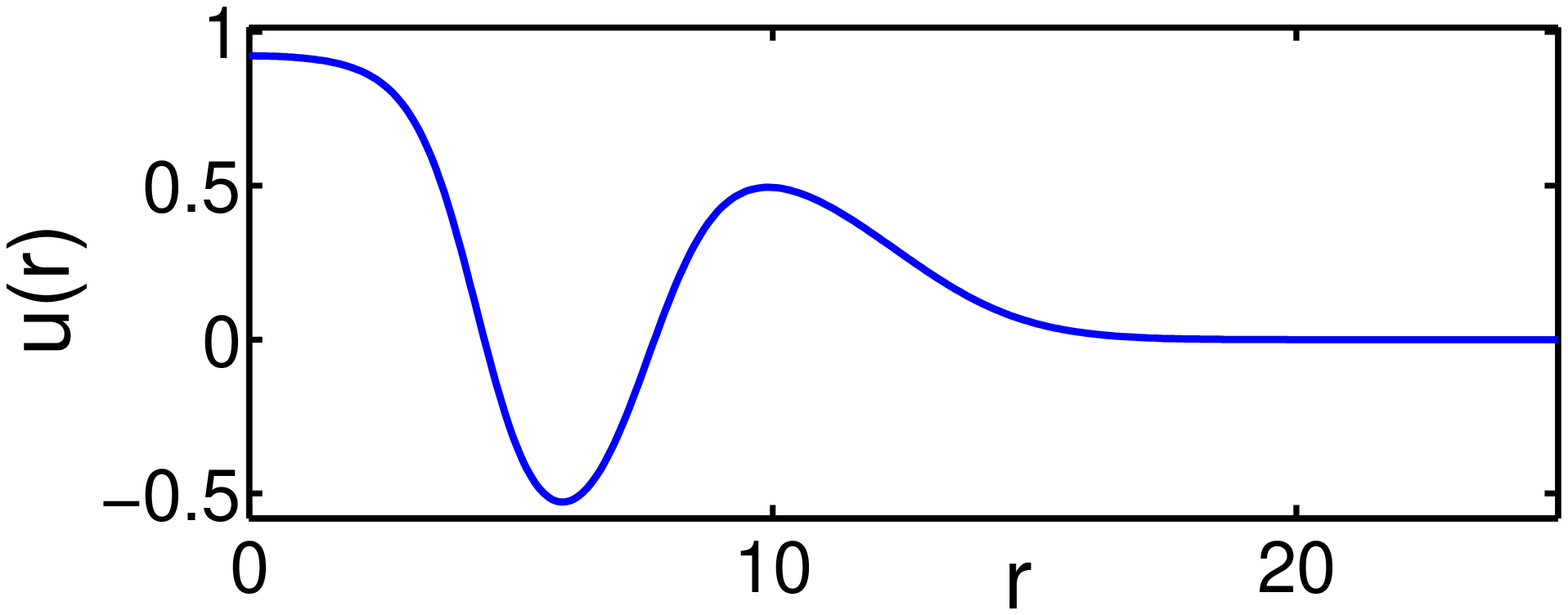}\\
\includegraphics[width=\wfig,angle=0,clip]{\rootfig 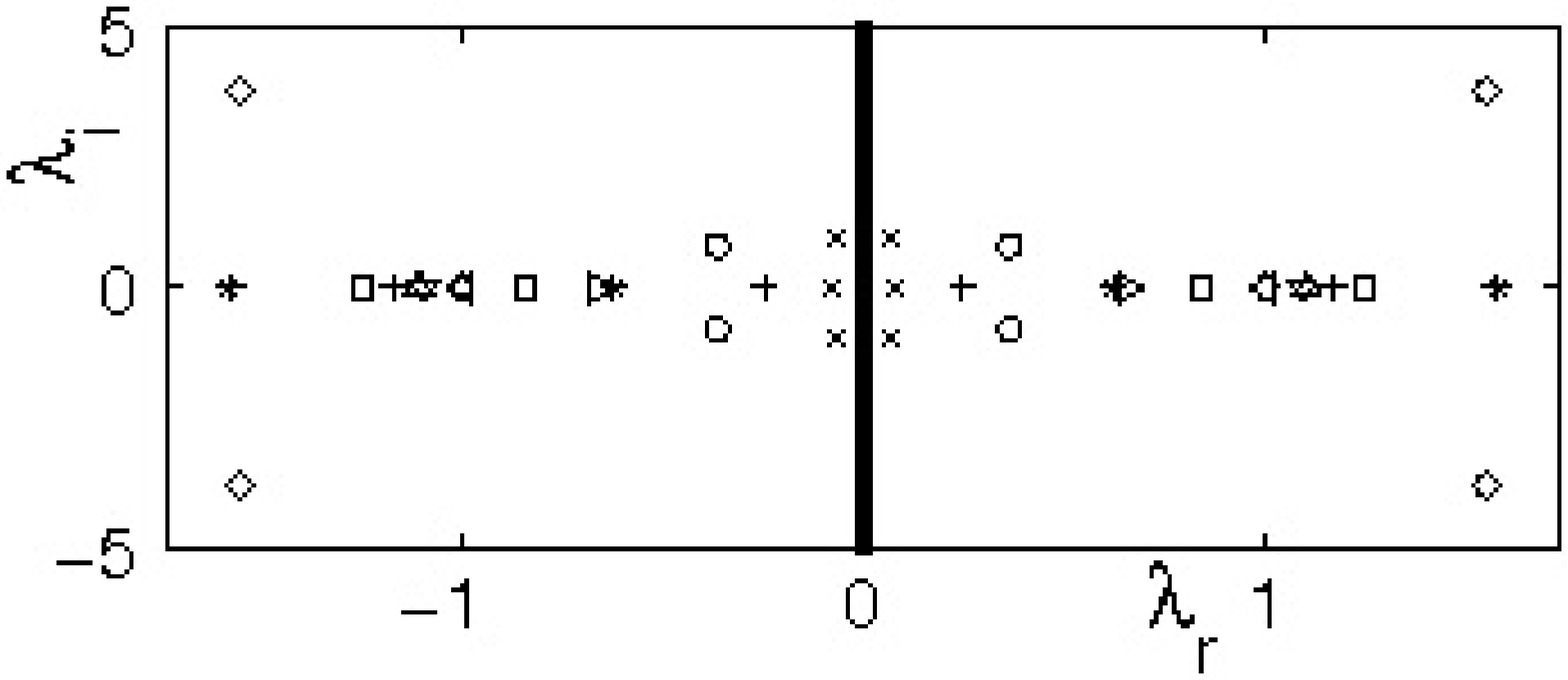}
\includegraphics[width=\wfig,angle=0,clip]{\rootfig 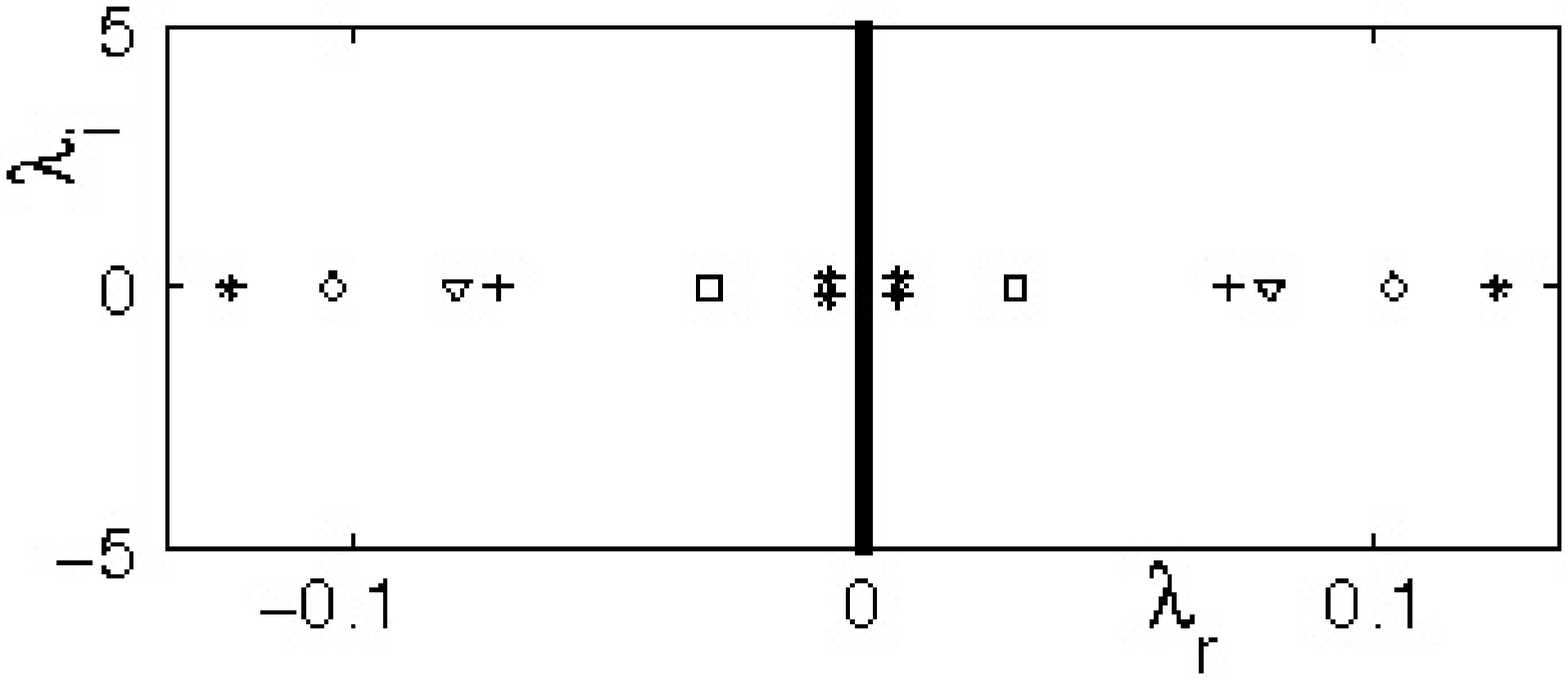}\\
~~\includegraphics[width=\wfigt,angle=0,clip]{\rootfig 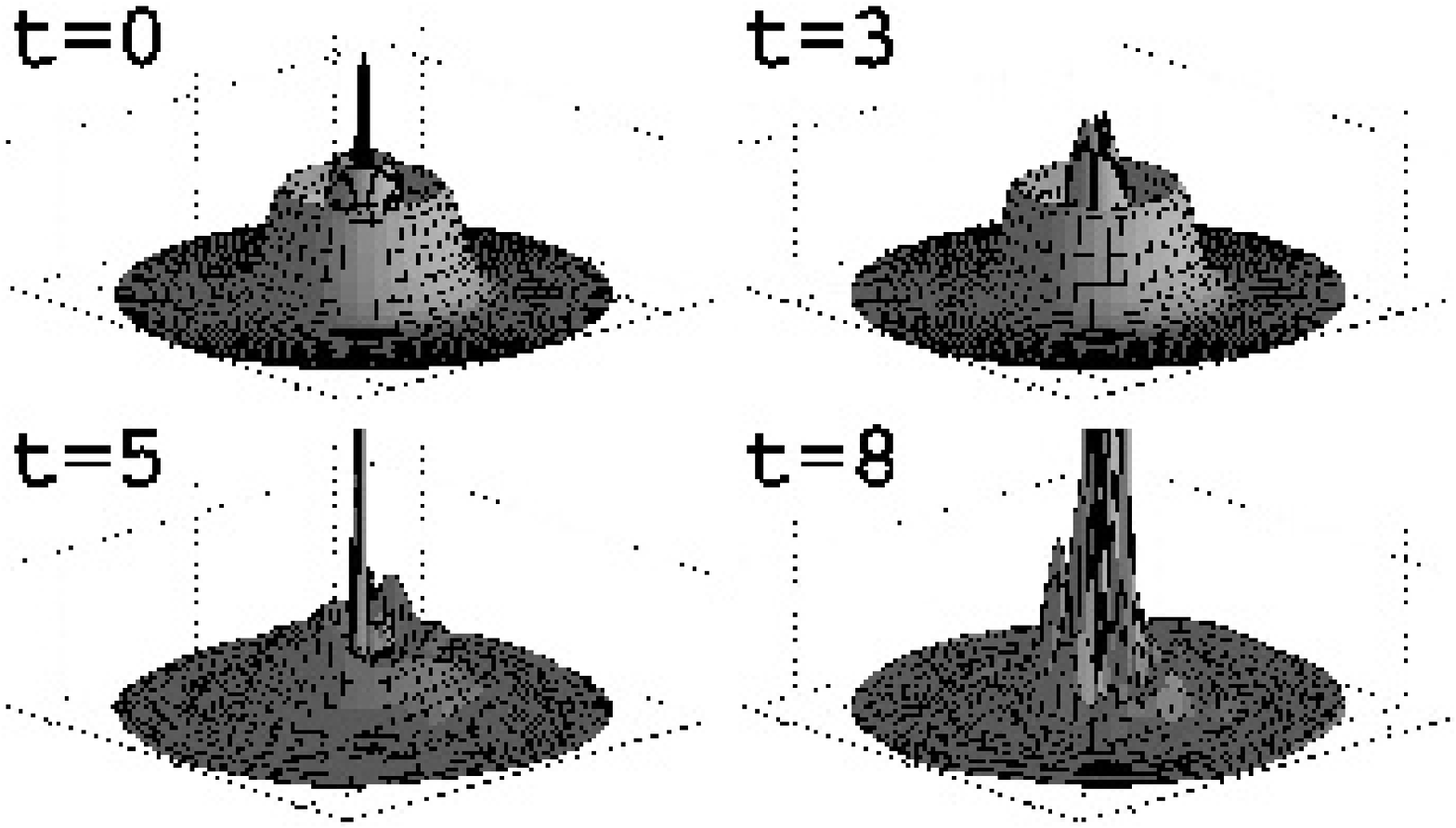}
~~~\includegraphics[width=\wfigt,angle=0,clip]{\rootfig 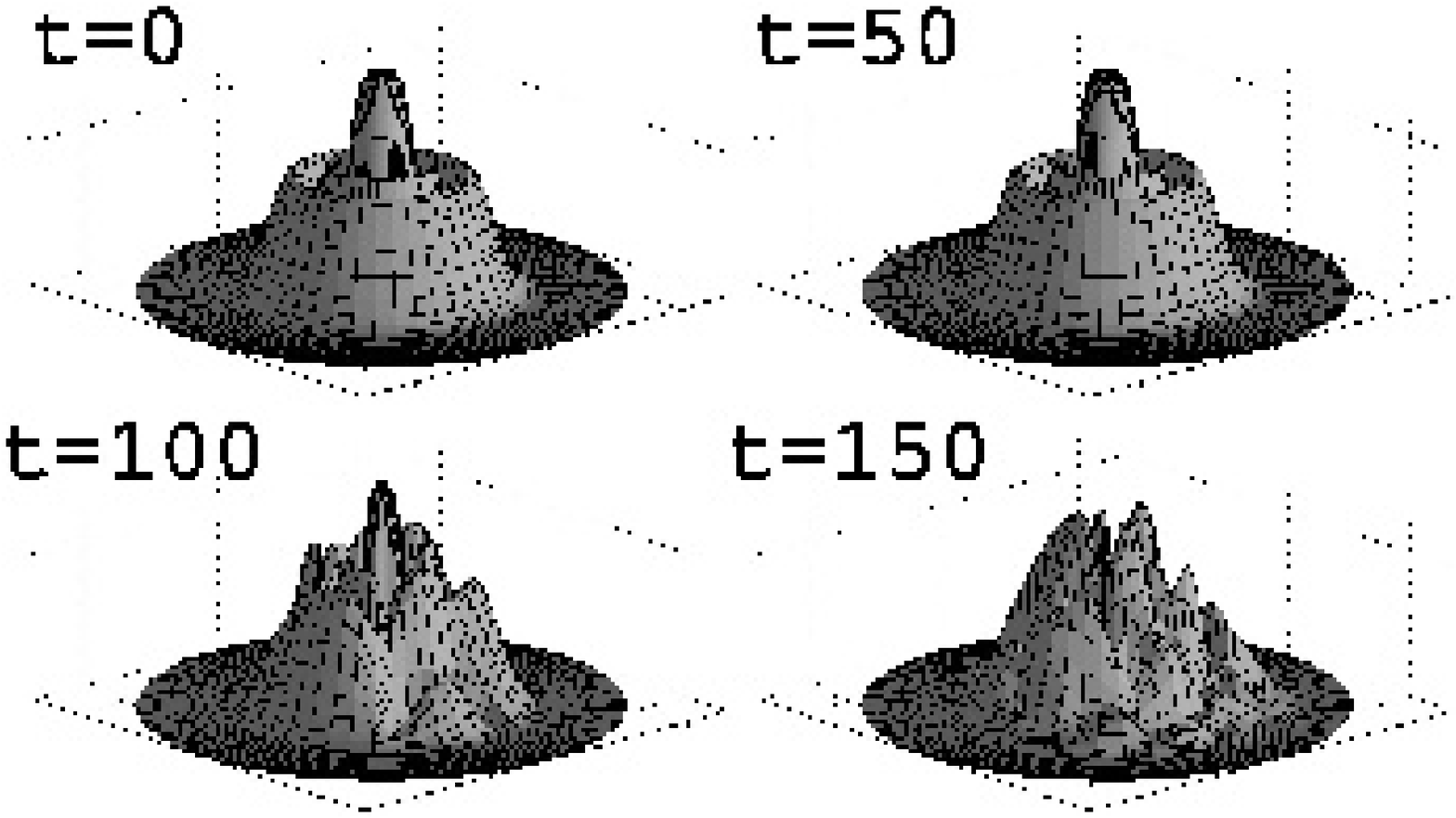}
\caption{Same as Fig.~\ref{Ground} ($m=0$) for the second excited state
($n_r=2$).}
\label{SecondEx}
\end{figure}

\begin{figure}[tbp]
\includegraphics[width=\wfig,angle=0,clip]{\rootfigsmall  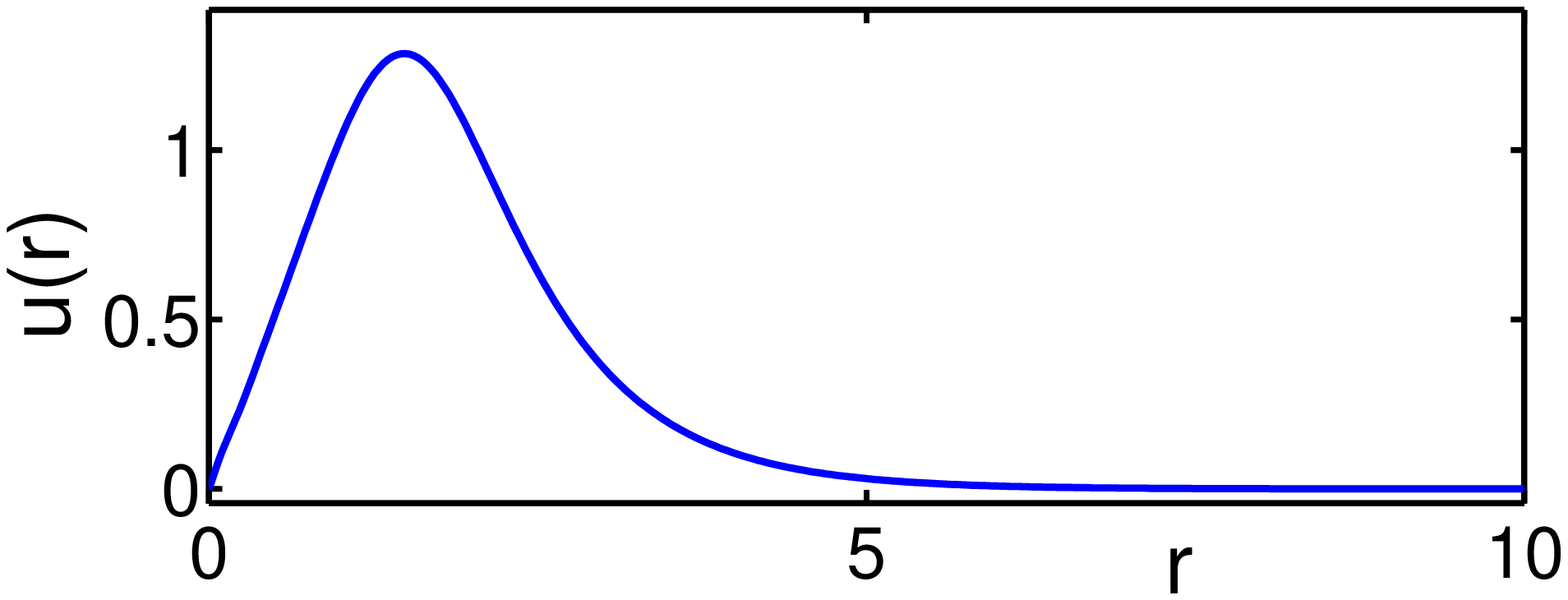}
\includegraphics[width=\wfig,angle=0,clip]{\rootfigsmall  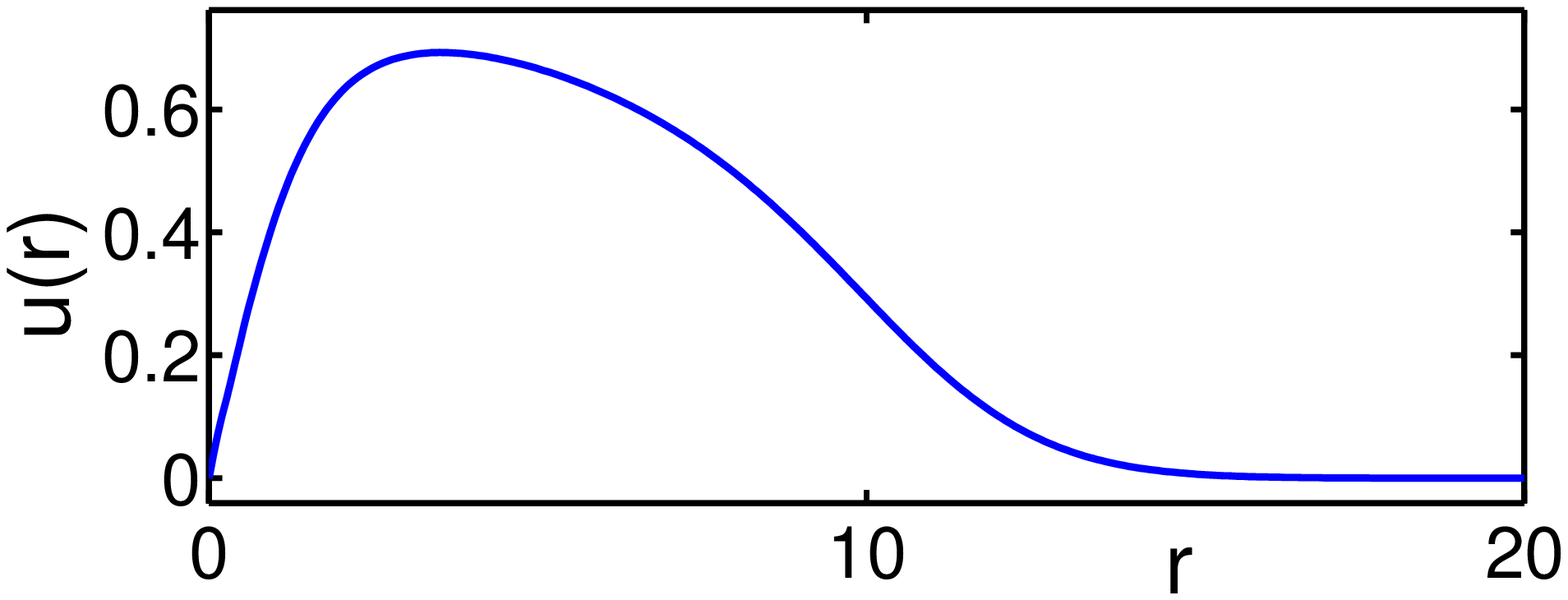}\\
\includegraphics[width=\wfig,angle=0,clip]{\rootfigsmall  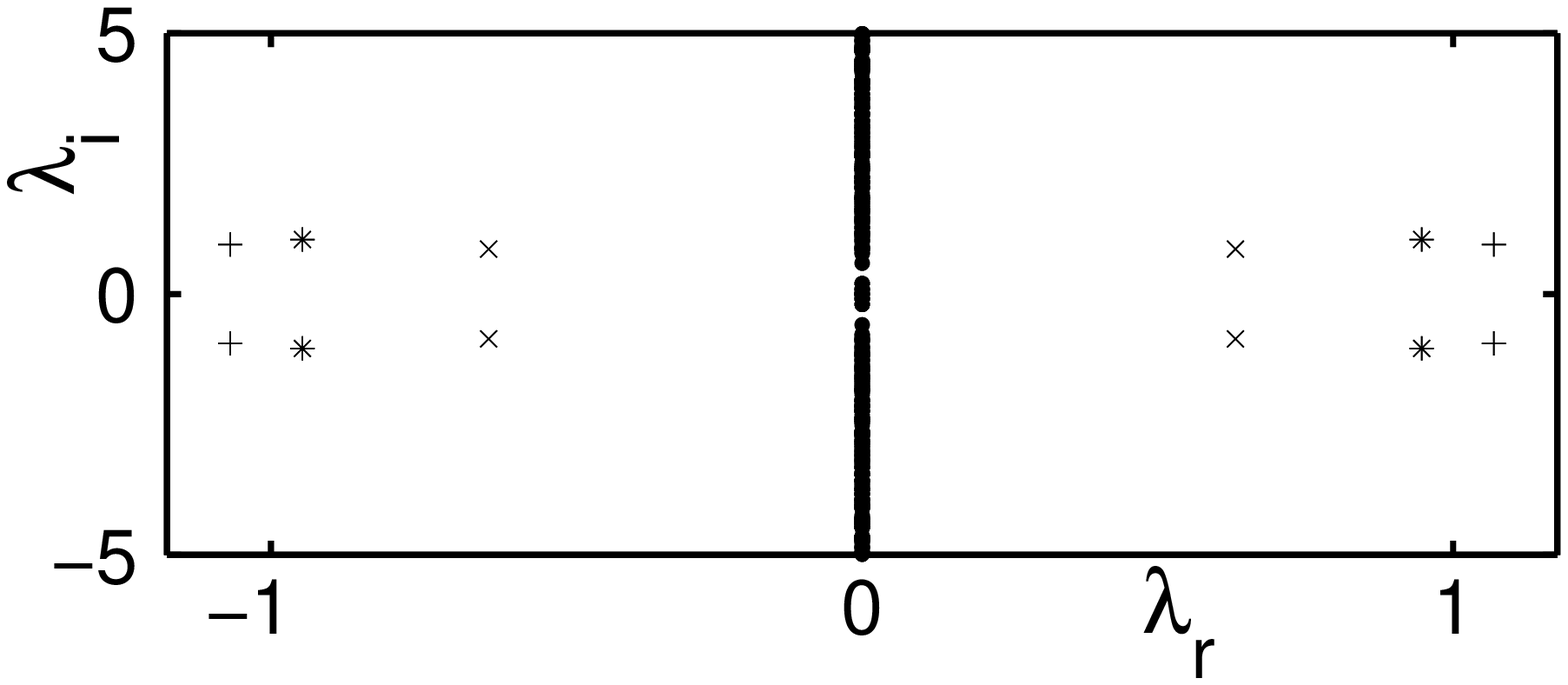}
\includegraphics[width=\wfig,angle=0,clip]{\rootfig 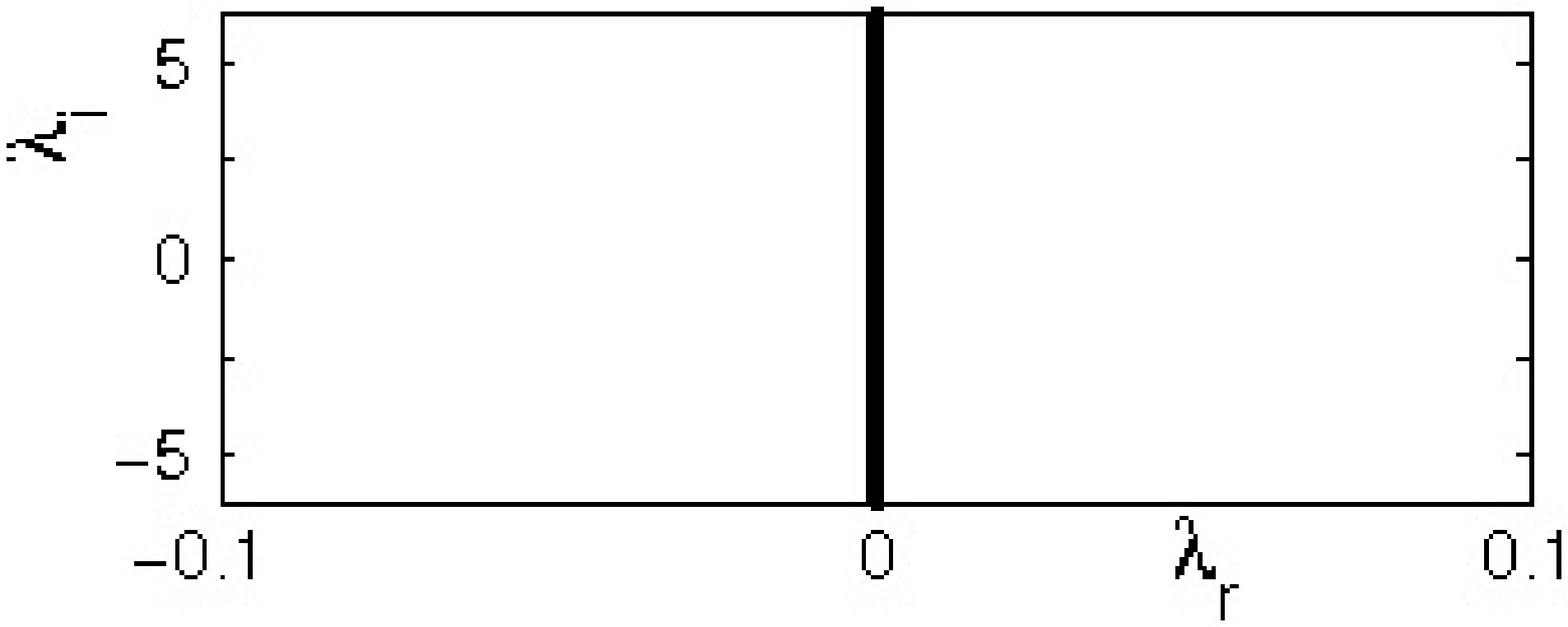}\\
~~\includegraphics[width=\wfigt,angle=0,clip]{\rootfig 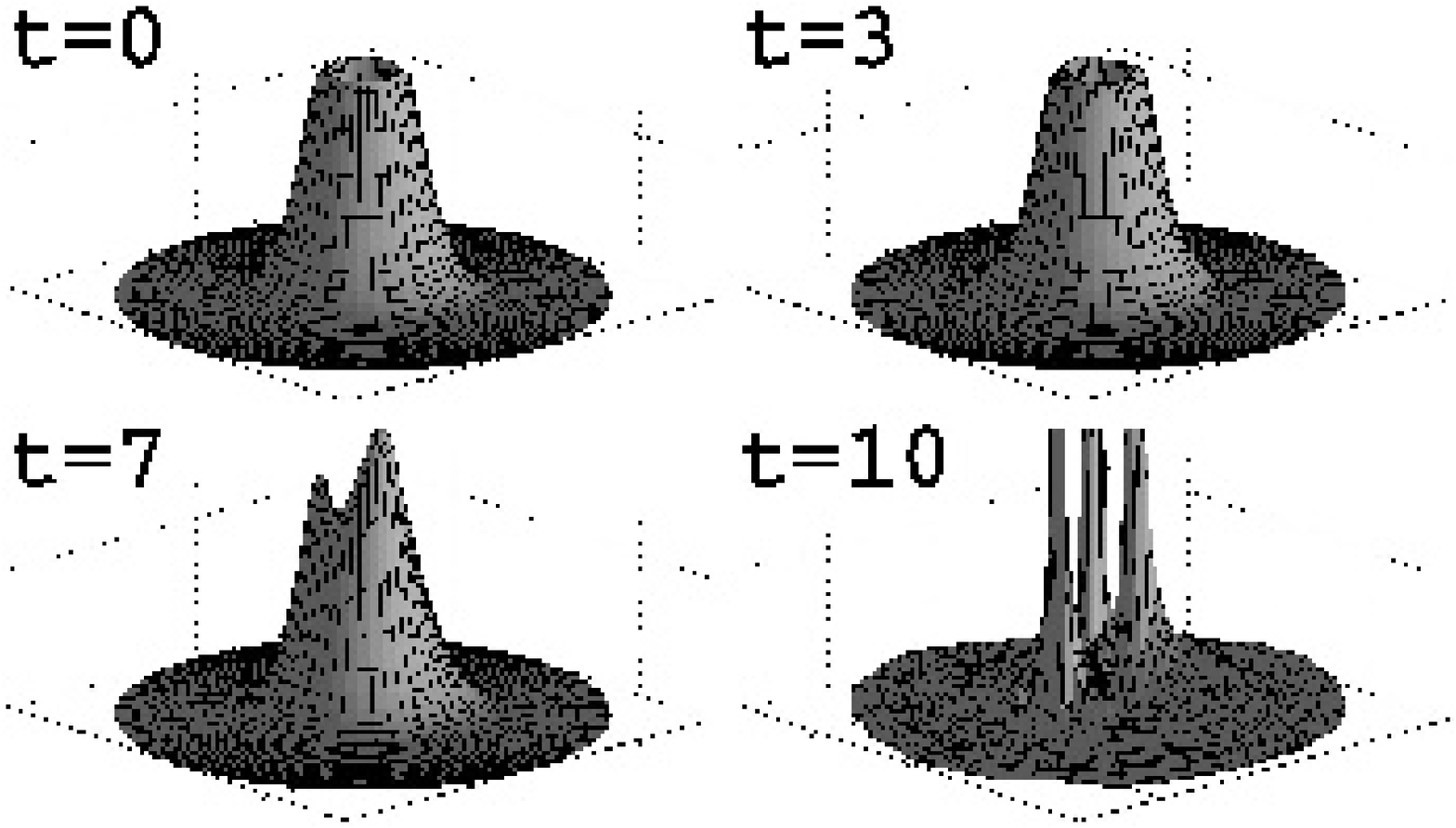}
~~~\includegraphics[width=\wfigt,angle=0,clip]{\rootfig 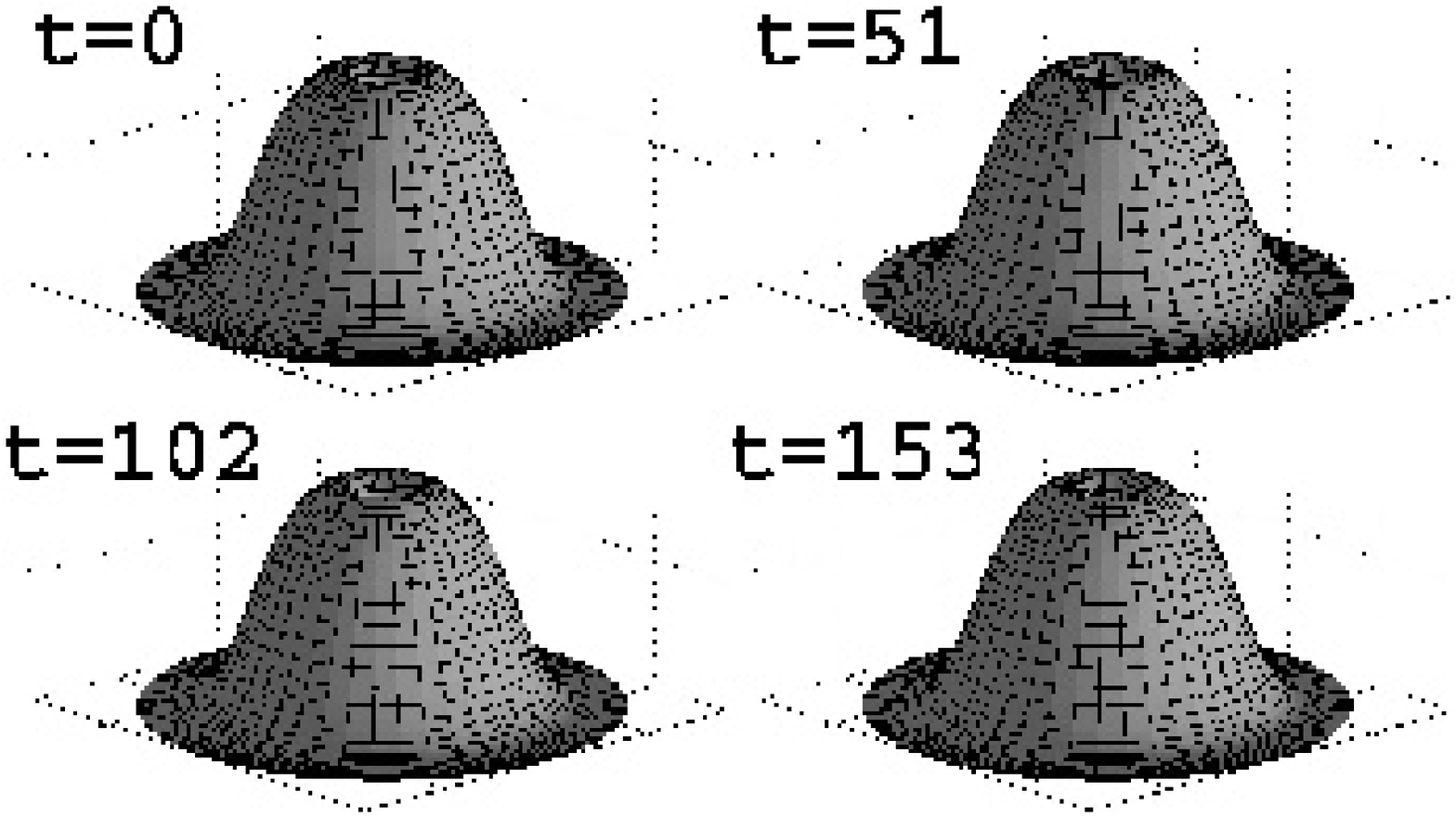}
\caption{
Same as Fig.~\ref{Ground} for $m=1$.
}
\label{Groundm1}
\end{figure}

\begin{figure}[tbp]
\includegraphics[width=\wfig,angle=0,clip]{\rootfigsmall  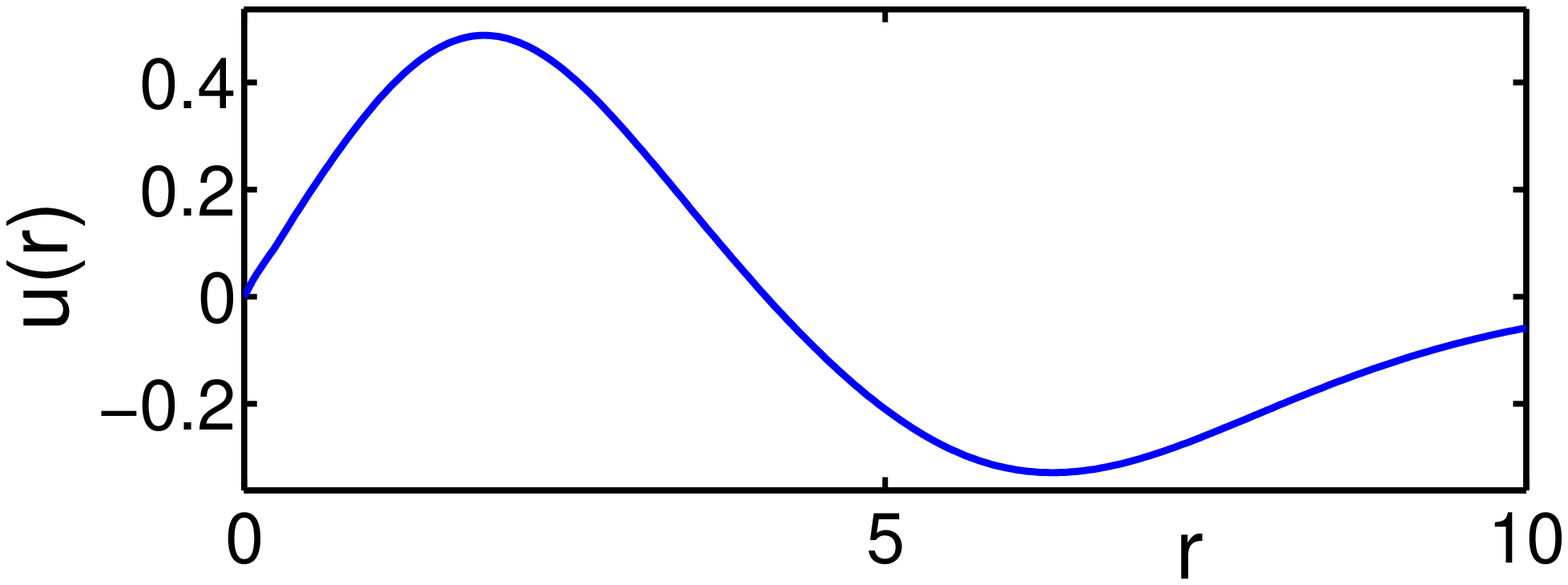}
\includegraphics[width=\wfig,angle=0,clip]{\rootfigsmall  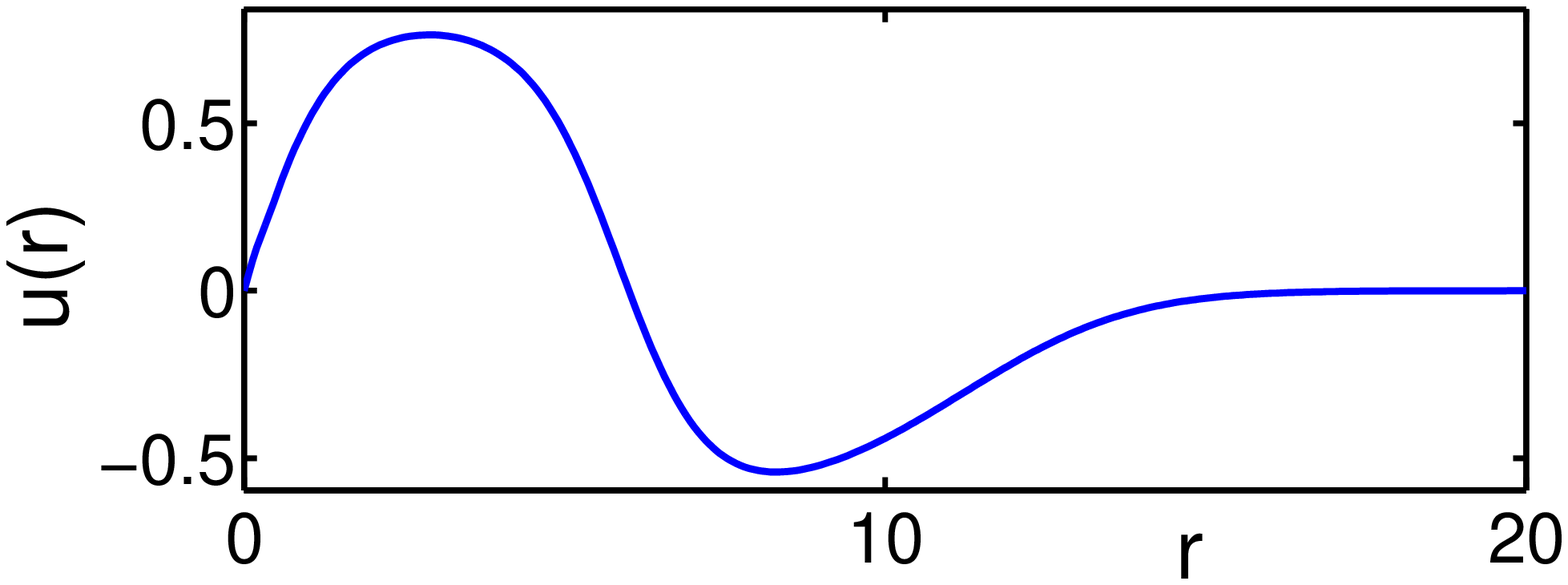}\\
\includegraphics[width=\wfig,angle=0,clip]{\rootfigsmall  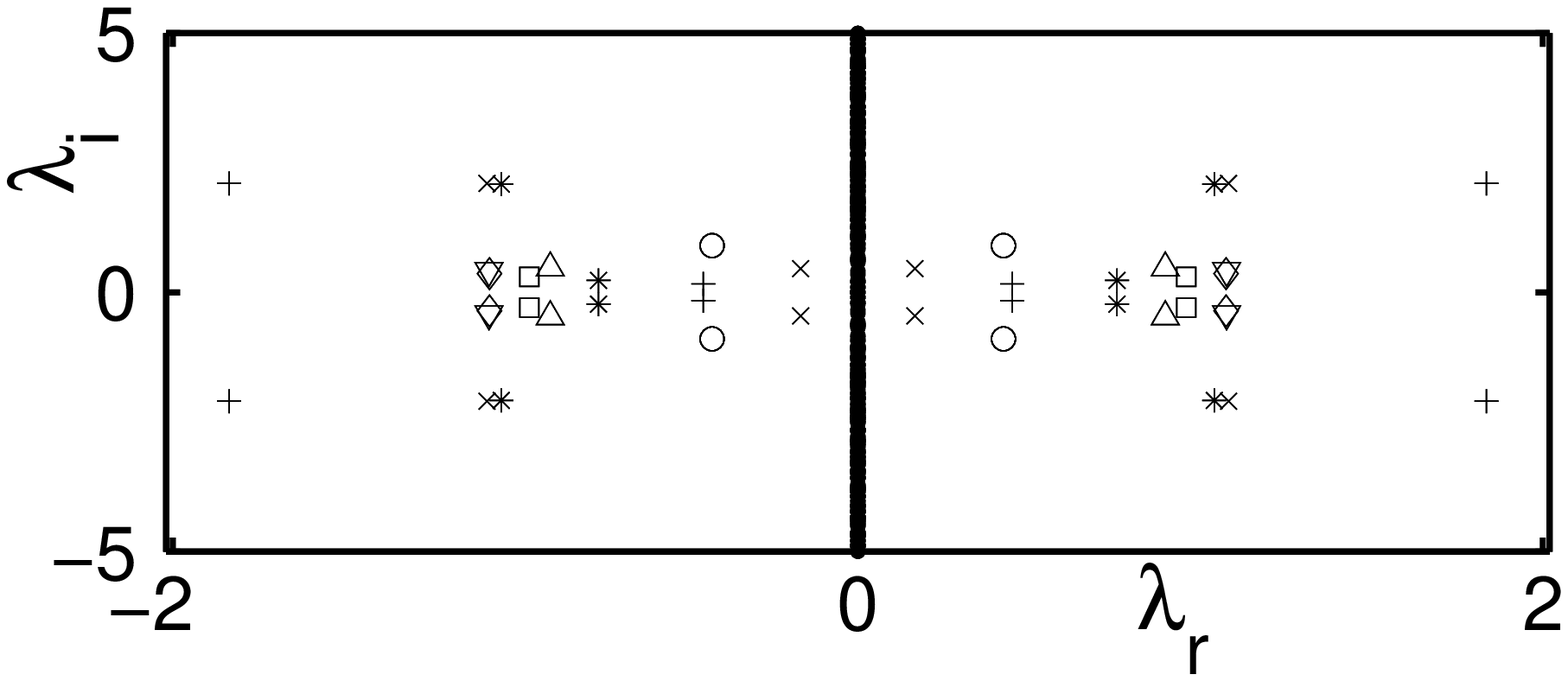}
\includegraphics[width=\wfig,angle=0,clip]{\rootfig 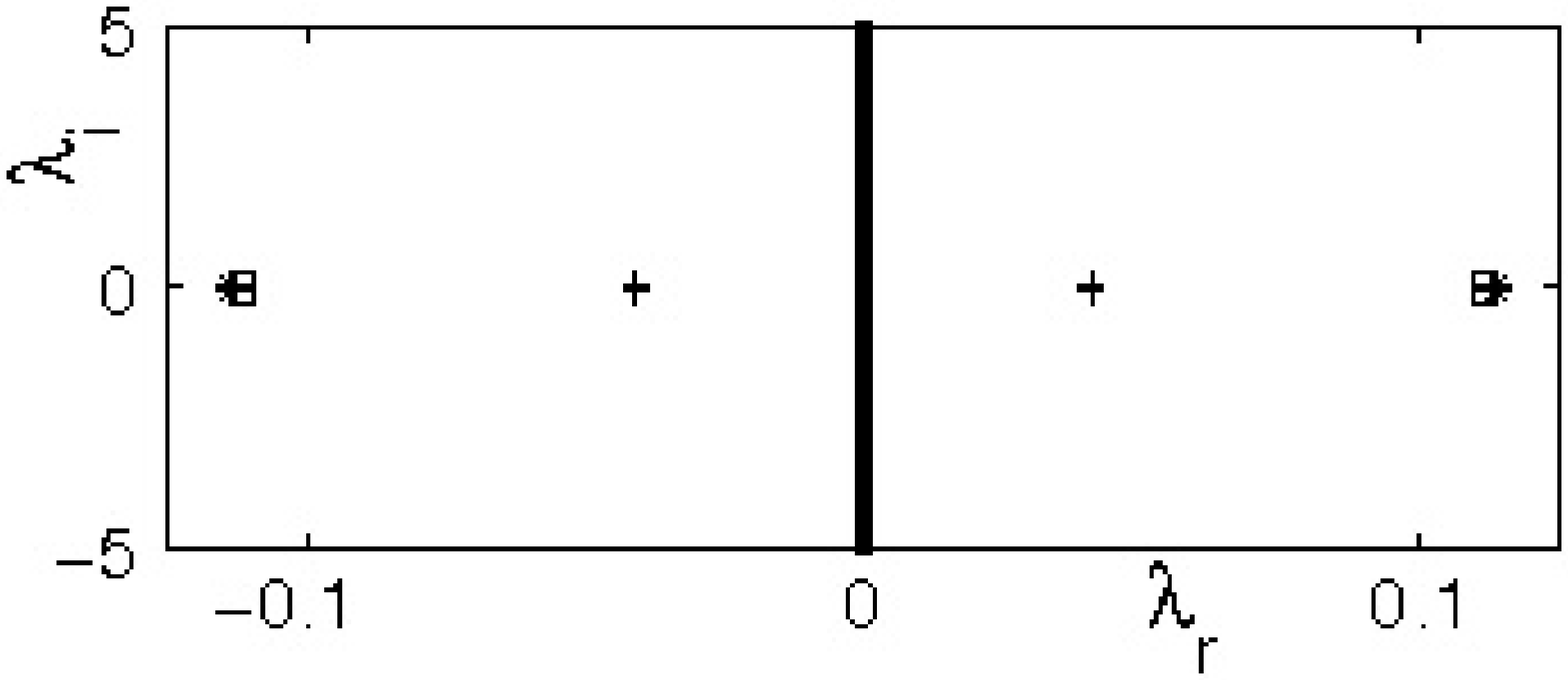}\\
~~\includegraphics[width=\wfigt,angle=0,clip]{\rootfig 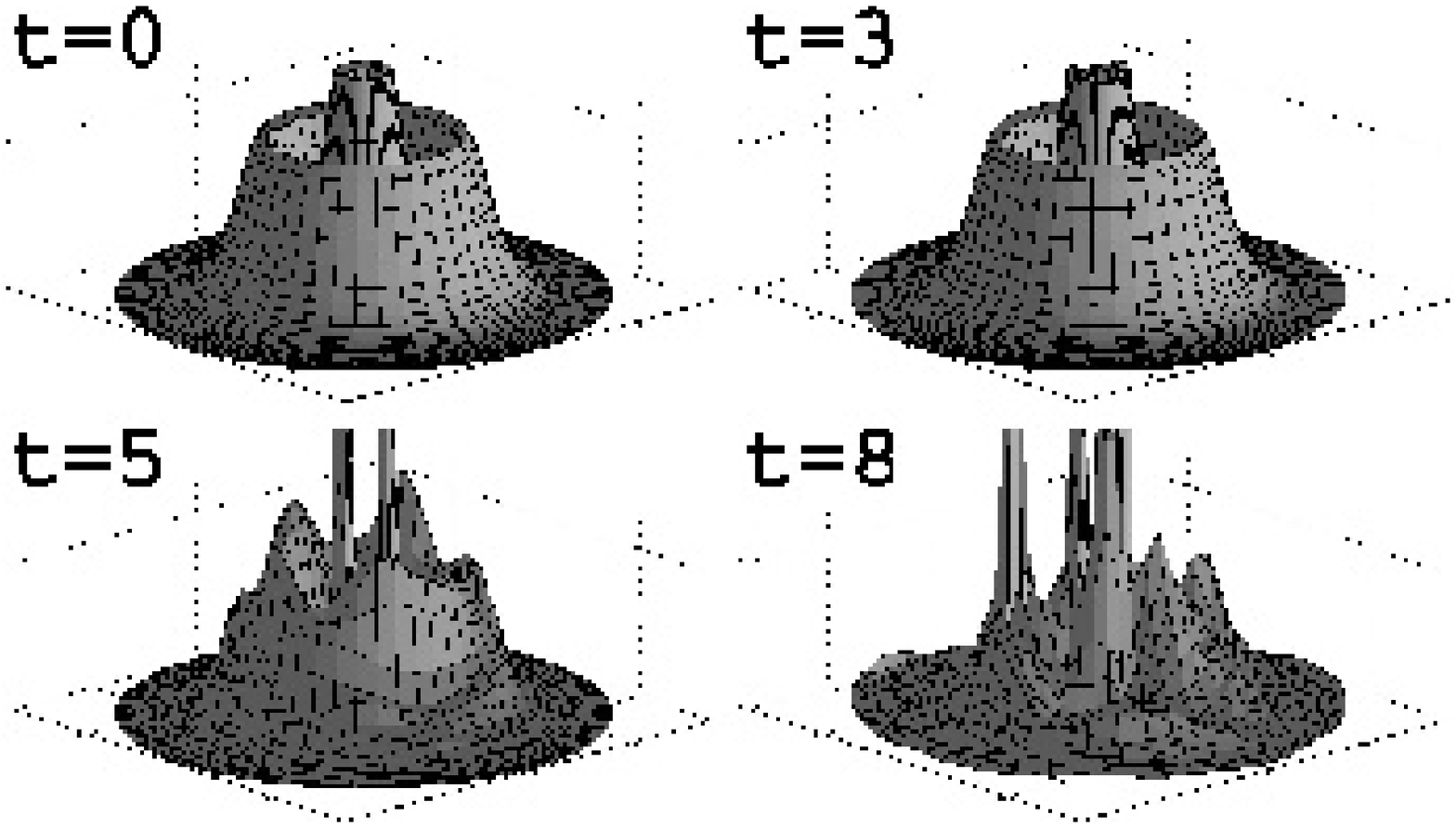}
~~~\includegraphics[width=\wfigt,angle=0,clip]{\rootfig 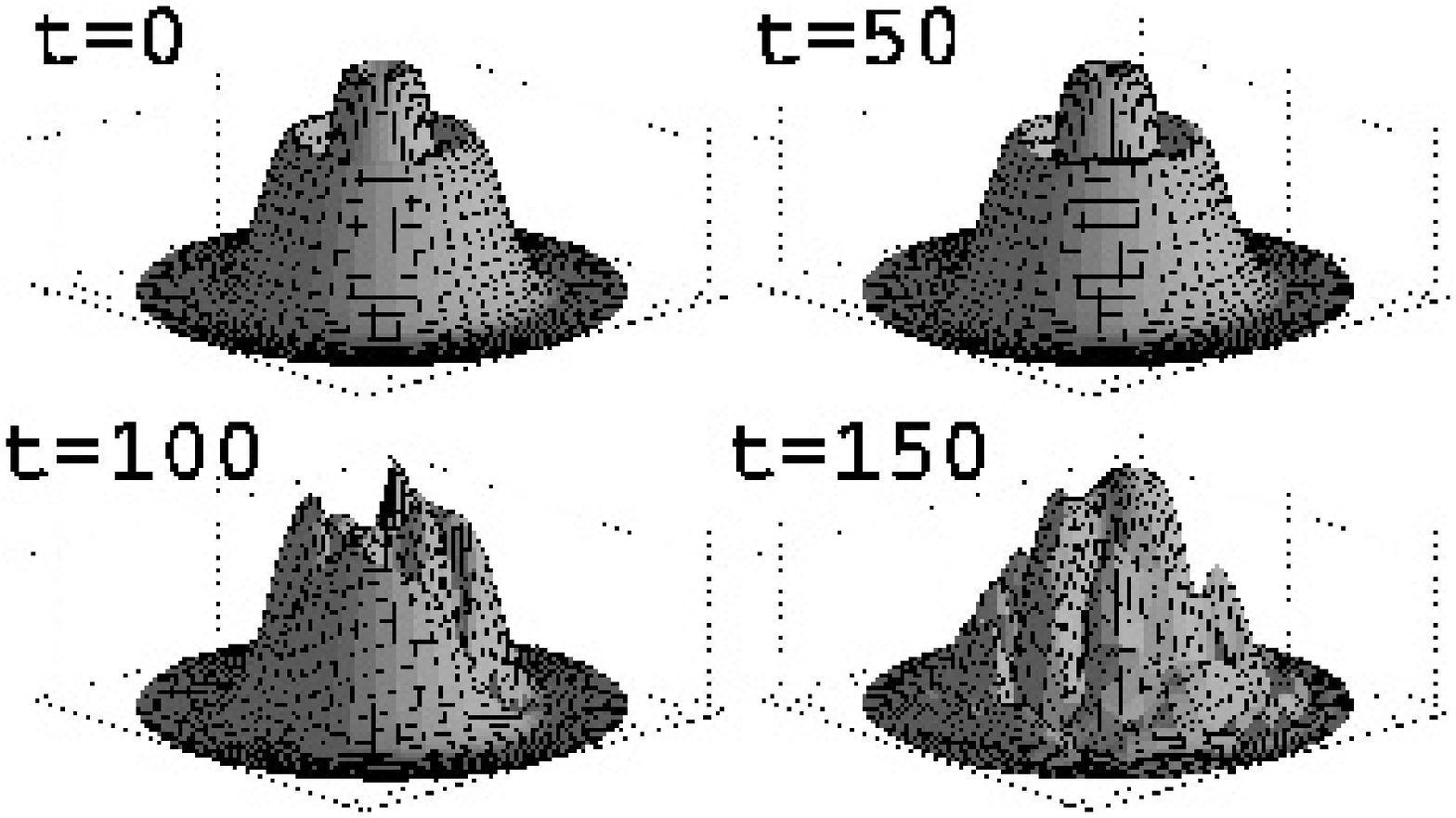}
\caption{Same as Fig.~\ref{Groundm1} ($m=1$) for the first excited state
$(n_r=1)$.
}
\label{FirstExm1}
\end{figure}

\begin{figure}[tbp]
\includegraphics[width=\wfig,angle=0,clip]{\rootfigsmall  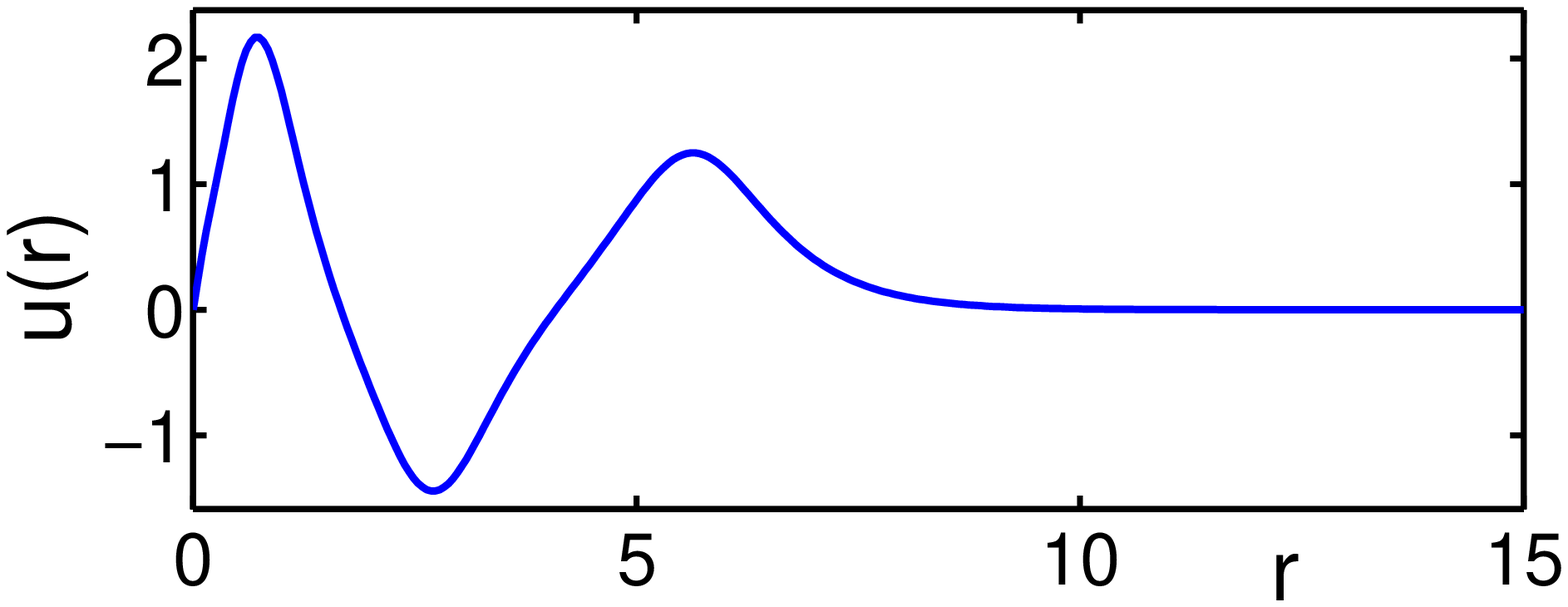}
\includegraphics[width=\wfig,angle=0,clip]{\rootfigsmall  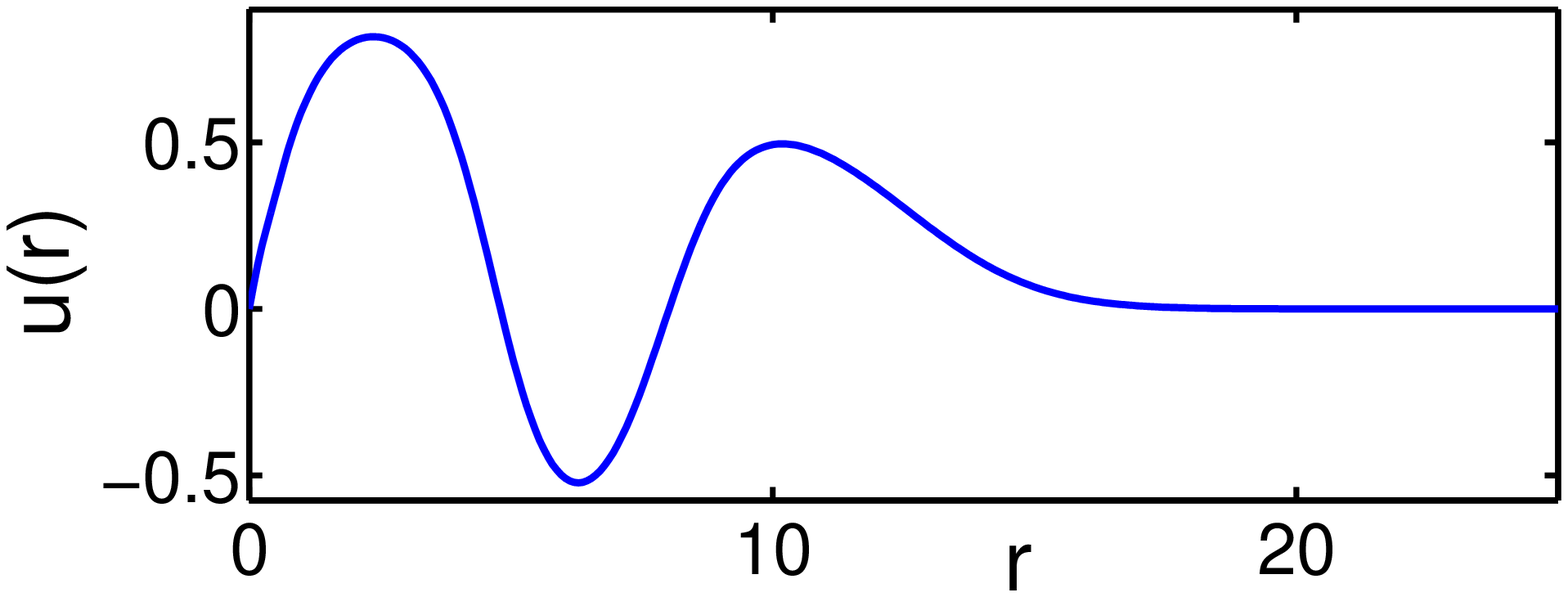}\\
\includegraphics[width=\wfig,angle=0,clip]{\rootfig 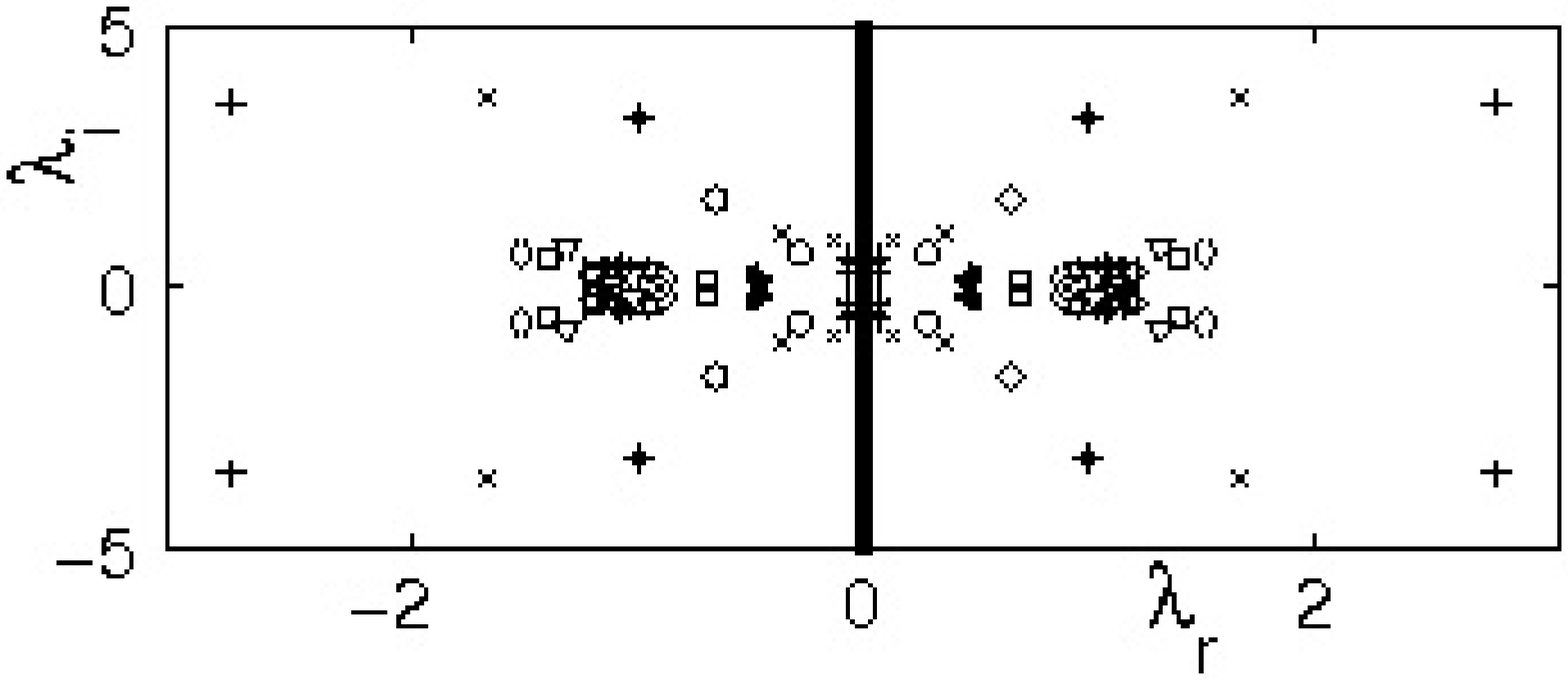}
\includegraphics[width=\wfig,angle=0,clip]{\rootfig 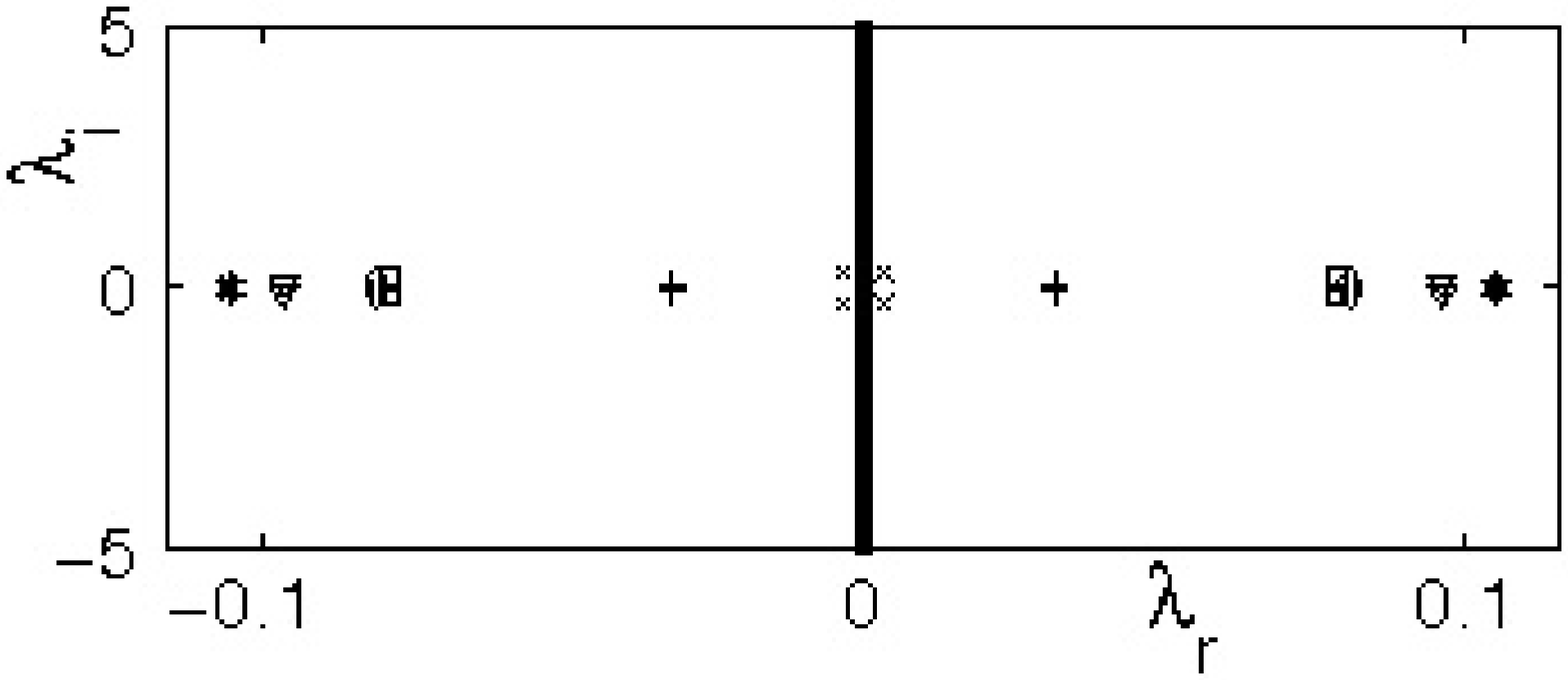}\\
~~\includegraphics[width=\wfigt,angle=0,clip]{\rootfig 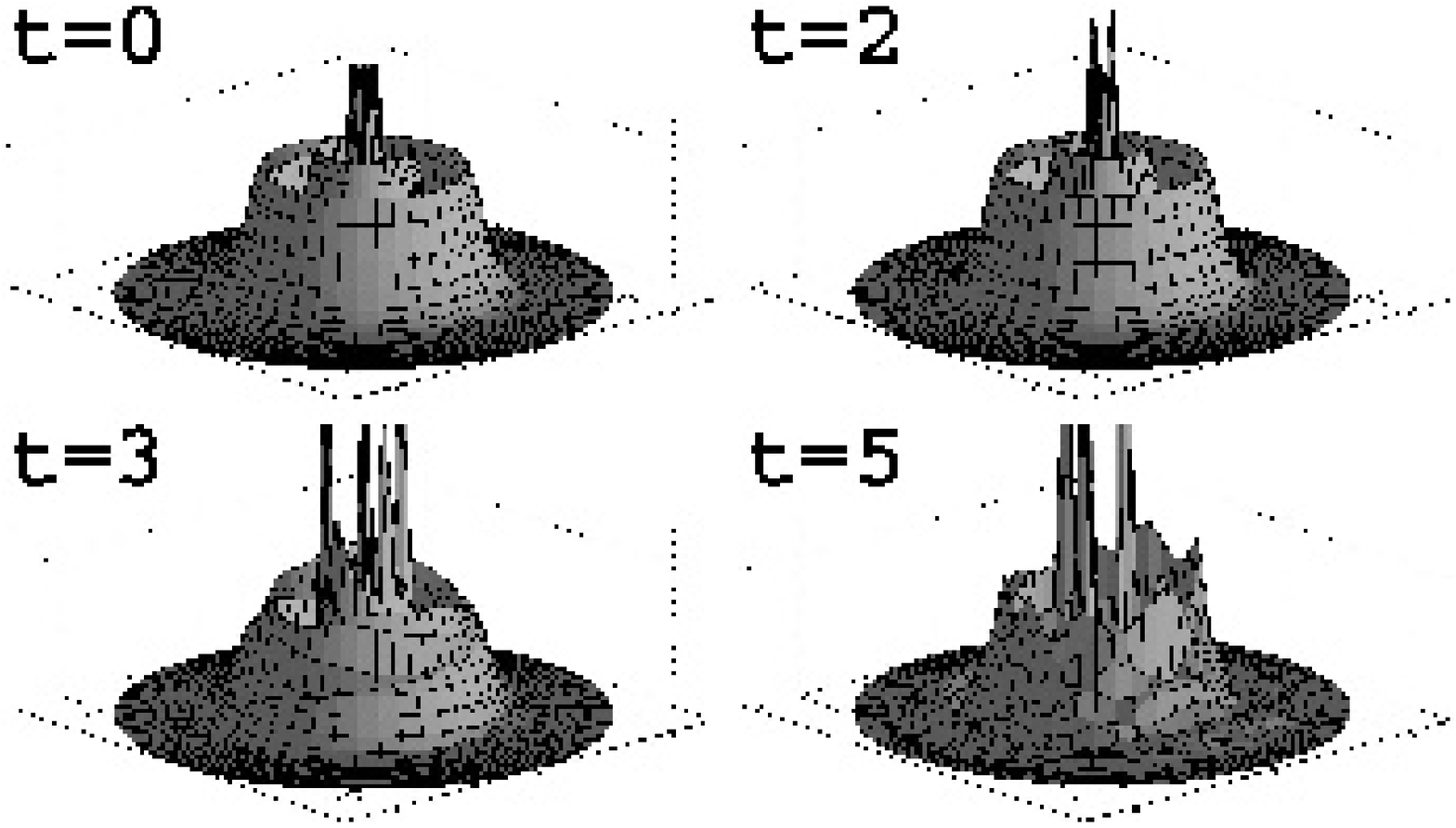}
~~~\includegraphics[width=\wfigt,angle=0,clip]{\rootfig 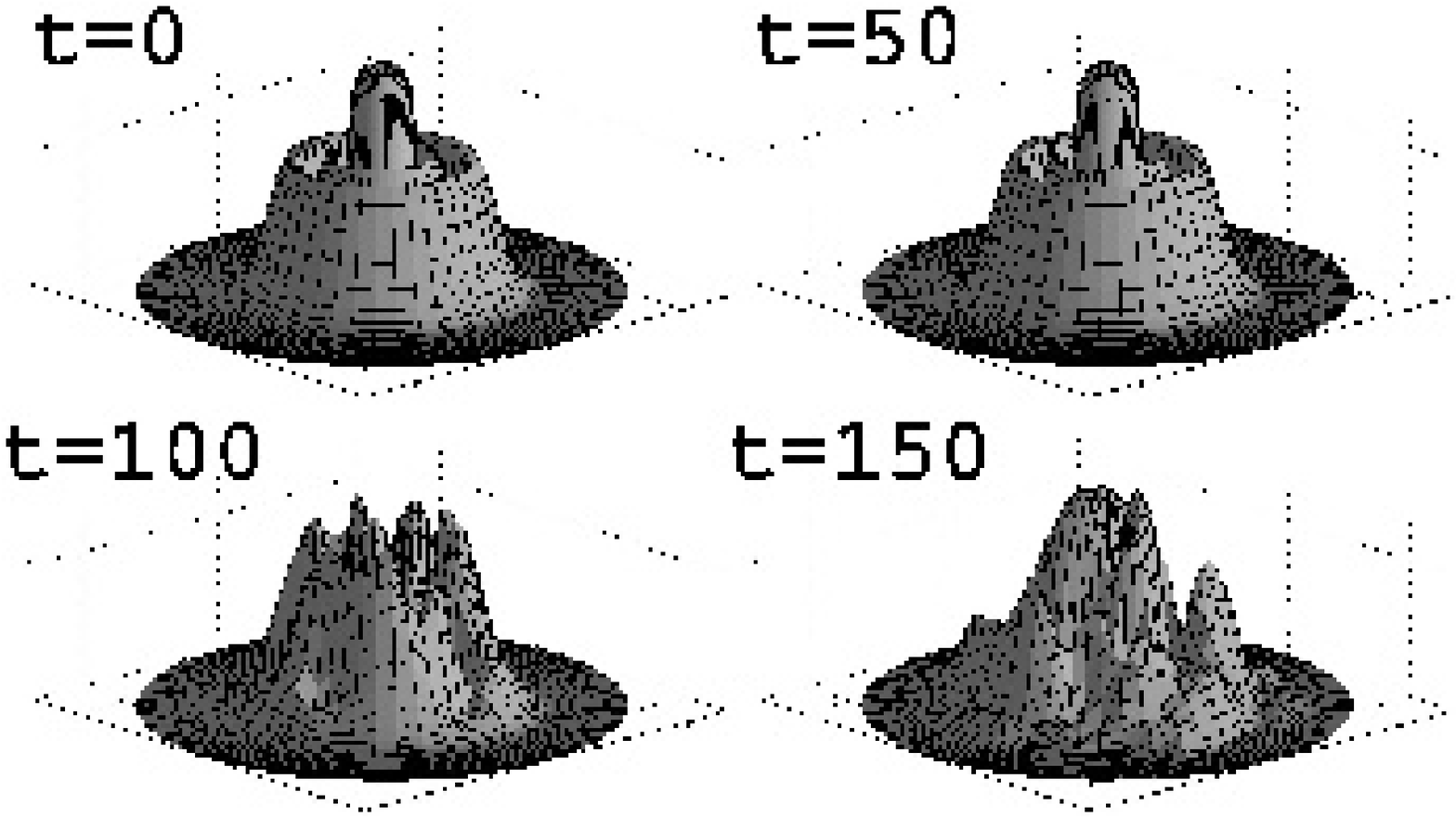}
\caption{Same as Fig.~\ref{Groundm1} ($m=1$) for the second excited state
($n_r=2$).}
\label{SecondExm1}
\end{figure}

\begin{figure}[tbp]
\includegraphics[width=\wfig,angle=0,clip]{\rootfigsmall  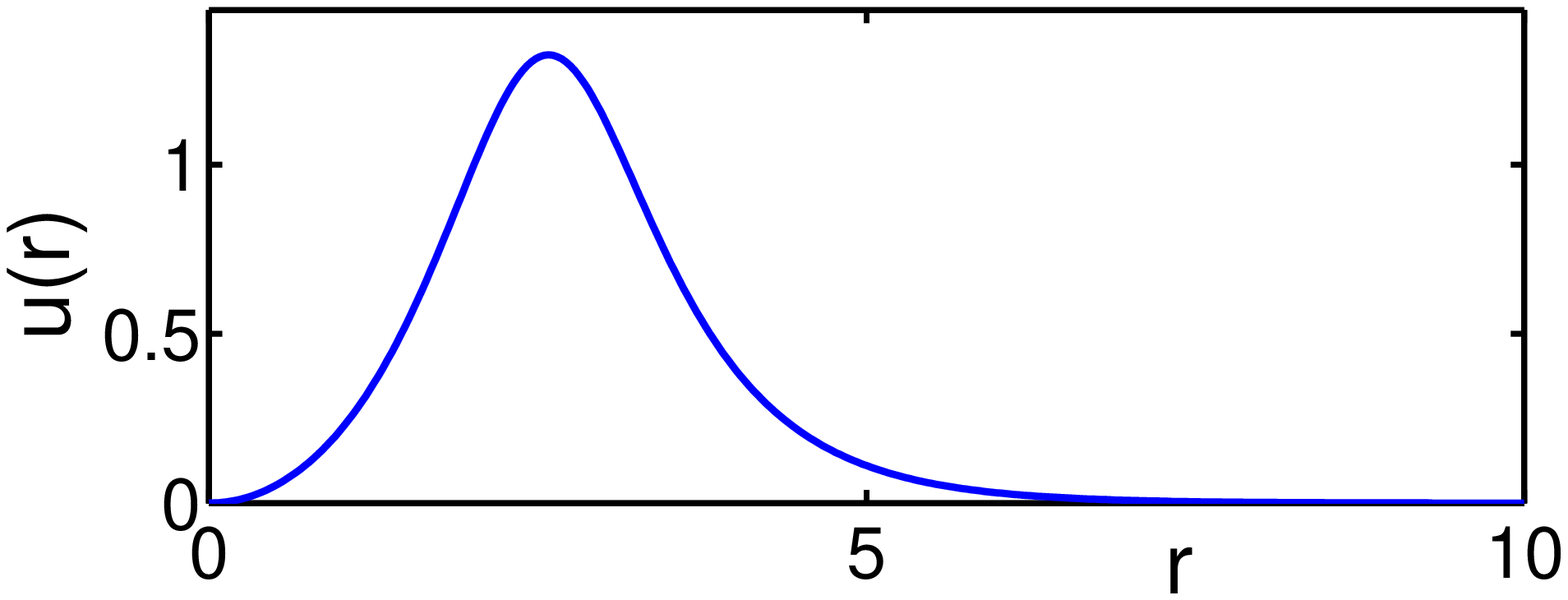}
\includegraphics[width=\wfig,angle=0,clip]{\rootfigsmall  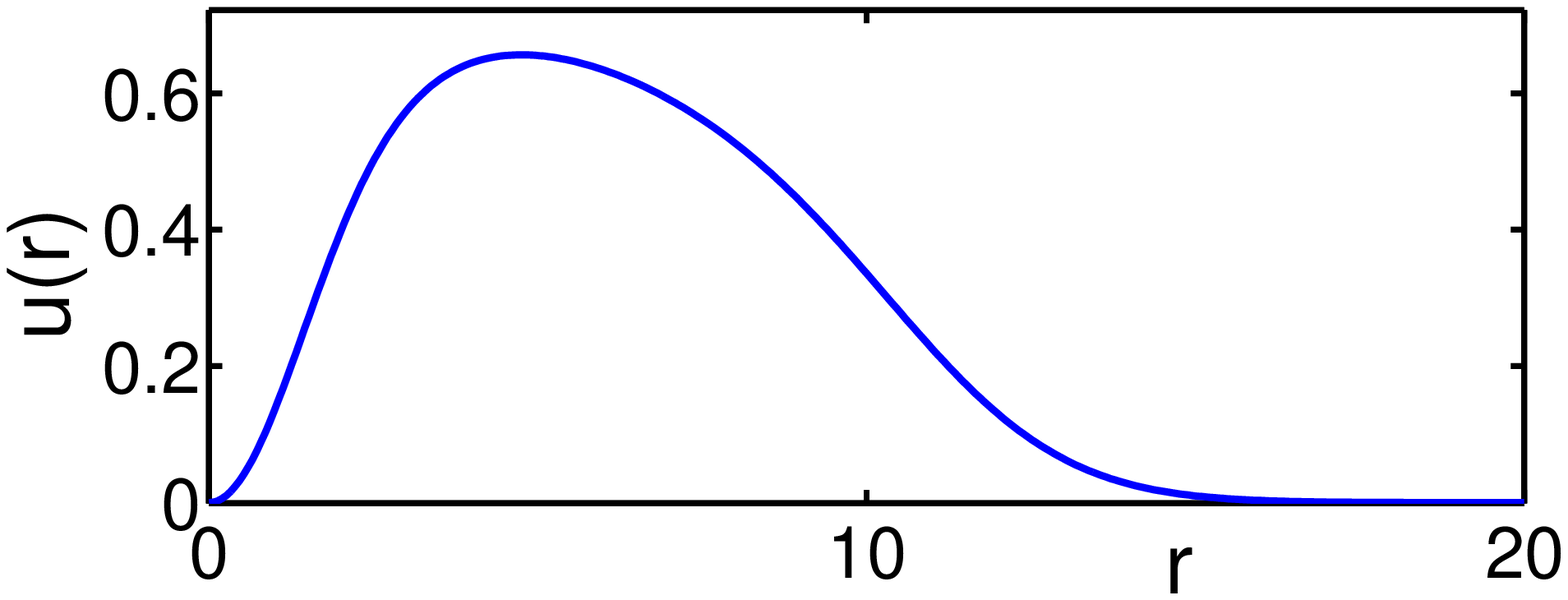}\\
\includegraphics[width=\wfig,angle=0,clip]{\rootfigsmall  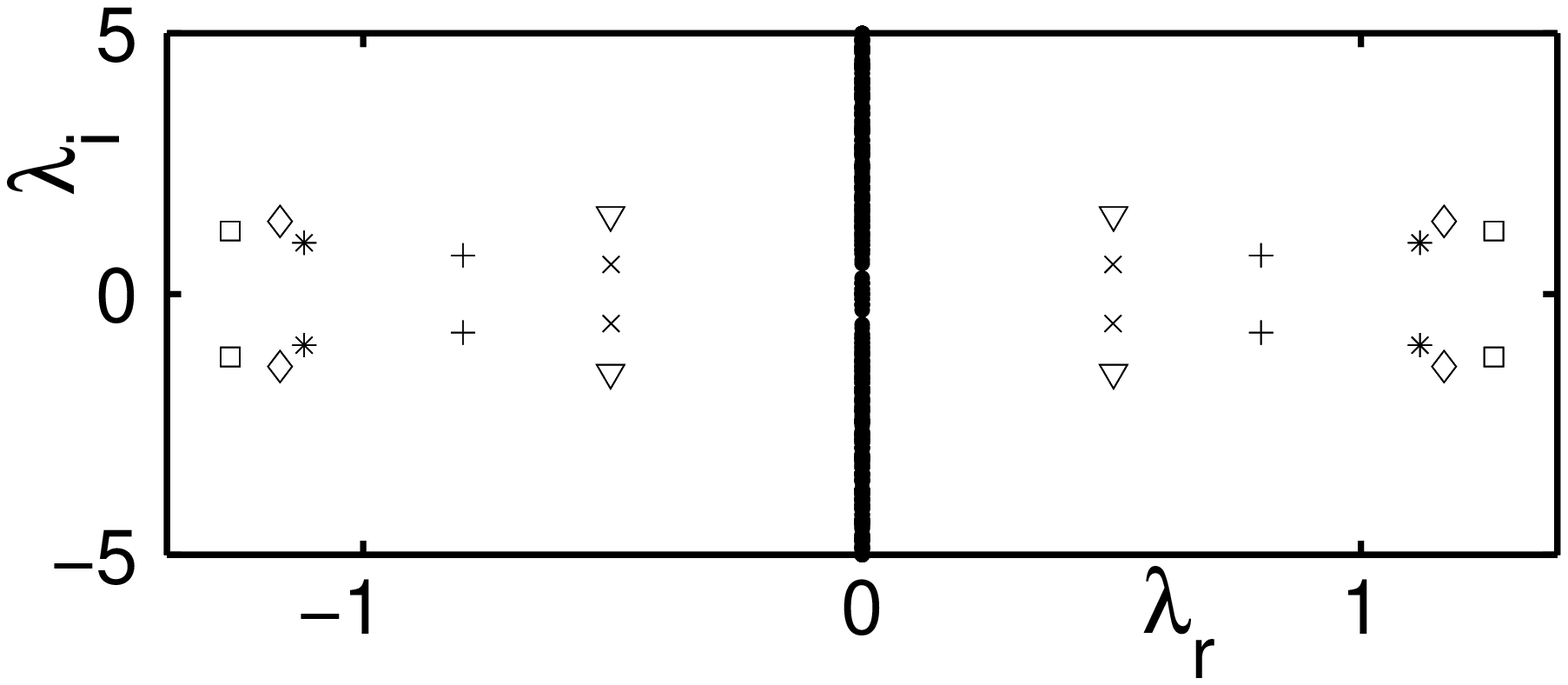}
\includegraphics[width=\wfig,angle=0,clip]{\rootfig 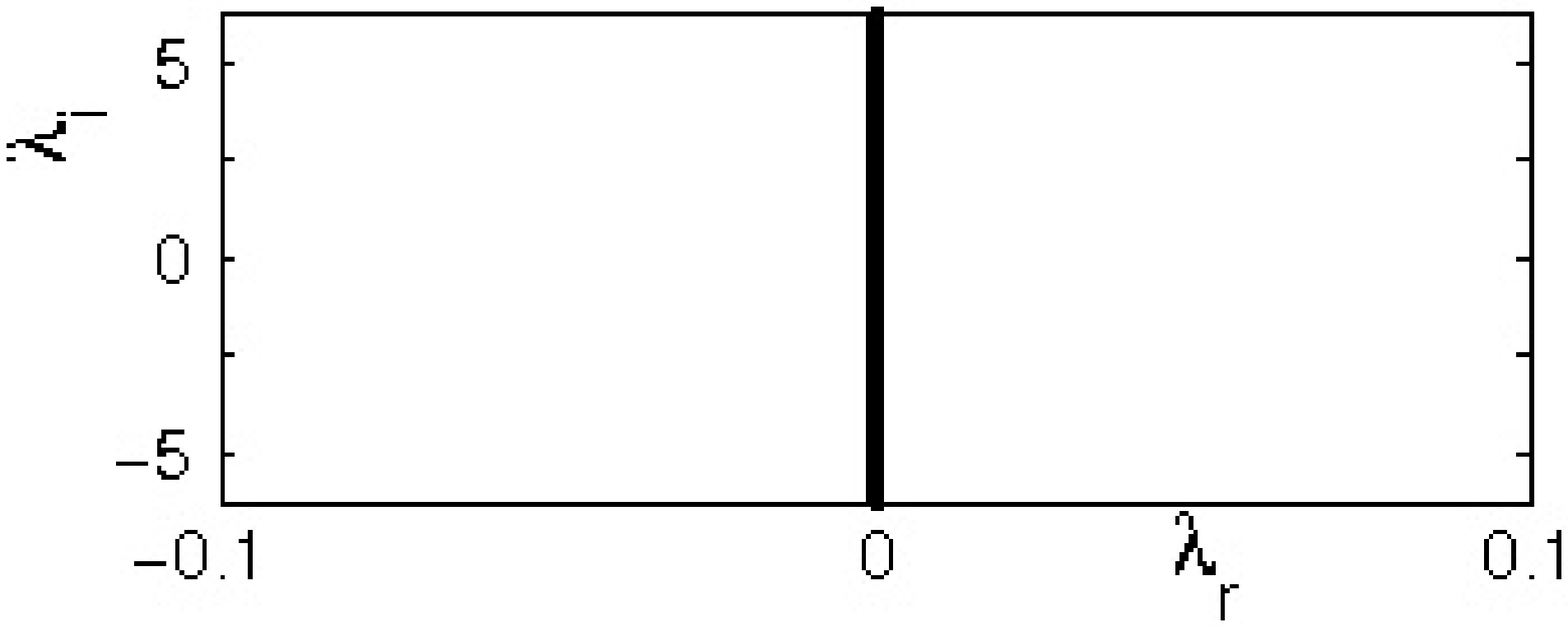}\\
~~\includegraphics[width=\wfigt,angle=0,clip]{\rootfig 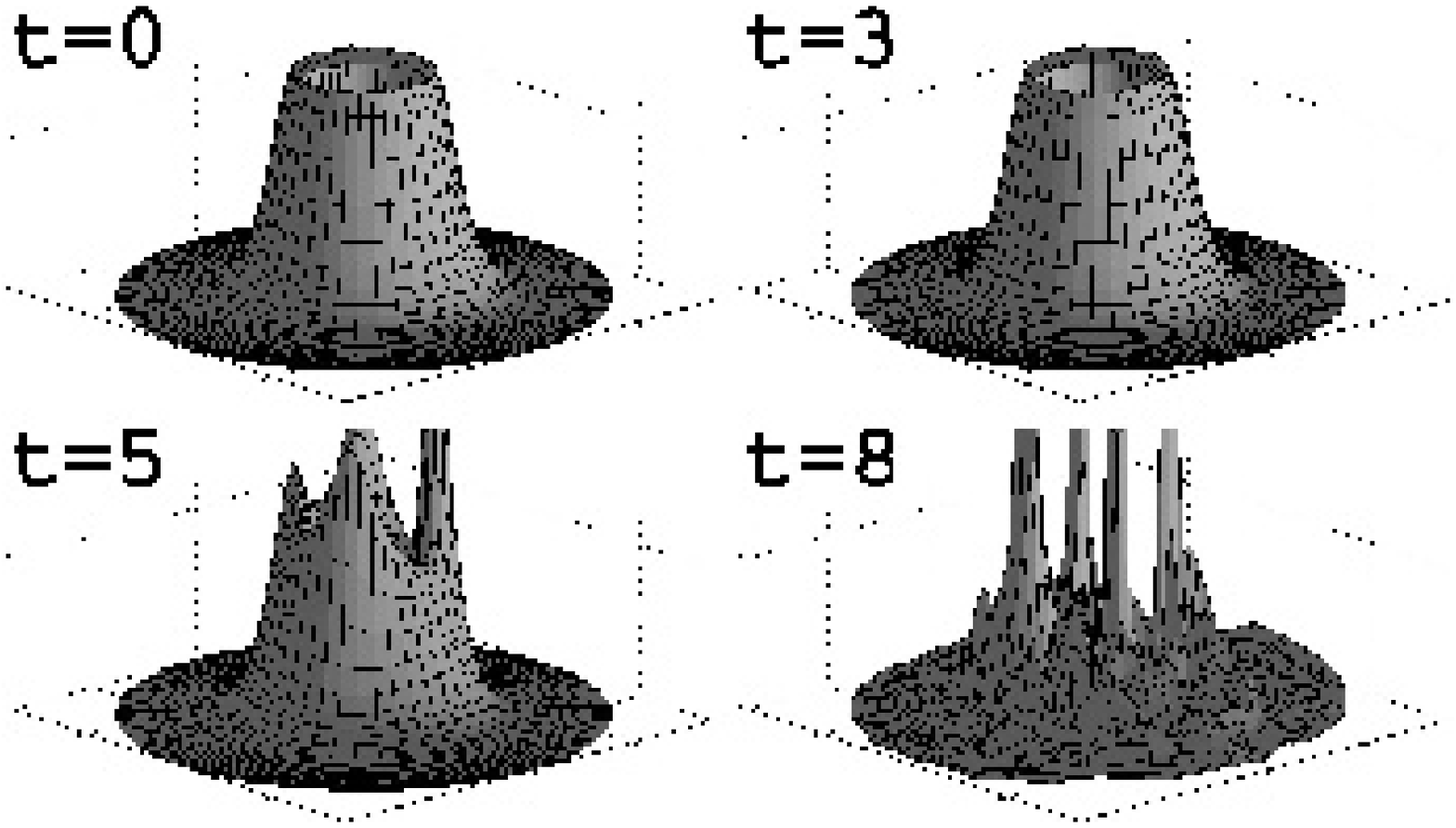}
~~~\includegraphics[width=\wfigt,angle=0,clip]{\rootfig 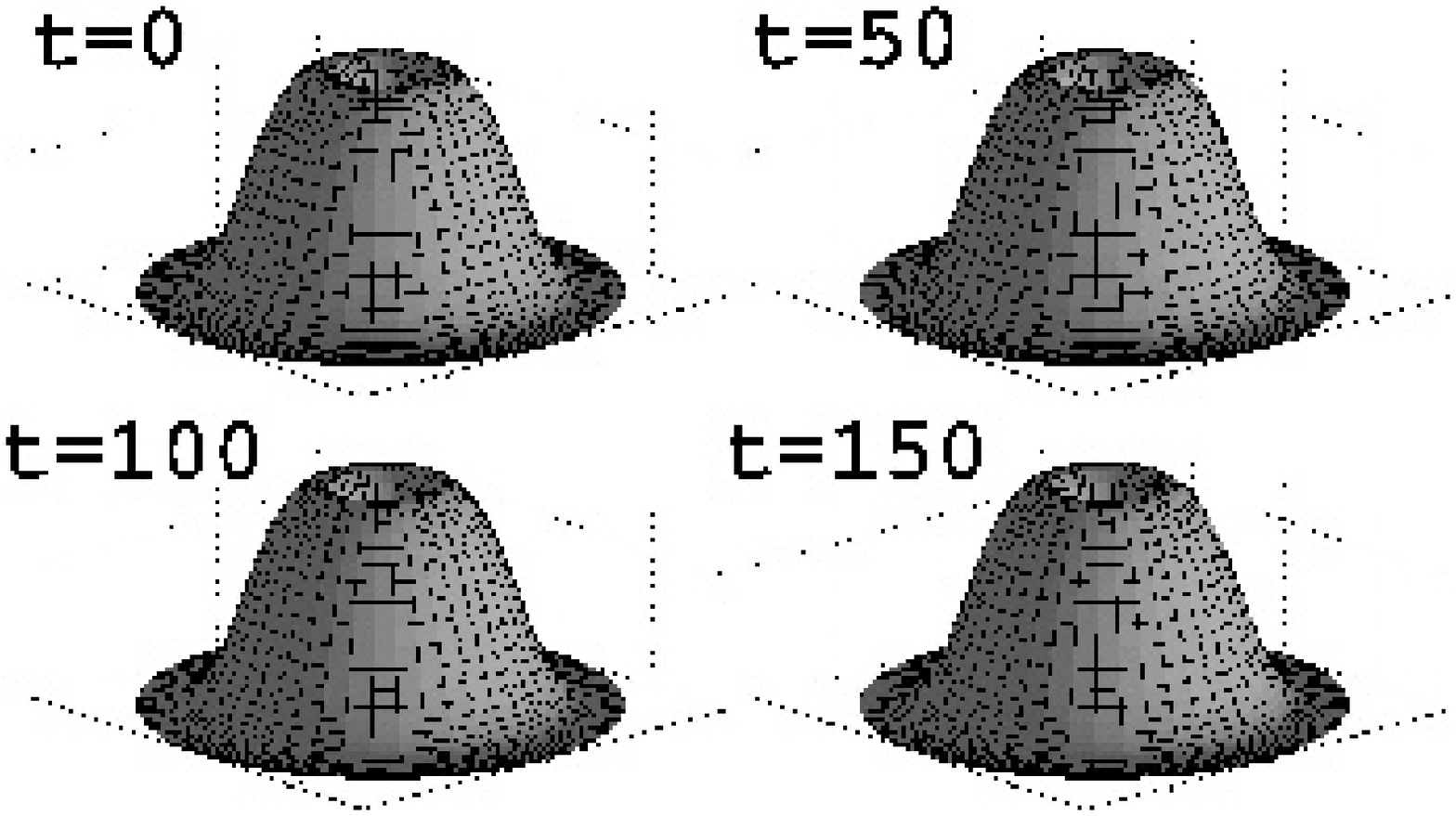}
\caption{
Same as Fig.~\ref{Ground} for $m=2$.
}
\label{Groundm2}
\end{figure}

\begin{figure}[tbp]
\includegraphics[width=\wfig,angle=0,clip]{\rootfigsmall  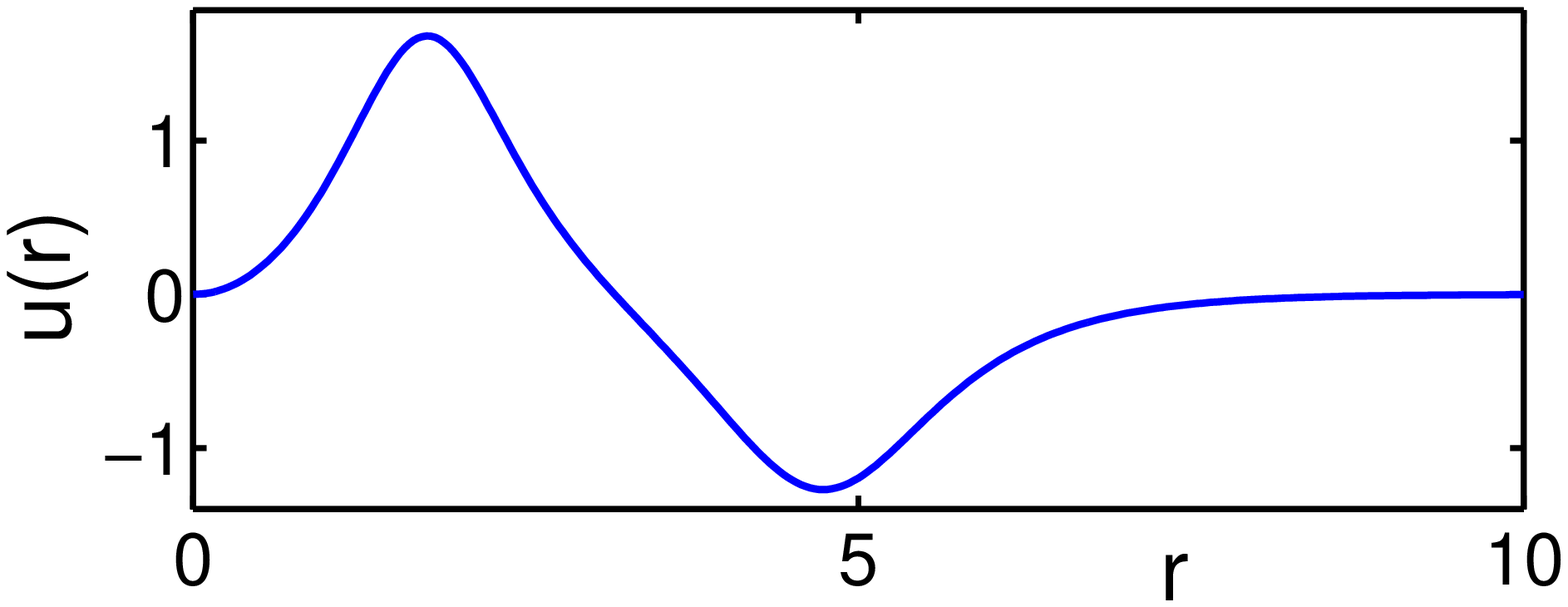}
\includegraphics[width=\wfig,angle=0,clip]{\rootfigsmall  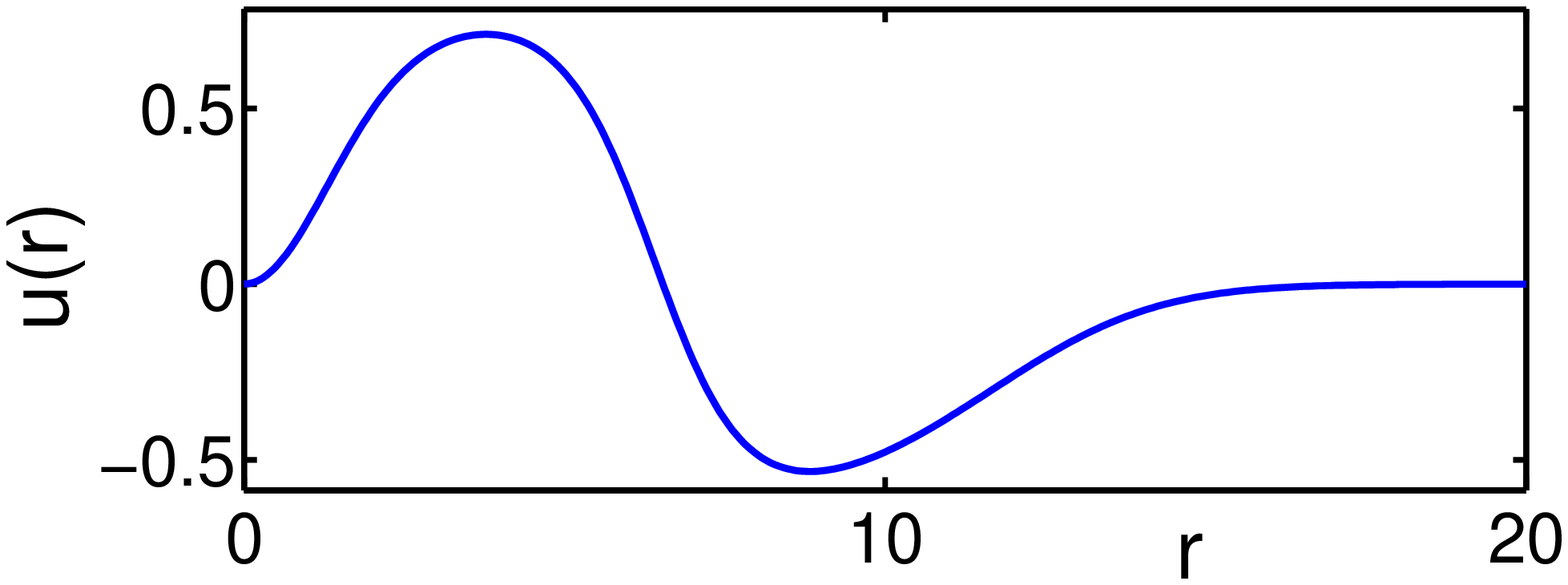}\\
\includegraphics[width=\wfig,angle=0,clip]{\rootfigsmall  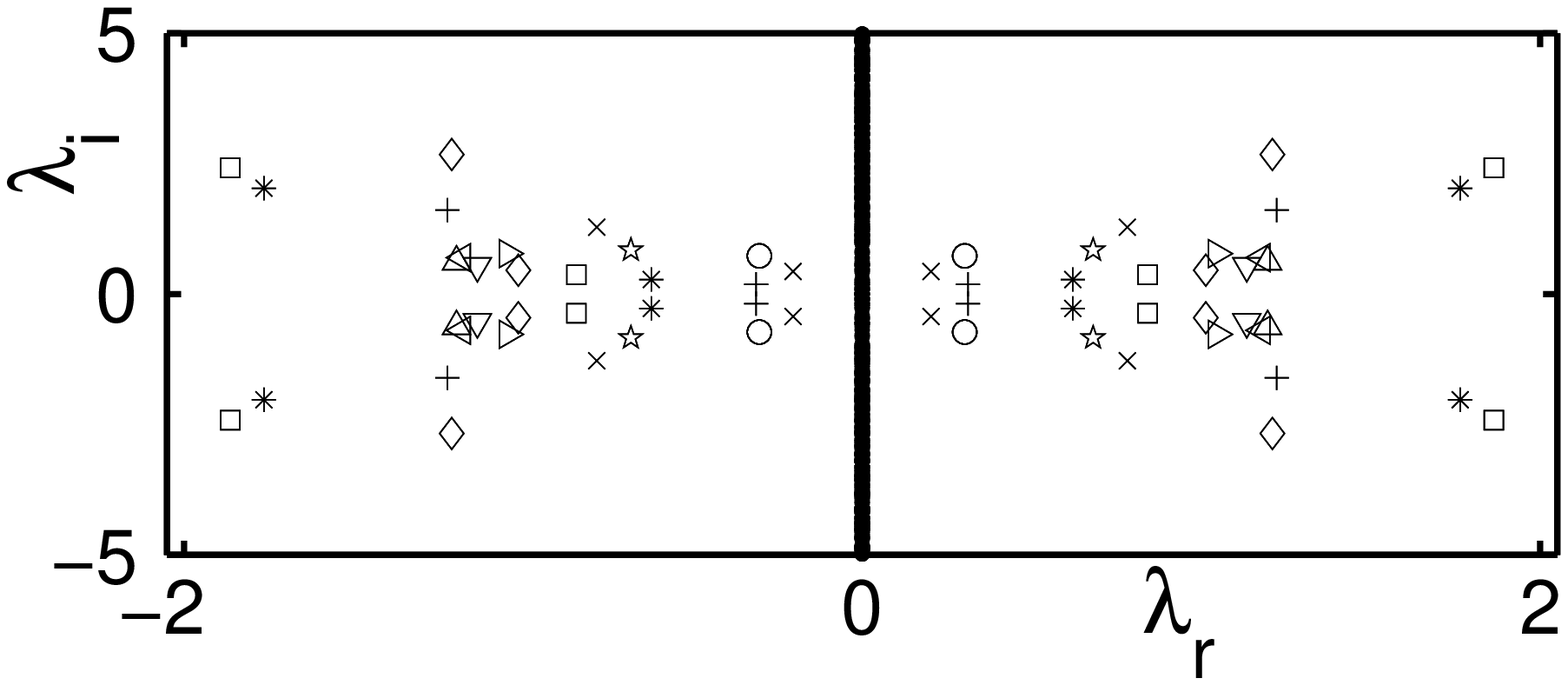}
\includegraphics[width=\wfig,angle=0,clip]{\rootfig 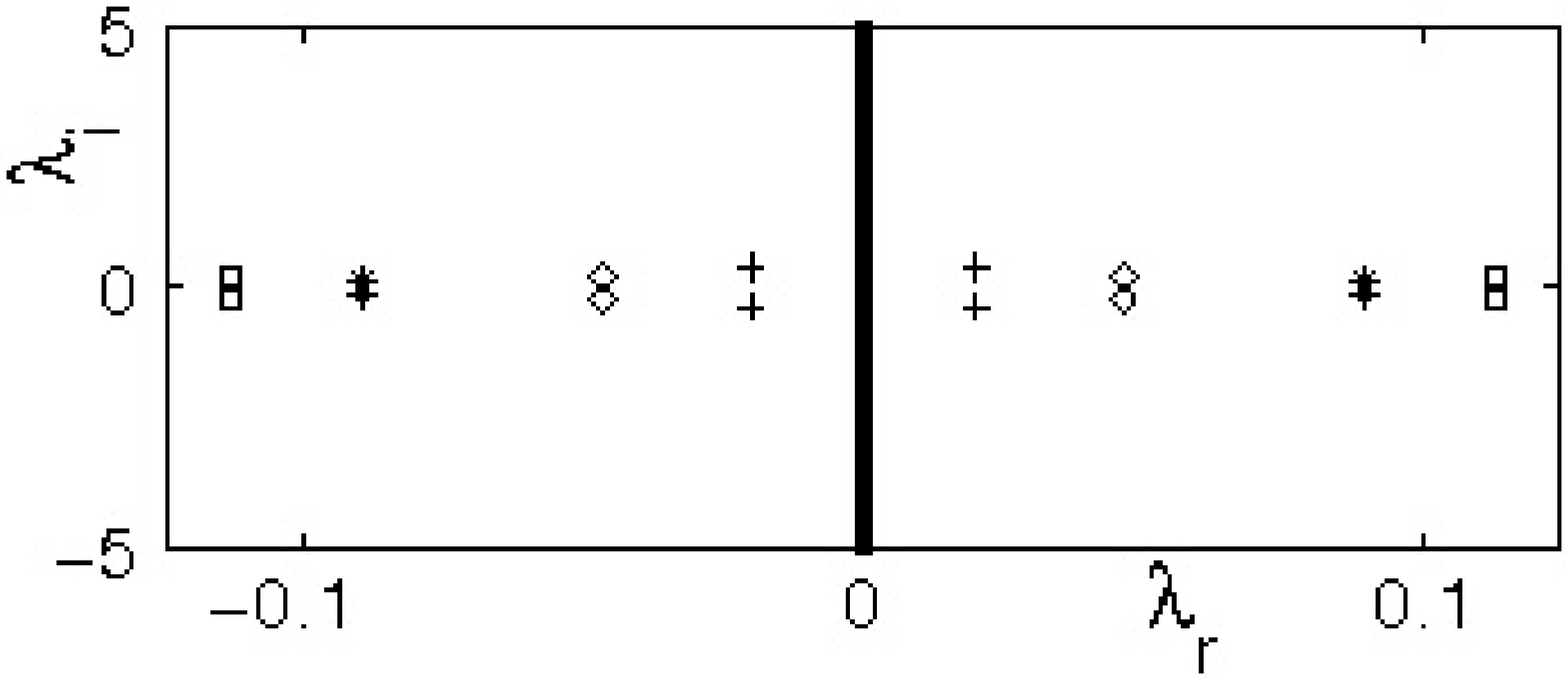}\\
~~\includegraphics[width=\wfigt,angle=0,clip]{\rootfig 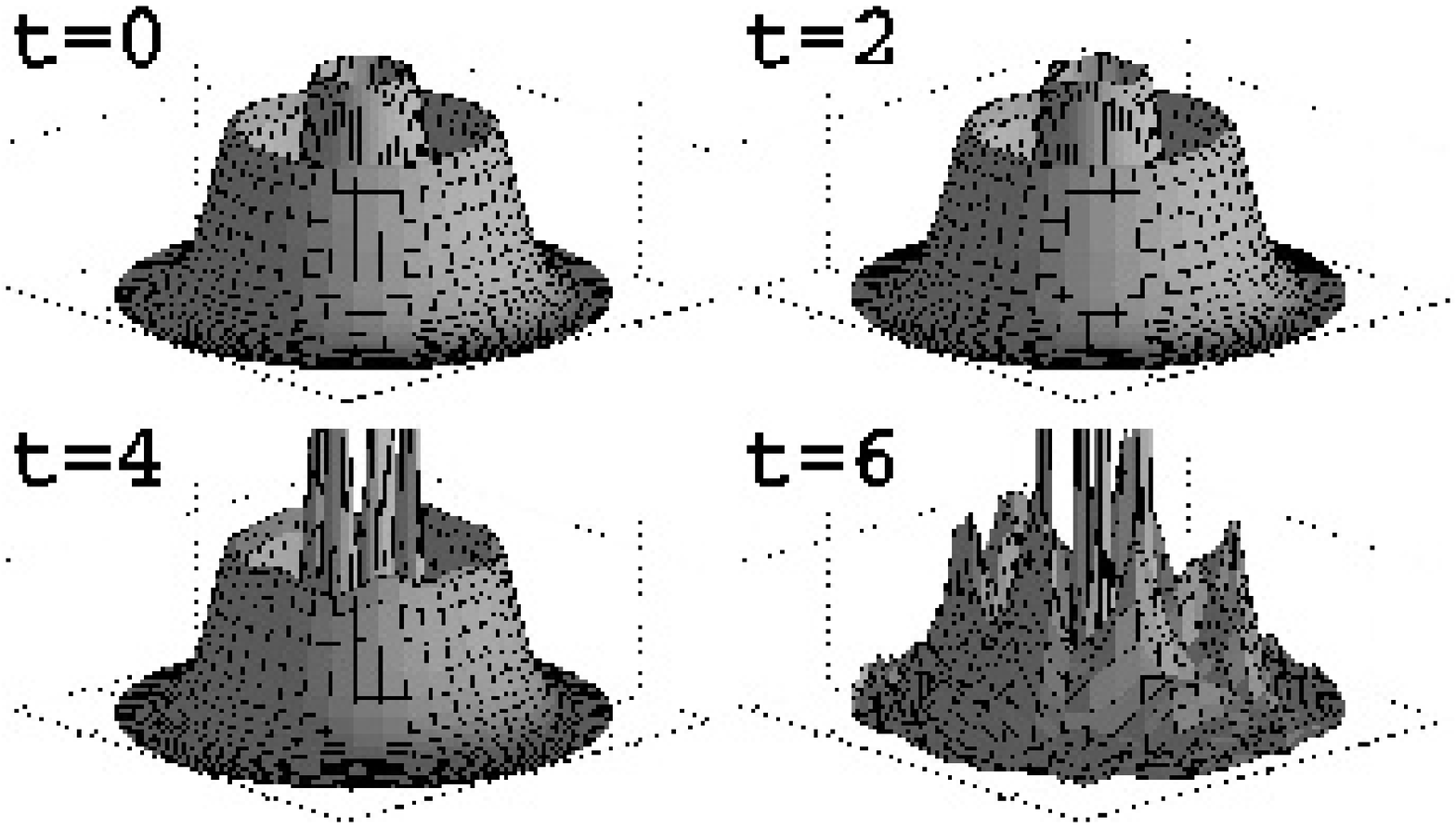}
~~~\includegraphics[width=\wfigt,angle=0,clip]{\rootfig 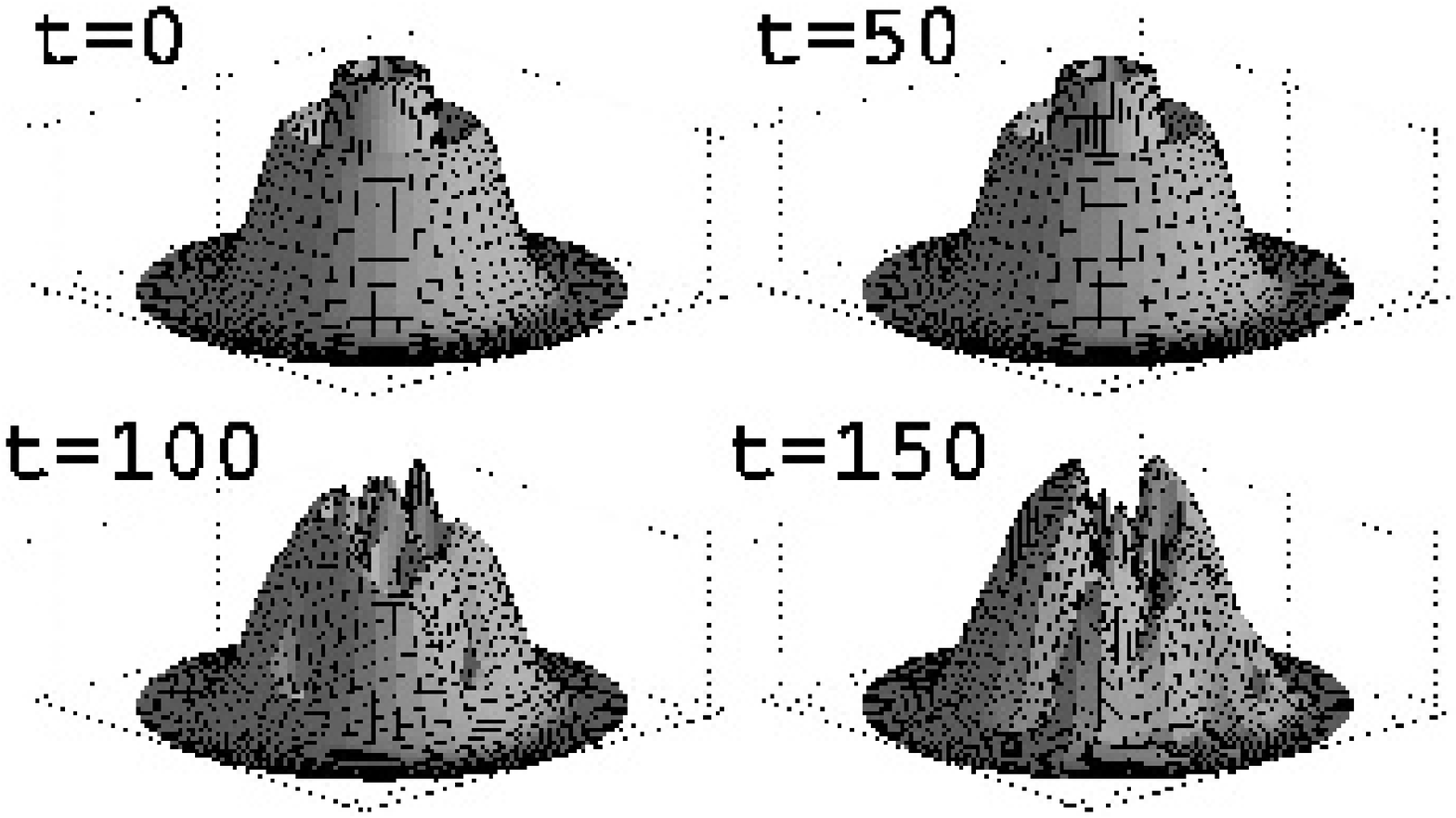}
\caption{Same as Fig.~\ref{Groundm2} ($m=2$) for the first excited state
($n_r=1$).}
\label{FirstExm2}
\end{figure}

\begin{figure}[tbp]
\includegraphics[width=\wfig,angle=0,clip]{\rootfigsmall  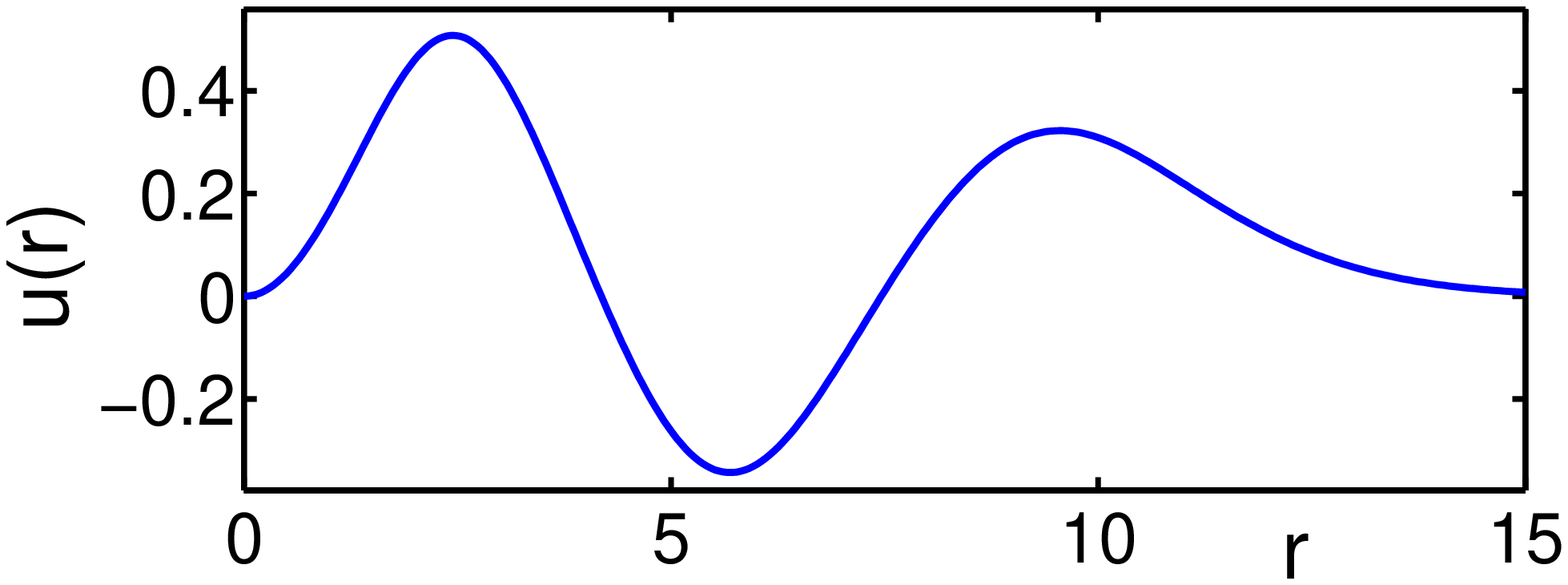}
\includegraphics[width=\wfig,angle=0,clip]{\rootfigsmall  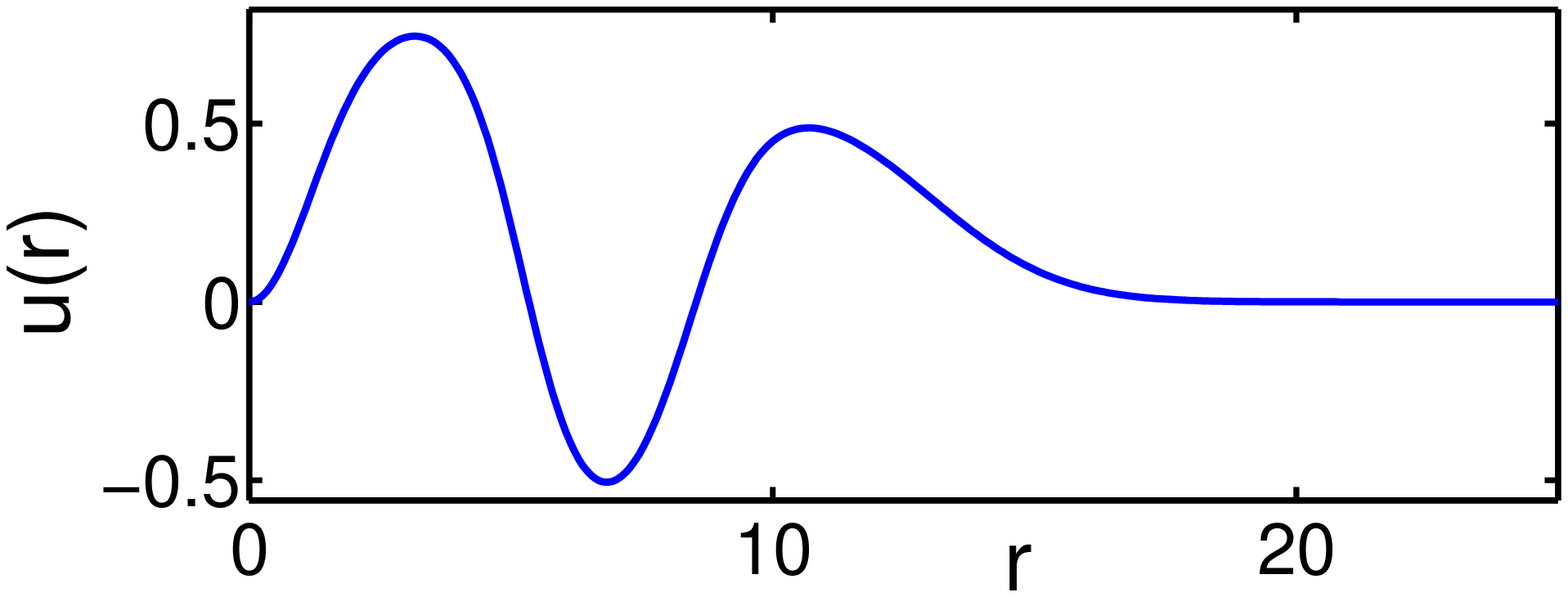}\\
\includegraphics[width=\wfig,angle=0,clip]{\rootfig 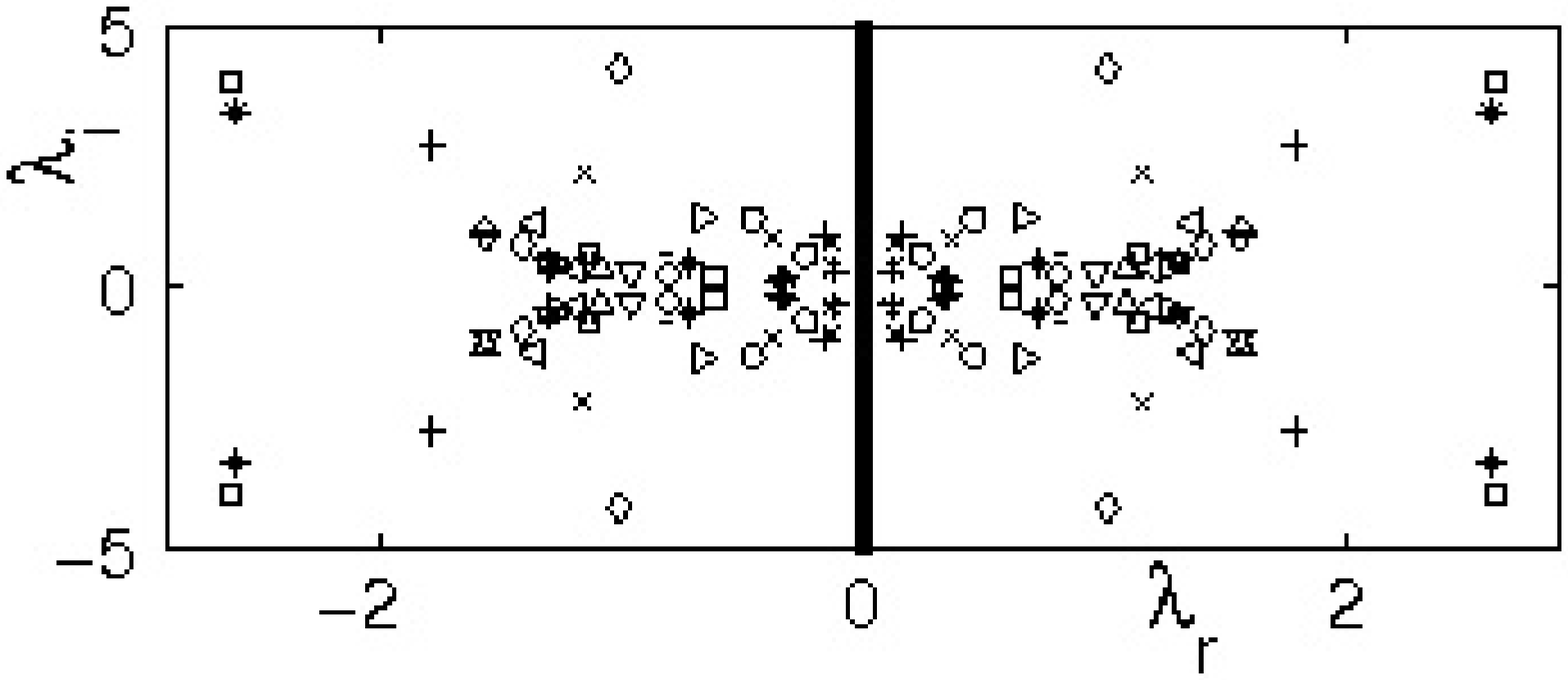}
\includegraphics[width=\wfig,angle=0,clip]{\rootfig 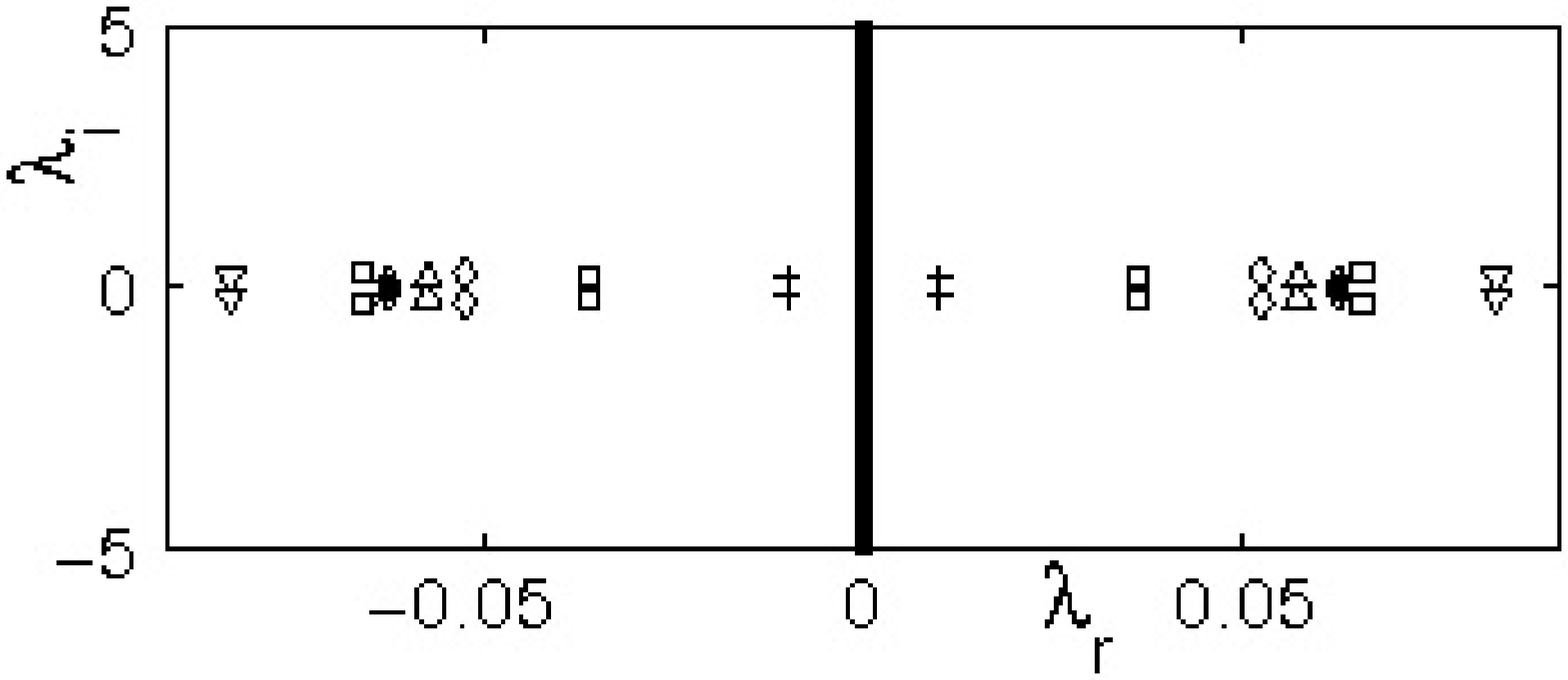}\\
~~\includegraphics[width=\wfigt,angle=0,clip]{\rootfig 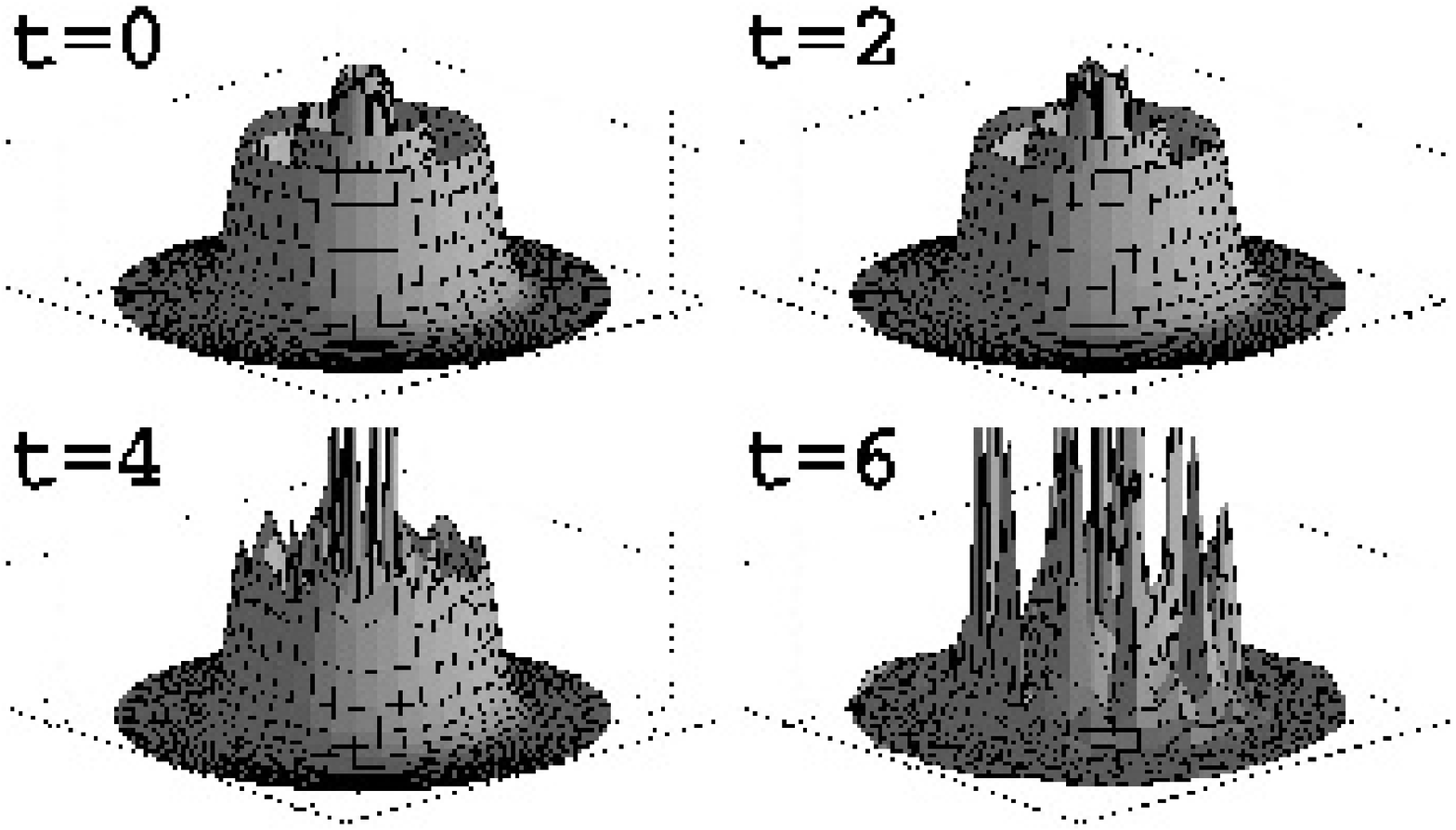}
~~~\includegraphics[width=\wfigt,angle=0,clip]{\rootfig 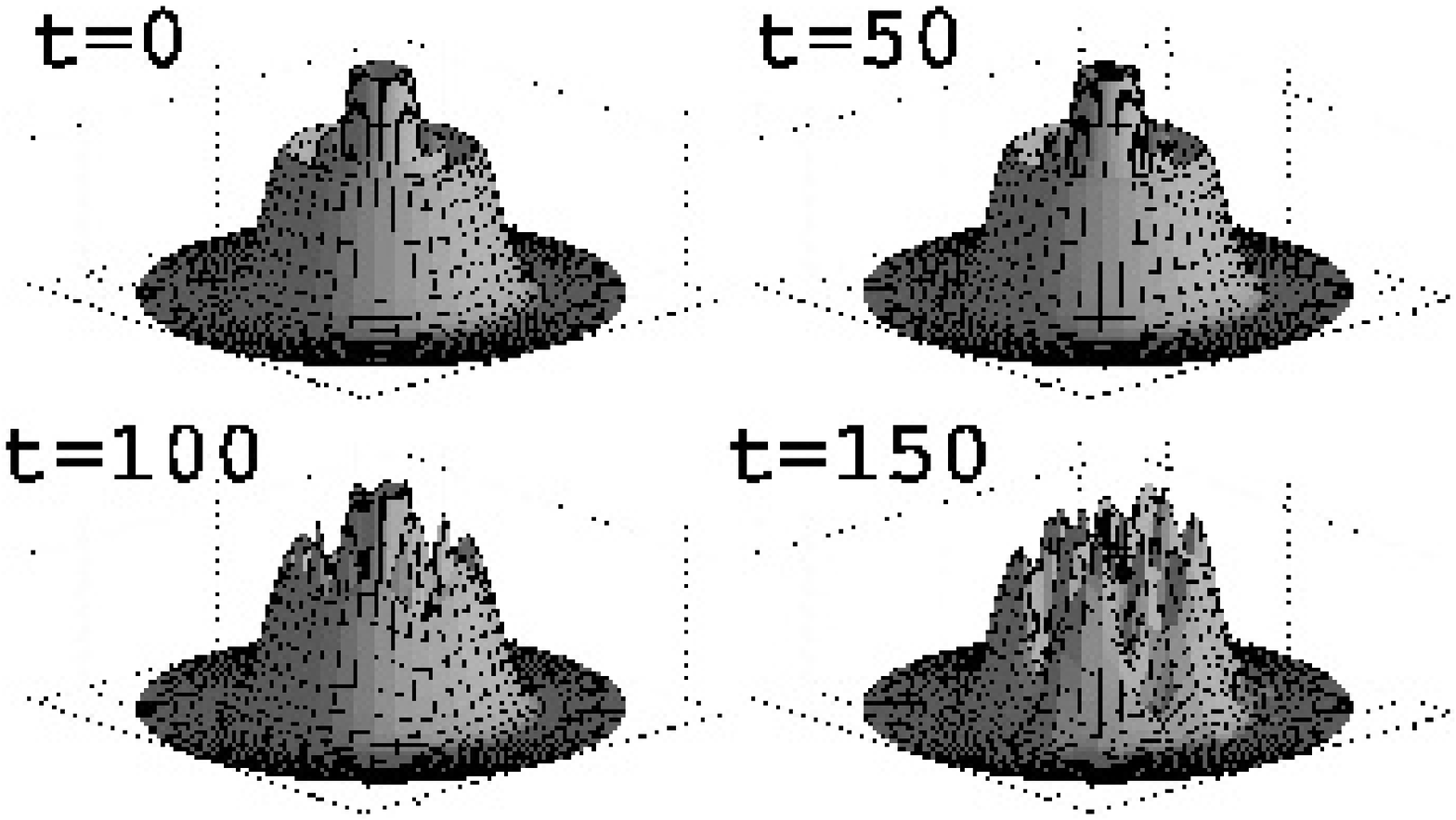}
\caption{Same as Fig.~\ref{Groundm2} ($m=2$) for the second excited state
($n_r=2$).}
\label{SecondExm2}
\end{figure}

Having offered an outline of the stability properties
of our solutions in Fig.~\ref{Branches},
we now examine the subject in detail in the following results.
In Figs.~\ref{Stability0}--\ref{Stability2} we depict the
real part of the primary ($q=0,1,2,3,4$) eigenvalues for
$m=0$, $m=1$, and $m=2$, respectively and for the states
with $n_r=0,1$ and $2$ in each case, i.e., the
ground state and first two excited
states in the top, middle and bottom panel of each figure.
The stability results showcase the presence of at least
one unstable eigenvalue.  In fact, there is only one
instability eigenvalue for the ground
state of the attractive case, and a larger number
of such eigenvalues for
excited states. The presence of such eigenvalues, whose larger
magnitude is tantamount to more rapid dynamical development
of the instability, indicates that the dynamical evolution
of such stationary states, when small perturbations are added
to them, will manifest the presence of such unstable eigenmodes
through the deformation and likely destruction of the initial structure.

This is shown in detail for $m=0$ in Figs.~\ref{Ground}--\ref{SecondEx}.
In the top row of panels in Fig.~\ref{Ground} we depict the radial profiles
of the solution for $\mu=-2$ and $\sigma=-1$ (left panel) and
$N|U|=100$ and $\sigma=+1$ (right panel) --- for all other figures
we use $\mu=-0.5$ when $\sigma=-1$. The middle row of panels
in the figure depicts their corresponding linear stability spectra
in the complex plane
$(\lambda_r,\lambda_i)\equiv({\rm Re}(\lambda),{\rm Im}(\lambda))$.
The unstable eigenvalues for different values of $q$ are given
different symbols as described in the table in Table~\ref{mytable}.
Eigenvalues with ${\rm Re}(\lambda)<10^{-7}$ or for $q>11$ are plotted
with a small dark dot.

Finally, the bottom row of panels in Fig.~\ref{Ground} depicts the
corresponding time evolution of the chosen profile after
a random perturbation of amplitude $10^{-2}$ was added to
the steady state profile at time $t=0$.
For the profile considered in Fig.~\ref{Ground}, namely the
ground state with $m=0$, it is clear that the solution
for the repulsive case ($\sigma=+1$) is stable
since it corresponds to the
Thomas-Fermi ground state.
On the other hand, the solution for the attractive case
($\sigma=-1$) is weakly unstable, due to the $q=0$ mode, as discussed
above.
The eigenvalues for this mode are depicted by
the circles in the middle-left panel. This
instability is due to the well-known collapse of the solution
for the attractive case, where the solution is seen to
tend to a thin spike carrying all the mass (see the time
evolution in the bottom-left panels).

In Figs.~\ref{FirstEx} and \ref{SecondEx} we present the
equivalent results for the {\em first} and {\em second}
excited states ($n_r=1$ and $2$ respectively) and for the same $m=0$ case.
In these cases the instability manifests itself
with the presence of a richer scenario of unstable
eigenvalues. Let us explain in detail the
first excited state in Fig.~\ref{FirstEx}. The first
excited state in the attractive case (left panels)
is still prone to collapse as evidenced by the
quartet of $q=0$ eigenvalues depicted by the
circles in the middle panel. This quartet
is responsible for the collapse of the central
spike of the solution as seen in the time series
evolution (bottom panels). Furthermore, the
$q=3$ and $q=4$ modes ($\varhexstar$ and $\Square$
symbols in the figure) are the most unstable ones
and are responsible for the azimuthal modulational
instability of the first ring of the solution
as evident in the time evolution
(bottom panels).

On the other hand, for the repulsive case (right panels)
of the first excited state with $m=0$ it is clear that
the $q=0$ eigenvalue is stable as there is no
collapse in the repulsive case. The dynamic evolution
of the instability in this case leads to the competition between
the unstable eigenvalues $q=3$, $q=4$ and $q=2$ (in order
of strength) that is seen to be dominated by the most unstable
mode $q=3$, as can be evidenced by the three humps
(surrounding the central peak) displayed at $t=150$.

Finally, in Fig.~\ref{SecondEx} we depict similar results
for the second excited state ($n_r=2$) for $m=0$. As can
be seen from the figures, the more excited the state, the
richer the (in)stability spectra.  It is worth
noticing again that the attractive case is, as before, prone
to collapse due to a strong unstable $q=0$ mode while,
naturally, the repulsive case lacks this $q=0$ collapsing mode.
The richer set of eigenvalues for higher excited states
is easy to interpret since higher excited states possess
more radial nodal rings.

In general, for the repulsive cases where collapse is absent,
we observe that in all cases the solution
is subject to azimuthal modulations
that produce coherent structures reminiscent of the
``azimuthons'' of Ref.~\cite{azimuthons}. On the other hand,
in the attractive
cases collapse is ubiquitous,
due to the mode with $q=0$;  in addition, azimuthal
modulations emerge.

\begin{figure}[tbp]
\includegraphics[width=\wfigthree,angle=0,clip]{\rootfig 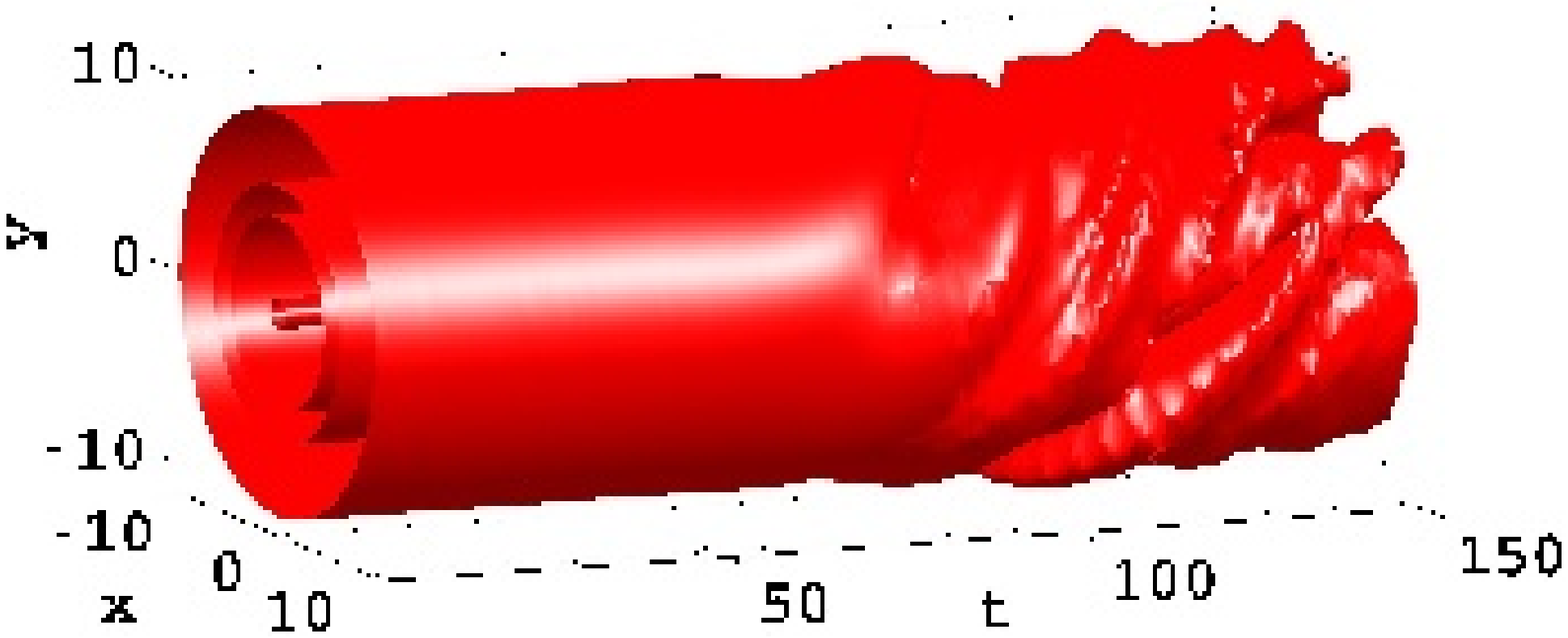}
\\[2.ex]
~~~\includegraphics[width=\wfigthree,angle=0,clip]{\rootfig 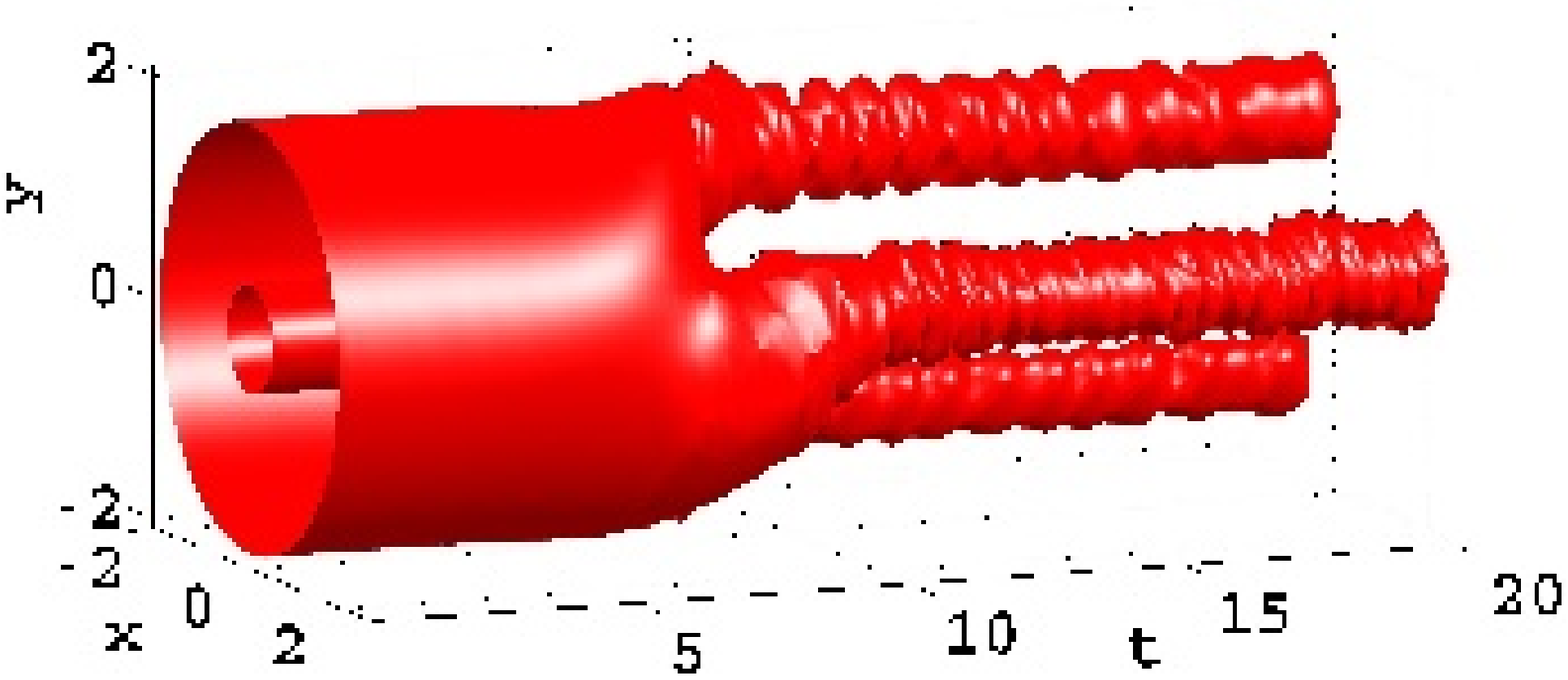}
\caption{(Color online)
Three-dimensional iso-density contour plots.
Top plot: iso-density contour for the
first excited state $n_r=1$ for $\sigma=+1$, $m=2$, i.e., the solution
corresponding to the right panel in Fig.~\ref{FirstExm2}.
Clearly observable is the rotation of the unstable $q=4$ mode
due to the intrinsic vorticity
($m=2$) of the solution.
Bottom plot: iso-density contour for the
ground state $n_r=0$ for $\sigma=-1$, $m=1$, i.e., the solution
corresponding to the left panel in Fig.~\ref{Groundm1}.
The rotation of the $q=3$ mode is quickly arrested and
the three spikes become stationary.
The contours correspond
to surfaces with density equal to half of the maximum density.
}
\label{LinSpM2B2Slice}
\end{figure}

The above discussion describes in detail the
stability and dynamics for the non-topologically
charged ($m=0$) solutions.
Let us now describe the case when the solutions have an
intrinsic topological charge, namely $m>0$.
In Figs.~\ref{Groundm1}--\ref{SecondExm1} we display the
results for the ground state, first and second excited
states ($n_r=0, 1$ and $2$ radial nodes)
with unit topological charge (or vorticity),
namely $m=1$. In fact,
Fig.~\ref{Groundm1} corresponds to the singly charged
vortex solution at the center of the harmonic
trap. As it is well known, this solution is stable in
the repulsive case (see right panels) and unstable
in the attractive case (see left panels). It is interesting
to note that the instability of the vortex solution in
the attractive case is not driven by collapse ($q=0$
eigenvalue) since the vorticity tends to push the
mass away from the center ($r=0$) of the trap.
This is a general feature of the cases with $m>0$, where
collapse appears to arise in the rings of the cloud,
rather than at its center.
Finally, in Figs.~\ref{Groundm2}-\ref{SecondExm2} we display
similar results for the doubly charged case $m=2$.

It is worth mentioning that the unstable modes for
vorticity-carrying solutions ($m>0$) tend to rotate as they are growing.
This effect can be clearly seen in the three-dimensional
iso-density contour plot depicted in the
top panel of Fig.~\ref{LinSpM2B2Slice}
corresponding to the first excited state with $m=2$ and
$\sigma=+1$. All modes tend to stop rotating after the
spikes created reach a certain height, as can
be seen clearly in the bottom panel in Fig.~\ref{LinSpM2B2Slice}.
This figure shows the ground state with $m=1$ and $\sigma=-1$.
Interestingly, a single solution with multiple radial nodes
(excited states) is able to pick out more than
one growing mode since each ring can be affected by a
different $q$-mode. This effect is seen
in the top-left panels in Fig.~\ref{FirstExm2} where
the first excited state for $m=2$ and $\sigma=-1$ is
seen to develop, at earlier times, the $q=4$ mode in the inner
ring while, at later times, the outer ring develops the
$q=6$ mode.

\section{Conclusions \label{sec:conc}}

In this paper, we have revisited the topic of nonlinear continuation of
linear, radially-dependent Laguerre-Gauss
states of the two-dimensional quantum harmonic oscillator model
in the presence of inter-particle interaction-induced nonlinearity.
We have systematically constructed such solutions starting from the
linear limit and, more importantly, we have detailed their linear
and nonlinear stability properties.
This was accomplished
by careful examination of the corresponding eigenvalue problem or,
more appropriately, the (infinite) one-parameter family 
of eigenvalue problems, in radial coordinates. We have also provided
a full numerical time evolution of the model on a radial-polar
grid. We have principally
observed that the ground state of the attractive case is unstable
due to collapse, although the growth rate of the instability may be
weak.
For the repulsive case the
relevant state is stable for topological charges $m=0$ and $m=1$; for
higher charges the stability depends on the atom number.
Excited states have been found to
be generically unstable in both attractive and repulsive
cases; the development of the instabilities produces collapse
in the former, while it results in the formation of azimuthally
modulated states in the latter.

It would certainly be of interest to extend the present techniques
to the
full 3D problem, again considering radial states and their continuation
from the linear limit, as well as their linear stability.
However, in the latter setting direct numerical computations
are significantly more intensive. Such studies are outside
of the scope of the present work and
will be considered in a future publication.

\section*{Acknowledgments}
P.G.K.~gratefully acknowledges the support of NSF-DMS-0204585 and
NSF-CAREER. P.G.K.~and R.C.G.~acknowledge the
support of NSF-DMS-0505663, and L.D.C.~acknowledges support under
NSF grant PHY-0547845 as part of the NSF CAREER program.
L.D.C.~acknowledges useful discussions with Charles Clark and Mark Edwards.

\section*{Appendix: Spectral Methods \label{sec:app}}

When dealing with the Laplacian in polar coordinates, the origin presents a significant challenge due to division by zero:
\begin{equation}
\Delta u = \frac{{\partial ^2 u}}{{\partial r^2 }}
 + \frac{1}{r}\frac{{\partial u}}{{\partial r}}
 + \frac{1}{{r^2 }}\frac{{\partial ^2 u}}{{\partial \theta ^2 }}.
\label{PolarLaplacian}
\end{equation}
This problem has been faced with in the past
\cite{tristram,tristram2,jpb,carr}, with respect to the GP
equation, but the methods used have so far limited the scope of
such investigations.  However, spectral methods can be used to avoid
the $r=0$ singularity to handle the Laplacian in studies of the GP
equation.

\begin{figure}[tbp]
\includegraphics[width=\wfigtwo,angle=0,clip]{\rootfigsmall 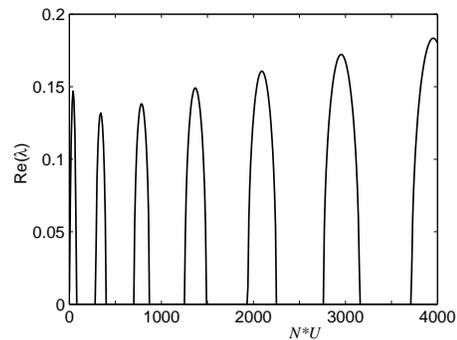}
\caption{
Recreated plot of Figure 1b in Ref.~\cite{pu} using spectral methods.
The figure shows the unstable part of the main eigenvalue
as a function of the nonlinear strength $N\times U$ for a doubly quantized
vortex ($m=2$).
}
\label{PuPic}
\end{figure}

The particular application  of spectral methods we used in our
calculations is described by Trefethen in Ref.~\cite{trefethen},
but is restated here for completeness.  This method avoids the
problem with the origin by avoiding the origin altogether through the use of
an even number of Chebyshev nodes.
In order to check for the accuracy of utilizing spectral methods for the purposes
of modeling the GP equation in polar coordinates, we recreated the results of
Fig.~1b in Ref.~\cite{pu}.
As can be seen from Fig.~\ref{PuPic}, our spectral method
code has generated a plot in close agreement with the original plot.

Spectrally discretizing the polar domain is achieved by discretizing the radial direction $r$ using polynomial interpolation, in particular the Chebyshev nodes:
\begin{equation}
\begin{array}{cc}
r_j=\cos(j\pi/N) & j=0,1,...,N,
\end{array}
\label{ChebPoints}
\end{equation}
meaning the domain must be normalized in order for the solution to be properly contained within $r = [0,1]$, while the angular direction $\theta$ is discretized using Fourier (or trigonometric) interpolation:
\begin{equation}
\begin{array}{cc}
\theta_j=2j\pi/N & j=0,1,...,N.
\end{array}
\label{FourierPoints}
\end{equation}
These choices for discretization are based upon the periodicity of the angular direction and the lack thereof in the radial direction (which causes the Gibbs phenomenon to occur if Fourier interpolation is used).  Additionally, the Chebyshev polynomials
\begin{equation}
T_k (x) = \cos (k\cos ^{ - 1} (x)),
\label{ChebPolys}
\end{equation}
can be thought of as a cosine Fourier series ($x = \cos z$), meaning they can easily be shown to possess similar results for accuracy and convergence as the Fourier method \cite{trefethen, SpecMethCFDs}.

Implementation is simply done by using the nodes described above along with their corresponding derivative matrices.  For the Chebyshev nodes, the Chebyshev differentiation matrix is given by:
\begin{equation}
D_N  = [D_{i,j} ] = \left\{ {
\begin{array}{ll}
\displaystyle
   {\frac{{c_i }}{{c_j }}\frac{{( - 1)^{i + j} }}{{(x_i  - x_j )}}} & {i \ne j},  \\[3.0ex]
\displaystyle
   {\frac{{ - x_j }}{{2(1 - x_j^2 )}}} & {i = j = 1,...,N - 1},  \\[3.0ex]
\displaystyle
   {\frac{{2N^2  + 1}}{6}} & {i = j = 0},  \\[3.0ex]
\displaystyle
   { - \frac{{2N^2  + 1}}{6}} & {i = j = N},
\end{array}
} \right.
\label{ChebDeriv}
\end{equation}
where
\begin{equation}
c_k  = \left\{ {
\begin{array}{ll}
   2 & {k = 0,N},  \\[2.0ex]
   1 & {\rm otherwise},
\end{array}
} \right.
\end{equation}
while the Fourier differentiation matrix (for an even number of Fourier nodes) is given by:
\begin{equation}
\begin{array}{rcl}
D_N &=& [D_{i,j} ] \\[2.0ex]
    &=& \left\{ {
\begin{array}{ll}
   0 & {(j - i) \equiv 0~(\bmod N)}  \\ [2.0ex]
   {\frac{( - 1)^{j - i}}{2} \cot \left( {(j - i)\frac{h}{2}} \right)} & {(j - i)\not  \equiv 0~(\bmod N)}.
\end{array}
} \right.
\end{array}
\label{FourierDeriv}
\end{equation}
It should be noted that the differentiation matrix for an odd number of Fourier nodes differs from the one above due to a correction which must be made in the derivation of the differentiation matrix for an even number of Fourier nodes \cite{trefethen}, but we have chosen to restrict ourselves to using even numbers of Fourier nodes.

Typically when using polar coordinates, the domain of the radial direction is restricted to $r \in [0,\infty]$ since any function on the unit disc must inherently satisfy the symmetry condition:
\begin{equation}
u(r,\theta ) = u( - r,(\theta  + \pi )(\bmod 2\pi )),
\label{SymCond}
\end{equation}
and since we have chosen to use an even number of Fourier nodes,
our grid must also satisfy this condition.  As a result, solution
values would be recorded and evaluated twice using this grid.
Initially, this may appear to be a significant disadvantage for
this discretization, but an elegant simplification can be arrived
at using a symmetry property of the Chebyshev spectral
differentiation matrix: $D_N (i,j) =  - D_{N} (N - i,N - j)$.

Using both symmetries, the derivative of $u$ in the radial direction can be shown to be exactly the same for both occurrences of the node:
\begin{align}
u'_r(r_i)
&= \sum\limits_{j = 0}^N {D_N (i,j)u(r_j ,\theta )} \notag \\
&= \sum\limits_{j = 0}^N { - D_N (N\! -\! i,N\! -\! j)u( - r_j ,(\theta  + \pi )(\bmod 2\pi ))} \label{DerivSym} \notag \\
&=  - u'_r ( - r_j ,(\theta  + \pi )(\bmod 2\pi )).
\end{align}

In addition, the derivative calculation can be restricted to using only the positive
$r$-axis by using (assuming $r_i > 0$):
\begin{align*}
u'_r (r_i ,\theta )
&= \sum\limits_{j = 0}^N {D_N (i,j)u(r_j ,\theta )} \\[2.0ex]
&= \sum\limits_{j = 0}^{\tfrac{{N - 1}}{2}} {D_N (i,j)u(r_j ,\theta )}  \\
&~~+ \sum\limits_{j = \tfrac{{N - 1}} {2} + 1}^N {D_N (i,j)u(r_j ,\theta )} \\[2.0ex]
&= \sum\limits_{j = 0}^{\tfrac{{N - 1}}{2}} {D_N (i,j)u(r_j ,\theta )}  \\
&~~+ \sum\limits_{j = \tfrac{{N - 1}}{2} + 1}^N {D_N (i,j)u( - r_j ,(\theta  + \pi )(\bmod 2\pi ))} \\[2.0ex]
&= \sum\limits_{j = 0}^{\tfrac{{N - 1}}{2}} {D_N (i,j)u(r_j ,\theta )}  \\
&~~+ \sum\limits_{j = 0}^{\tfrac{{N - 1}}{2}} {D_N (i,N - j)u(r_j ,(\theta  + \pi )(\bmod 2\pi ))}.
\end{align*}

The same simplification can be inductively shown to work for higher derivatives since they are calculated by multiple applications of the derivative matrix
\begin{equation}
D_2 u = D^2 u = D (Du),
\nonumber
\end{equation}
and Eq.~(\ref{DerivSym}) has already established the symmetry of the first derivative, thus by induction, all higher derivatives can be calculated using only the positive $r$-axis.  Consequently, any simulations based upon a Chebyshev-Fourier polar grid can be written to strictly use the positive $r$-axis, resolving the computational time and storage problems, but still benefiting from the advantages of spectral methods.  (See Chapter 11 of Ref.~\cite{trefethen} for an implementation of this method in Matlab.)

Spectral methods can also be used to evaluate integrals with greater accuracy. In particular, this was used for calculating the power of the numerically determined steady state solutions. To further explore how spectral methods can be used for integration, take a generic integral
\begin{equation}
I(x) = \int\limits_{ - 1}^x {f(y)dy},
\label{GenIntegral}
\end{equation}
for some sufficiently smooth function $f$.  This integral can then be restated in the form of an initial value ODE
\begin{equation}
\begin{array}{ll}
\displaystyle
   {\frac{{dI}}{{dx}} = f(x)}, ~ & ~{I( - 1) = 0},  \\
\end{array}
\label{IntegralODE}
\end{equation}
and discretized using Chebyshev nodes, thus replacing the derivative by the Chebyshev spectral differentiation matrix $D_N$
\begin{equation}
\tilde D_N I = f,
\label{DiscIntODE}
\end{equation}
where $f$ here is the discrete version of the integrand above, $f
= (f(x_0),...,f(x_{N-1}))^T$, and the boundary condition,
$I(-1)=0$, is imposed by stripping off the last row and column of
$D_N$ to obtain the $N \times N$ matrix $\tilde{D}_N$.  The value
of the integral at every Chebyshev node is then easily found by
inverting the derivative matrix to get $I = \tilde{D}_N^{-1} f$.
However, the only value we are interested in is $I(1)$ (i.e. the
integral over the entire domain) and this can easily be found by
using only the first row of $\tilde{D}_N$, call it $w$: $I(1) =
wf$.

\end{document}